\documentclass[fleqn,usenatbib]{mnras}

\usepackage{newtxtext,newtxmath}
\usepackage{amsmath}
\usepackage{float}
\usepackage{graphicx}

\usepackage[T1]{fontenc}
\DeclareRobustCommand{\VAN}[3]{#2}
\let\VANthebibliography\thebibliography
\def\thebibliography{\DeclareRobustCommand{\VAN}[3]{##3}\VANthebibliography}

\usepackage{graphicx}	
\usepackage{amsmath}	
\usepackage{xcolor}
\usepackage{comment}

\title[Hidden variability in accreting X-ray binaries]{Unveiling hidden variability components in accreting X-ray binaries using both the Fourier power and cross spectra}

\author[M. M\'endez et al.]
{Mariano M\'endez$^{1}$\thanks{mariano@astro.rug.nl}, 
Valentina Peirano$^{1}$, 
Federico Garc\'{\i}a$^{2}$,
Tomaso Belloni$^{3}$,  
Diego Altamirano$^{5}$,
\newauthor Kevin Alabarta$^{6}$ 
\\
$^{1}$Kapteyn Astronomical Institute, University of Groningen, P.O. Box 800, 9700 AV Groningen, The Netherlands\\
$^{2}$Instituto Argentino de Radioastronom\'{\i}a (CCT La Plata, CONICET; CICPBA; UNLP), C.C.5, (1894) Villa Elisa, Buenos Aires, Argentina\\
$^{3}$INAF - Osservatorio Astronomico di Brera, Via E. Bianchi 46, I-23807, Merate, Italy\\
$^{4}$School of Physics and Astronomy, University of Southampton, Southampton, Hampshire SO17 1BJ, UK\\
$^{5}$Center for Astrophysics and Space Science (CASS), New York University
Abu Dhabi, PO Box 129188, Abu Dhabi, UAE\\
$^{\dagger}$Deceased
}

\pubyear{2023}

\begin{document}
\label{firstpage}
\pagerange{\pageref{firstpage}--\pageref{lastpage}}
\maketitle

\begin{abstract}
We present a novel method for measuring the lags of (weak) variability components in neutron-star and black-hole low-mass X-ray binaries (LMXBs). For this we assume that the power and cross spectra of these sources consists of a number of components that are coherent in different energy bands, but are incoherent with one another. The technique is based on fitting simultaneously the power spectrum (PS) and the Real and Imaginary parts of the cross spectrum (CS) with a combination of Lorentzian functions. We show that, because the PS of LMXBs is insensitive to signals with a large Imaginary part and a small Real part in the CS, this approach allows us to uncover new variability components that are only detected in the CS. We also demonstrate that, contrary to earlier claims, the frequency of the type-C quasi-periodic oscillation (QPO) in the black-hole binary GRS 1915+105 does not depend on energy. Rather, the apparent energy dependence of the QPO frequency can be explained by the presence of a separate QPO component with a slightly higher frequency than that of the QPO, whose rms amplitude increases faster with energy than the rms amplitude of the QPO. From all the above we conclude that, as in the case of the PS, the CS of black-hole and neutron-star binaries can be fitted by a combination of Lorentzian components. Our findings provide evidence that the frequency-dependent part of the transfer function of these systems can be described by a combination of responses, each of them acting over relatively well-defined time scales. This conclusion challenges models that assume that the main contribution to the lags comes from a global, broadband, transfer function of the accreting system.
\end{abstract}


\section{Introduction}
\label{introduction}

The X-ray light curves of accreting black-hole and neutron-star X-ray binaries show a complex pattern of variability with time scales ranging from milliseconds to years \citep[e.g.][and references therein]{Levine-2006,Belloni-2011,Ingram-2019a}. In the last four decades our understanding of the properties of these sources has advanced significantly due to the realisation that their power spectra can be decomposed into several variability components with well-defined properties \citep[e.g.,][]{vanderKlis-1994,Nowak-2000} that correlate with each other \citep{Belloni-2002}, and other source properties (see below). 

An effective approach to explore the geometry of the accretion flow in these systems is through the energy- and frequency-dependent phase lags of the variability \citep{vanderKlis-1987,Nowak-1999a}. The phase lags measure the phase angle in the complex Fourier plane of the cross vector of correlated signals measured in two energy bands as a function of Fourier frequency \citep{Vaughan-1997,Nowak-1999a}. 

In the last decade, several models have been proposed to constrain the physical and geometrical properties of the accretion flow from measurements of the rms amplitude and lag spectra of these sources \citep[e.g.][]{Ingram-2011,Ingram-2016,Mastroserio-2018,Kylafis-2020,Karpouzas-2020,Bellavita-2022}. While the rms amplitude of the different variability components can be obtained from fits to the power spectra of these sources with a combination of Lorentzian functions \citep{Nowak-2000,Belloni-2002}, the lags are obtained from measurements that do not separate the contribution of those individual components \cite[e.g.,][but see \citealt{Nowak-1999b}]{vanderKlis-1987,Reig-2000}. 

Specifically, until now the method of measuring the lags of a broadband noise (BBN) component or a quasi-periodic oscillation (QPO) in the power spectrum of low-mass X-ray binaries (LMXBs) was the following: 

\begin{enumerate}
    \item Compute the power spectrum of the light curve of the source \citep[e.g.,][]{vanderKlis-1989} in the broadest energy band available. 
    
    \item Find the interesting components in the power spectrum \citep[e.g.,][]{Psaltis-1999}. In the case of a BBN component, identify the minimum and maximum frequency over which one wants to measure the lags \citep[e.g.][]{Reig-2000,Altamirano-2015,Kara-2019}. In the case of a QPO, find the centroid frequency, $\nu_0$, and the full width at half maximum (FWHM), $\Delta$, of the QPO to measure the lags in the range from $(\nu_0 - x\Delta)$ to $(\nu_0 + x\Delta)$, where usually $x=1/2$ \citep[e.g.,][]{Wijnands-1999} 
    
    \item Compute the cross spectrum using light curves in two energy bands \citep[e.g.,][]{vanderKlis-1987,Vaughan-1998,Nowak-1999a,Uttley-2014}. 
    
    \item Compute the average of the Real and Imaginary parts of the cross spectrum in the selected frequency range \citep[e.g.,][]{vanderKlis-1987,Lewin-1988,Reig-2000,Belloni-2021}. 
    
    \item Compute the ratio of the Imaginary to the Real parts and take the inverse tangent function of that to get the phase lag, $\Delta\phi$. 
    
    \item If one is interested in the time lag, $\Delta\tau$, in that frequency range, take some representative frequency in the range, e.g., $\langle \nu \rangle = (\nu_{\rm min} + \nu_{\rm max})/2$ in the case of the BBN, or $\nu_0$ in the case of the QPO, and divide $\Delta\phi$ by $2\pi$ times that frequency to get $\Delta\tau$ \citep[e.g.,][]{Nowak-1999a,deAvellar-2013,Barret-2013}. Here we call this the traditional method.

\end{enumerate}

When using this method, the (underlying) assumption is that the component of interest dominates in the power and the cross spectra over the frequency range of interest. This is usually true for kilohertz QPOs in neutron-star systems \citep[kHz QPOs;][]{Mendez-2021} and high-frequency QPOs in black-hole systems \citep{Morgan-1997,Belloni-2012,Mendez-2013}, since these QPOs appear in a frequency range in which the only other component present in the power spectrum is the Poisson noise and the QPO dominates. In many occasions this is also the case for low-frequency QPOs \citep[see][and references therein]{Belloni-2014,Ingram-2019a} that appear in a region of the power spectrum in which other components are present (e.g., a BBN component, sub-harmonics and second or higher harmonics of the QPO, or other timing components not harmonically related to the QPO of interest), when the QPO itself is significantly stronger than those other components \citep[e.g.,][]{Zhang-2017,Zhang-2020}. However, this is not always the case \citep[e.g.,][]{Nowak-1999a,vanStraaten-2002}, and if another component contributes significantly to the power and cross spectra in the frequency range of interest, the above procedure fails to give the lags of the component of interest. The situation is even worse when this other component dominates the variability in that frequency range \citep[e.g.,][]{Ma-2021,Alabarta-2022}.

In the past this problem has been tackled in different ways; for instance \cite{Ma-2021,Ma-2023} find the ``net'' lag spectrum of a QPO by subtracting from the lags of the QPO the average lags over a frequency range just outside the QPO. This method is mathematically incorrect: The two measured lags give the angles of the cross vectors in Fourier space, whereas the angle of the net cross vector is not the difference of the angles of the two individual cross vectors, but depends also on their magnitudes. Alternatively, one could subtract the Real and Imaginary parts of the cross spectrum outside the QPO instead of the lags, which would be mathematically correct (as far as we are aware this has not been done in the literature). However, unless the Real and Imaginary parts of the unwanted components are constant with frequency, their contribution over the frequency range of interest is not necessarily the same as what one measures over the adjacent frequency range. Furthermore, this procedure would over-correct the lags whenever the wings of the component of interest contribute significantly to the power in the adjacent frequency range. 

Another approach that has been used in the literature is to ignore the frequency range in which the unwanted components contribute significantly to the variability \citep[e.g.,][]{Wang-2021,Wang-2022}. This, however, does not work when parts of the unwanted components contribute significantly to the frequency range of interest \citep[see, for instance, Fig. 4 in][and the discussion in \S\ref{hidden}]{Wang-2021}. In all these cases, because the power and cross spectra of these sources contain many overlapping components  \citep[e.g.,][]{Nowak-2000,Zhang-2020}, it is often impossible to find a ``clean'' frequency interval in which one can measure the variability of the component of interest without contamination from other unwanted components. 

If one has a model of the cross spectrum both as a function of energy and Fourier frequency, instead of extracting the lags one can fit the model directly to the Real and Imaginary parts of the cross spectrum as a function of energy in a number of frequency bands \citep[e.g.,][]{Mastroserio-2018} or vice versa \citep[e.g.,][]{Rapisarda-2014}. For this, one uses the transfer function of the system \citep[e.g., between the corona and the accretion disc,][we expand further on the transfer function in \S\ref{TF}]{Reynolds-1999} as a function of photon energy and Fourier frequency, which is a mathematically valid approach as long as the model of that transfer function includes all the components that contribute to the variability. An advantage of this is that one can build the possible interactions of the individual components in the model of the transfer function \citep[e.g.,][]{Ingram-2013, Rapisarda-2014}. Some models, however, only consider the global (broadband in Fourier frequency) transfer function of the system, and explicitly exclude the frequency range in which the variability is dominated by narrow components, like QPOs \citep[e.g.,][]{Mastroserio-2018,Ingram-2019c}.

It is traditional to fit power spectra of LMXBs with a linear combination of Lorentzian functions \citep[usually called a multi-Lorentzian model; see, e.g.,][]{Nowak-2000,Belloni-2002}. The centroid frequency, $\mathrm{FWHM}$, rms fractional amplitude, and rms and lag spectra of each of these Lorentzians correlate with each other \citep[e.g.,][]{vanStraaten-2002,vanStraaten-2003,Casella-2004,Altamirano-2005,Altamirano-2008,Motta-2011}, with properties of other Lorentzians in the power spectrum \citep[e.g.,][]{Mendez-2001}, with the total intensity and different hardness ratios of the source \citep[e.g.,][]{Mendez-1999,Jonker-2002,vanStraaten-2005}, with the spectral parameters of the different components used to fit the total energy spectrum of the source \citep[e.g.,][]{DiMatteo-1999,Homan-2001,Remillard-2002,Pottschmidt-2003,Vignarca-2003,Zdziarski-2005,Shaposhnikov-2009,Motta-2009,Motta-2010,Motta-2011,Reig-2013,Stiele-2013,Shidatsu-2014,Grinberg-2014,Altamirano-2015,Kalamkar-2015,Reig-2018,deMarco-2015,deMarco-2017,deMarco-2021}, with the fluxes of those components \citep[e.g.,][]{Markwardt-1999,Sobczak-2000} and the total flux \citep[e.g.,][]{Sobczak-2000,Remillard-2002} or luminosity \citep{Ford-2000} and with radio properties of the source \citep[e.g.,][]{Muno-2001,Fender-2004,Mendez-2022,Garcia-2022}. These correlations (the list of references in this paragraph is certainly incomplete, but it is impossible to do justice to all the papers that discuss this topic) offer compelling evidence that those Lorentzian components are not just a useful empirical description of the power spectrum, but they represent physical phenomena in the system with (rather) well-defined characteristic time scales. 

In the case of GX 339-4, \cite{Nowak-1999b} observed that whenever one Lorentzian component dominates the power spectrum (PS) coherence function (see \S\ref{mathematical}) approaches unity and the phase lags exhibit a relatively flat `shelf’; whenever two Lorentzian components intersect in the PS, there is a decline in the coherence function and a transition from one characteristic phase-lag shelf to another \citep[c.f., for instance, Figs. 1 and 10 in][]{Nowak-1999a}. This result indicates that, at least in this source, there is a relation between the individual components that fit the PS and the lags in the cross spectrum (CS).

Here we propose a new method to measure the lags of these components in the cross spectrum of LMXBs. In \S\ref{mathematical} we describe the mathematical foundation of this approach. In \S\ref{examples} we give examples of the application of the new proposed method to a number of cases, and demonstrate the advantages of this method in comparison with the traditional one. In fact, in doing this we unveil new variability components that have gone undetected before, and we show that ignoring these components led to wrong conclusions regarding some characteristics of the QPOs. In \S\ref{discussion} we summarise our findings and discuss some unexpected consequences of our results that have significant impact upon models of the variability that have been presented in the literature. Finally, in \S\ref{conclusions} we sum up our conclusions.

\section{Mathematical formalism}
\label{mathematical}

Suppose we have two noiseless\footnote{Although we discuss the case of noiseless continuous functions, the same formalism applies to discrete data that include noise, as shown in \cite{Bendat-2010} and \cite{Vaughan-1997}.} time series, $x(t)$ and $y(t)$, that represent the X-ray intensity of a variable source measured in two different energy bands, with corresponding complex Fourier transforms $X(\nu)$ and $Y(\nu)$. We define\footnote{Here we mostly use the notation of \cite{Bendat-2010}. In the notation of \cite{Vaughan-1997} $\nu=f$, $G_{xx}(\nu) = \langle\left|S_1(f) \right|^2\rangle$, $G_{yy}(\nu) = \langle\left|S_2(f) \right|^2\rangle$, and $G_{xy}(\nu) = \langle C(f) \rangle$.} the power spectrum of each of these series, $G_{xx}(\nu) = \langle X(\nu) X^*(\nu) \rangle = \langle |X(\nu)|^2 \rangle$, $G_{yy}(\nu) = \langle Y(\nu) Y^*(\nu) \rangle = \langle |Y(\nu)|^2 \rangle$, and the cross spectrum between the two series, $G_{xy}(\nu) = \langle X(\nu) Y^*(\nu) \rangle = \langle |X(\nu)||Y(\nu)| e^{i\Delta\phi_{xy}(\nu)} \rangle$, where $\Delta\phi_{xy}(\nu)$ is the phase lag between the two series at frequency $\nu$, and the angle brackets indicate averaging over an ensemble of measurements of $X(\nu)$ and $Y(\nu)$ \citep[see][for details]{Bendat-2010}. We can then define the coherence function between the two series \citep{Bendat-2010,Vaughan-1997,Nowak-1999a}:
\begin{equation}
    \gamma^2_{xy}(\nu) = \frac{|G_{xy}(\nu)|^2}{G_{xx}(\nu) G_{yy}(\nu)}.
    \label{eq1}
\end{equation}

From the above definition it is clear that if the two time series are related by a linear transformation, $G_{yy}(\nu) = |H(\nu)|^2 G_{xx}(\nu)$ and $G_{xy}(\nu) = H(\nu) G_{xx}(\nu)$, the coherence function is unity at all frequencies. The complex function $H(\nu) = |H(\nu)| e^{-i\Delta\phi(\nu)}$ is the frequency response function, sometimes also called the transfer function, of the system \citep[c.f.,][see \S\ref{TF} for more details of the possible interpretations of the transfer function]{Bendat-2010}.

Let us consider that the time series $x(t)$ and $y(t)$ can be decomposed in a finite number of individual components that are significant over a limited frequency range. It is traditional to fit the power spectrum of LMXBs with a linear combination of Lorentzian functions \citep{Nowak-2000,Belloni-2002}, 
\begin{equation}
\;L(\nu;\nu_0,\mathrm{\Delta}) = 
\frac{\Delta}{\pi + 2\tan^{-1} {\left(\frac{2\nu_0}{\Delta}\right)}}
\frac{1}{\left(\nu-\nu_0\right)^2 + \left(\frac{\Delta}{2}\right)^2},
\label{eq2}
\end{equation}
where $\nu_0$ and $\Delta$ are, respectively, the centroid frequency and the FWHM of each Lorentzian. (With the above definition the integral of the Lorentzian function from zero to infinity is one.) Some of these Lorentzians have a central frequency equal to zero and are therefore called zero-centred Lorentzians, while others have a central frequency that is different from zero and, depending on their quality factor, $Q=\nu_0/\Delta$, they are called either peaked noise if $Q<2$ or QPOs if $Q>2$ \citep[e.g.][]{vanderKlis-1994}. Since this is an arbitrary definition \citep{Belloni-2002}, we call these components QPOs regardless of their $Q$ values.

We will now work out the expression of the Real and Imaginary parts of the cross spectrum assuming, as in \cite{Nowak-1999b}, that:\\
(i) there are a number of input processes that are coherent with the corresponding output processes;\\
(ii) the individual input processes (as well as the individual output processes\footnote{This is ensured if the input processes are incoherent with one another, but each output process is perfectly coherent with its corresponding input process.}) are incoherent with one another. We note that we only require that two input components are incoherent with each other if they overlap in Fourier frequency. This assumption is not necessary if the components are very far apart from each other in frequency, as is for instance the case of harmonics/sub-harmonics of a QPO.

Assumption (i) is justified by the fact that, as far as this was measured, strong QPOs, which dominate the variability over a certain frequency range, have a coherence that is consistent with unity. The cleanest case is that of the kHz QPOs, which appear in a part of the power spectrum in which no other component contributes to the variability. This is important because the coherence drops if two or more components with 
different amplitudes and phases of their cross vectors contribute to the variability over the same frequency range \citep{Vaughan-1997}.
For instance, \cite{deAvellar-2013} showed that the coherence of one of the strongest kHz QPO in the neutron-star system 4U 1608--52 is consistent with unity across the QPO profile (see their Fig. 1). On the other hand, \cite{Troyer-2018} showed that for the kHz QPOs in 14 neutron-star LMXBs the total rms amplitude in the power spectrum and the cross amplitudes in the cross spectra are consistent with being the same. For this to be the case the coherence at the QPO frequency must be unity.

The coherence function of strong low-frequency QPOs in black-hole LMXBs is also consistent with being unity. For instance, in XTE J1550--564, while over a broad range of Fourier frequencies the coherence is less than one, at the frequency of the type-C QPO the coherence always climbs back to one \citep[see, e.g., Fig. 1 of][and Fig. 6 of \citealt{Rapisarda-2017a}]{Cui-2000}. The same is true for the type-C QPO in GRS 1915+105 \citep{Muno-2001}.

This holds true not only for narrow QPOs, but also for broader variability components. For instance, the power spectrum of Cyg X--1 consists (basically) of two broad Lorentzian components \cite[e.g.][]{Grinberg-2014}, while the coherence is about unity over the whole frequency range in which either of these components dominates, and it drops slightly when the two components cross \citep{Nowak-1999a}. In fact the power spectrum of Cyg X-1 consists of more than just two components \citep[see, e.g., Fig. 3 in][]{Nowak-2000}, and the coherence drops whenever two of those components cross \citep[see, e.g., Fig. 5 in][and Fig. 5 in \citealt{Rapisarda-2017b}]{Nowak-1999a}. Interestingly, in Cyg X--1 the frequencies at which the coherence drops coincide with the frequencies at which the power spectrum shows breaks, and the lag spectrum shows shelves of more or less constant lags \citep[compare, for instance, the bottom-left panel of Fig. 1, the bottom-right panel of Fig. 5 and the bottom-right panel of Fig. 10 in][]{Nowak-1999a}.

Assumption (i) implies that we can write, for each Lorentzian, $G_{xy,i} = H_i(\nu) G_{xx,i}(\nu)$, and therefore each Lorentzian will have its own lag spectrum, $\arg{[H_i(\nu)]} = \Delta\phi_{xy,i}(\nu)$. We can therefore write that:
\begin{equation}
\begin{aligned}
&G_{xx}(\nu) = \sum_{i=1}^{n} G_{xx,i}(\nu) \coloneq \sum_{i=1}^{n} A_i L(\nu;\nu_{0,i},\Delta_i)\\
&G_{yy}(\nu) = \sum_{i=1}^{n} G_{yy,i}(\nu) \coloneq \sum_{i=1}^{n} B_i L(\nu;\nu_{0,i},\Delta_i),
\label{eq3}
\end{aligned}
\end{equation}
where $A_i,B_i \in \mathbb{R}$ are the integrated power, from zero to infinity, of each Lorentzian component in each of the two energy bands. 

Assumption (ii), which allows us to decompose the power spectra as in eq.~\ref{eq3}, is that the $n$ Lorentzians are uncoupled and their contribution to the power spectrum are additive. While it has been used repeatedly in the literature to fit the power spectra of LMXBs \citep[e.g.][etc.]{Nowak-2000,Belloni-2002,vanStraaten-2003,Altamirano-2008}, this assumption would appear to contradict the linear relation between the broadband absolute rms amplitude and the X-ray flux \cite[e.g.][]{Uttley-2005} in LMXBs and Active Galactic Nuclei; this rms-flux relation suggests that the variability should be due to a multiplicative rather than an additive process, favouring models of propagating mass accretion rate fluctuations \citep[][see \citealt{Ingram-2013} and \citealt{Rapisarda-2016} for the mathematical description and an implementation of the model of propagating fluctuations to produce the power and cross spectra of LMXBs]{Arevalo-2006}. However, \cite{Heil-2011} showed that the low-frequency QPO in the black-hole candidate XTE J1550--564 does not follow a single linear rms-flux relation, but that the slope of this relation depends on the frequency of the QPO. \cite{Rapisarda-2017a}, on the other hand, demonstrated that propagation models fail to reproduce the data, and concluded that the discrepancies between data and model are a generic issue inherent to these models, rather than being specific to any particular implementation of that scenario. Finally, since one computes power spectra of stationary time series, there must be a linear, additive, model that describes the data \citep{Scargle-2020}. In any case, we can test the predictions of our hypothesis that the Lorentzians are mutually incoherent, as we explain below.

If we accept that assumption (ii) above holds, it is enough to work out the case of a signal with a power spectrum consisting of only one Lorentzian, and combine the results to get the power and cross spectra of a combination of Lorentzians at the end:
\begin{equation}
\begin{aligned}
&    G_{xx}(\nu) = A\; L(\nu;\nu_0,\Delta)\\
&    G_{yy}(\nu) = B\; L(\nu;\nu_0,\Delta).\\
\end{aligned}   
\label{eq4}
\end{equation}

From eq.~\ref{eq1}, since for each Lorentzian $\gamma^2_{xy}=1$, we can write:
\begin{equation}
\begin{aligned}
        |G_{xy}(\nu)|^2 = \mathrm{Re}[G_{xy}(\nu)]^2 + \mathrm{Im}[G_{xy}(\nu)]^2 = \\
        = G_{xx}(\nu) G_{yy}(\nu)=
        A B L^2(\nu;\nu_0,\Delta).\\
\end{aligned}
\label{eq5}
\end{equation}

The frequency-dependent phase lag between the two signals represented
by the Lorentzian functions can be written in the usual way:
\begin{equation}
\Delta\phi_{xy}(\nu) = 
\tan^{-1}{\left(\frac{\mathrm{Im}[G_{xy}(\nu)]}{\mathrm{Re}[G_{xy}(\nu)]}\right)} = g(\nu;p_j),
\label{eq6}
\end{equation}
where $g(\nu;p_j)$ is a function of frequency with $m$ parameters $p_j$.  It is easy to see that we can combine eqs~\ref{eq5} and \ref{eq6} to get 
\begin{equation}
\begin{aligned}
& \mathrm{Re}[G_{xy}(\nu)] = C\;L(\nu;\nu_0,\Delta) \cos{[g(\nu;p_j)]}\\
& \mathrm{Im}[G_{xy}(\nu)] = C\;L(\nu;\nu_0,\Delta) \sin{[g(\nu;p_j)]},\\
\end{aligned}
\label{eq7}
\end{equation}
where $C = \sqrt{A\;B}$.

In the general case, if we assume that the power spectra of two signals $x(t)$ and $y(t)$ can be decomposed into a finite number, $n$, of Lorentzian functions\footnote{In reality this argument is valid for any function, not just Lorentzians, as long as the same linear combination of those functions, except perhaps for the normalisation factors, fits the power spectra of the two signals, $x(t)$ and $y(t)$.} that are uncorrelated with each other, because the individual Lorentzians are coherent between the two energy bands at all Fourier frequencies, the Real/Imaginary part of the cross spectrum between the two signals is also a linear combination of Lorentzian functions multiplied by a cosine/sine function of the, in principle frequency-dependent, phase lag between the two corresponding Lorentzian signals:
\begin{equation}
\begin{aligned}
& \mathrm{Re}[G_{xy}(\nu)] = \sum_{i=1}^n C_i\;L(\nu;\nu_{0,i},\Delta_i) \cos{[\Delta\phi_{xy,i}(\nu)]}\\
& \mathrm{Im}[G_{xy}(\nu)] = \sum_{i=1}^n
C_i\;L(\nu;\nu_{0,i},\Delta_i) \sin{[\Delta\phi_{xy,i}(\nu)]},\\
\end{aligned}
\label{eq8}
\end{equation}
with $\Delta\phi_{xy,i}(\nu) = g_i(\nu;p_{j,i})$ and $C_i = \sqrt{A_i B_i}$.

In practice, one fits the PS in two energy bands using eq.~\ref{eq3} and the CS between the same two bands using eq.~\ref{eq8}, for an arbitrary choice of $g(\nu;p_{j,i})$ (more on this below). The goal is to find, for each Lorentzian in the model, the parameters $A_i$, $B_i$, $\nu_{0,i}$, $\Delta_i$ and $p_{j,i}$, $j=1...m$. Instead of the PS in two bands one can use the full-band PS, $G_{zz}$, which can also be written as a a combination of the same Lorentzian functions, $G_{zz}(\nu) = \sum_{i=1}^n D_i L(\nu;\nu_{0,i},\Delta_i)$ (because, except for the normalisations, the Lorentzians that fit the QPOs in the individual bands also fit the QPOs in the full band); in that case, for each Lorentzian, the parameters of the model are $D_i$, $\nu_{0,i}$, $\Delta_i$, $C_i$ and $p_{j,i}$.

From here we can write the  \textit{total}, frequency-dependent, phase lags, $\Delta\phi(\nu)$, as:
\begin{equation}
\begin{aligned}
& \Delta\phi(\nu) = \tan^{-1}{\left(\frac{\sum_{i=1}^n C_i\;L(\nu;\nu_{0,i},\Delta_i) \sin{[g_i(\nu;p_{j,i})]}}{\sum_{i=1}^n C_i\;L(\nu;\nu_{0,i},\Delta_i) \cos{[g_i(\nu;p_{j,i})]}}\right)}.
\end{aligned}
\label{eq9}
\end{equation}
(Notice that the total phase lags given in eq.~\ref{eq9} are different from the phase lags of each Lorentzian, $\Delta\phi_{xy,i}(\nu) = g_i(\nu;p_{j,i})$ defined in eq.~\ref{eq6}.) One prediction of our proposal is that the same model that fits the power and cross spectra should correctly reproduce the phase-lag vs. frequency spectra, and in particular yield the same ``shelves'' observed in the plots of the phase-lag vs. Fourier frequency \citep[see, for instance, Fig. 10 of][especially the bottom-right panel]{Nowak-1999a}.

If we use two PS and the CS, we can also write the \textit{total} intrinsic coherence function of the two signals as a function of frequency:
\begin{equation}
\begin{aligned}
& \gamma^2_{xy}(\nu) = \frac{|G_{xy}(\nu)|^2}{G_{xx}(\nu) G_{yy}(\nu)} = \\
& \left[\left(\sum_{i=1}^n C_i\;L(\nu;\nu_{0,i},\Delta_i) \cos{[g_i(\nu;p_{j,i})]}\right)^2 +\right.\\
& \left.\left(\sum_{i=1}^n
C_i\;L(\nu;\nu_{0,i},\Delta_i) \sin{[g_i(\nu;p_{j,i})]}\right)^2\right] \times \\
& \frac{1}{\left(\sum_{i=1}^n A_i\;L(\nu;\nu_{0,i},\Delta_i)\right)\left(\sum_{i=1}^n B_i\;L(\nu;\nu_{0,i},\Delta_i)\right)}.\\
\end{aligned}
\label{eq10}
\end{equation}
(This is the same as eq. 10 in \citealt{Vaughan-1997} but for $n$ components.) 
From this, the second prediction of our proposal is that, if indeed the Lorentzian functions are incoherent with one another, the same model that fits the power and cross spectra should correctly reproduce the coherence function vs. Fourier frequency. Specifically, if two or more Lorentzian functions contribute significantly at some Fourier frequency, and the Lorentzians have different phase lags and different ratios of the Fourier amplitude in each energy band, the observed intrinsic coherence function should drop at those frequencies. Conversely, if the observed intrinsic coherence function drops at some Fourier frequency, the best-fitting model of the power and cross spectra should contain two or more Lorentzians with different properties of the cross vectors that contribute significantly to the variability at that Fourier frequency  \citep[see][for details]{Vaughan-1997}. Finally, if a strong Lorentzian dominates the variability over a frequency range, and if our assumption (i) is correct, the coherence should tend to unity in that frequency range.

\subsection{Phase-lag model}
\label{phaselagmodel}

Since \textit{any} function $g(\nu;p_j)$ will satisfy eqs.~\ref{eq5} and \ref{eq7}, as we mention above, to fit data we need to specify these functions for each Lorentzian $i$ in the model. The most basic case is when each of the functions depends only on one parameter, $p_i=k_i$, and from that family of functions the two simplest cases are: (i) The phase lags are constant\footnote{This is equivalent to assuming that the time lags are proportional to $\nu^{-1}$.}, $\Delta\phi_{xy,i}(\nu) = g_i(\nu;p_i) = 2\pi k_i$. In this case, for all Fourier frequencies, the cross vector of each individual Lorentzian function points in the same direction, at an angle $2\pi k_i$ radians with respect to the Real axis. (ii) The time lags are constant, and hence the phase lags increase linearly with frequency, $\Delta\phi_{xy,i}(\nu) = g_i(\nu;p_i) = 2\pi k_i\nu$. This could be the case if the lags are produced by delays of photons propagating in the accretion flow of LMXBs. In this case the cross vector  rotates, always in the same direction, at a constant rate of $2\pi k_i$ radians per Hz in the Fourier plane as a function of Fourier frequency. 

From the above it is apparent that, unless $g(\nu)$ is constant, the cross vector will rotate in some more or less complex way, perhaps changing the direction of the rotation, as a function of frequency in the Fourier plane. If this happens, the phase lags may wrap as the angle changes abruptly from $\pi$ at some Fourier frequency to $-\pi$ at the next frequency, or vice versa. For the phase lags not to wrap it must be true that $-\pi \leq g(\nu) < \pi$ over the frequency range in which the signal contributes significantly to the variability. (We would not be able to observe a phase wrap if it happened at frequencies where the power spectrum was dominated by noise.) For instance, if a component in the power spectrum has a time lag of $\sim100$ ms that is constant with Fourier frequency, the cross vector associated with that component will rotate in the Fourier plane at $\sim0.6$ radians per Hz, such that if the component extends over a frequency range of $\gtrsim10$ Hz its phase lags will wrap at least once. Since, until now, such a wrap has not been reported, and since in LMXBs the time lags go as $\nu^{\sim -0.7}$ over a relatively broad frequency range \citep[at least $\sim 100$ Hz; e.g.,][]{Nowak-1999a}, such that the phase lags increase very slowly,  $\Delta\phi \sim \nu^{\sim 0.3}$, from the two simplest cases mentioned above the one of constant phase lags appears more natural. In what follows we will assume that the phase lags of individual components in the power spectrum are independent of Fourier frequency (constant phase lags) or increase linearly with frequency (constant time lags).

\smallskip
\subsection{Average phase and time lags over a frequency range}
\label{averages}
In the above discussion the functions $g_i(\nu;p_{j,i})$ are the frequency-dependent phase lags of the individual Lorentzian components. A quantity sometimes given in the literature \citep[e.g.][]{Reig-2000} is the total phase lag over the frequency range $\Delta\nu=\nu_2 - \nu_1$:

\begin{equation}
\begin{aligned}
& \Delta\phi_{\Delta\nu} =
\tan^{-1}
{
\left(\frac{\int_{\nu_1}^{\nu_2} \mathrm{Im}[G_{xy}(\nu)] d\nu}
{\int_{\nu_1}^{\nu_2} \mathrm{Re}[G_{xy}(\nu)] d\nu} 
\right)} =
\\
= & ~ 
\tan^{-1}{\left(
\frac{\int_{\nu_1}^{\nu_2}
\sum_{i=1}^n C_i\;L(\nu;\nu_{0,i},\Delta_i) \sin{[g_i(\nu;p_{j,i})]}
d\nu}
{\int_{\nu_1}^{\nu_2}
\sum_{i=1}^n C_i\;L(\nu;\nu_{0,i},\Delta_i) \cos{[g_i(\nu;p_{j,i})]} d\nu}\right)
}.\\
\end{aligned}
\label{eq11}
\end{equation}

A related, but different, quantity is the average phase lag over the same frequency range:
\begin{equation}
\begin{aligned}
&\langle \Delta\phi\rangle_{\Delta\nu}= 
\frac{1}{\Delta\nu}\int_{\nu_1}^{\nu_2} \Delta\phi(\nu) d\nu =\\
&\frac{1}{\Delta\nu}\int_{\nu_1}^{\nu_2}
\tan^{-1}{\left(
\frac{\mathrm{Im}[G_{xy}(\nu)]}{\mathrm{Re}[G_{xy}(\nu)]}
\right)}
d\nu =
\\
&= \frac{1}{\Delta\nu} \int_{\nu_1}^{\nu_2}
\tan^{-1}{\left(
\frac{
\sum_{i=1}^n
C_i\;L(\nu;\nu_{0,i},\Delta_i) \sin{[g_i(\nu;p_{j,i})]}
}{
\sum_{i=1}^n
C_i\;L(\nu;\nu_{0,i},\Delta_i) \cos{[g_i(\nu;p_{j,i})]}}\right) d\nu
}.
\label{eq12}
\end{aligned}
\end{equation}
It is easy to show that when the cross spectrum consists of a single Lorentzian with $g(\nu;k) = 2\pi k$ (constant phase lags), $\langle \Delta\phi\rangle_{\Delta\nu} = 2\pi k = \Delta\phi_{\Delta\nu}$, but for any other form of $g(\nu,p_j)$, or if the cross spectrum consists of a linear combination of Lorentzians, $\langle \Delta\phi\rangle_{\Delta\nu} \neq \Delta\phi_{\Delta\nu}$, and hence $\langle \Delta\phi \rangle_{\Delta\nu}$ should not be reported.

It is not possible to define the total time lag over the frequency range $\Delta\nu$, $\Delta\tau_{\Delta\nu}$, in a similar way as we defined $\Delta\phi_{\Delta\nu}$ in eq.~\ref{eq11}, but a quantity commonly given in the literature is $\Delta\tilde{\tau}_{\Delta\nu} = \Delta\phi_{\Delta\nu} / (2\pi \langle\nu\rangle)$, where $\langle\nu\rangle = (\nu_1+\nu_2)/2$. In the case of a narrow component, this is more or less similar to the usual definition for a pure sinusoidal function with frequency $\nu_{0}$, $\Delta\tau(\nu_0) = \Delta\phi(\nu_0)/(2\pi\nu_0)$, but for a broad component, e.g., the BBN \citep[e.g.][]{Kara-2019,Wang-2022}, these two expressions are not mathematically equivalent and hence $\Delta\tilde{\tau}_{\Delta\nu}$ is not a well-defined quantity and should not be used.

As with the phase lags, we can define the average time lags over a frequency range $\Delta\nu$:
\begin{equation}
\begin{aligned}
\langle \Delta\tau\rangle_{\Delta\nu} = 
\frac{1}{\Delta\nu}\int_{\nu_1}^{\nu_2} \Delta\tau(\nu) d\nu = \frac{1}{\Delta\nu}\int_{\nu_1}^{\nu_2} \frac{\Delta\phi(\nu)}{2\pi\nu} d\nu,
\end{aligned}
\label{eq13}
\end{equation}
which is equal to $\langle \Delta\phi\rangle_{\Delta\nu} / (2\pi \langle\nu\rangle)$ {\em only} when the time lags are constant with frequency, $\Delta\tau(\nu) =k$. For instance, if $\Delta\phi(\nu) = 2\pi k$ (constant phase lags), $\langle \Delta\tau\rangle_{\Delta\nu} = \left(k/\Delta\nu \right) \ln{\left(\nu_2/\nu_1\right)}$, and the average time lags will reflect the (arbitrary) boundaries of the selected frequency range. Since in LMXBs it is generally true that the time lags extending over a broad frequency range are not constant \citep[see, e.g., Fig. 1 in \citealt{deMarco-2021}, or Fig. 3h in][]{Wang-2022}, contrary to what is sometimes mentioned in the literature \citep[e.g.][]{Belloni-2022} one should not report $\langle \Delta\tau\rangle_{\Delta\nu}$ to characterise those lags either. \citep[Notice that rebinning the time-lag spectrum as, e.g., in][ is equivalent to this, and should therefore not be done.]{Wang-2022} 
%
%
\begin{table*}
    \centering
    \caption{Definitions and acronyms}
    \begin{tabular}{|l|l|}
     \hline
    \textbf{Acronym/text shortcut} & \textbf{Explanation}\\ \hline
    PS     &  Power density spectrum\\
    CS     &  Either the cross spectrum, or the Real and Imaginary parts of the CS, depending on the context\\
    CV      &  Cross vector in the Fourier plane with phase $\Delta\phi(\nu)$ and modulus $|\mathrm{CV}(\nu)|$ \\
    Phase-lag frequency spectrum & Phase lags between light curves in two energy bands vs. Fourier frequency \\
    Derived model & Model of the phase-lag frequency spectrum derived from the model fitted to the Real and Imaginary parts of the CS\\
    & Model of the intrinsic coherence function derived from the model fitted to PS in two energy bands and the Real and \\
    & Imaginary parts of the CS in those same two bands.\\ 
    & These models are \textit{not} fitted to the lags or the intrinsic coherence function\\
    Phase lags of a QPO & Phase lags of the Lorentzian that fits the QPO, computed as $\tan^{-1}{\left(N_{\rm IM}/N_{\rm RE}\right)}$, where $N_{\rm RE}$ and $N_{\rm IM}$ are the integrals\\
    &  from zero to infinity of the Lorentzian functions used to fit, respectively, the Real and Imaginary parts of the CS\\
    Traditional phase lags of a QPO & $\tan^{-1}{\left(\langle \mathrm{IM} \rangle / \langle \mathrm{RE} \rangle \right)}$, where $\langle \mathrm{RE} \rangle$ and $\langle \mathrm{IM} \rangle$ are the averages of, respectively, the Real and Imaginary parts of the \\
    & CS over a fixed frequency range across the QPO profile\\
    Constant phase-lags model & Model of the CS assuming constant phase lags with Fourier frequency for each individual Lorentzian component, \\
    & $\Delta\phi_i(\nu) = g_i(\nu;p_{j,i}) = 2\pi k_i$ (see \S\ref{phaselagmodel})\\
    Constant time-lags model & Model of the CS assuming constant time lags with Fourier frequency for each individual Lorentzian component, \\
    & $\Delta\tau_i(\nu) = k_i$ such that $\Delta\phi_i(\nu) = g_i(\nu;p_{j,i}) = 2\pi k_i \nu$  (see \S\ref{phaselagmodel})\\
\hline
\end{tabular}
    \label{tab:definitions}
\end{table*}

\subsection{Transfer/response function}
\label{TF}

Any linear time-invariant system is characterised by an impulse response function, $h(\tau)$, which gives the output of the system at any time, $y(t)$, to an input, $x(t)$, applied a time $\tau$ before, $y(t) = \int_{-\infty}^{\infty} h(\tau) x(t-\tau) d\tau$. The Fourier transform of the impulse response function, $H(\nu)$, is the frequency response or transfer function of the system \citep{Bendat-2010}. If $H(\nu)$ is the same for all realisations of the two processes, the processes are said to be coherent at frequency $\nu$ (see \S\ref{mathematical}).

In a manner similar to \cite{Bendat-2010}, one can consider $x(t)$ and $y(t)$ as correlated light curves of a source in two energy bands, as discussed earlier. Here, $H(\nu)$ represents the transfer function that yields the light curve in band $y$ based on the light curve in band $x$. Alternative, one can think of $x(t)$ and $y(t)$ as the signals of two physical components of the system, e.g., the accretion disc and the corona, such that the transfer function gives the signal of one in terms of the signal of the other \citep[e.g.,][]{Reynolds-1999,Mastroserio-2018,Ingram-2019c}. In the first case, for light curves in two energy bands, the transfer function depends only upon Fourier frequency (for the given energy bands), and it refers to two observed quantities, each of them (possibly) being produced by a combination of more than one physical mechanism (e.g., the accretion disc, the corona, etc.). In the second case, the transfer function refers to the physical mechanisms themselves, and it depends both on Fourier frequency and energy; if one wants to compute light curves or power and cross spectra in certain energy bands to compare with observations, one needs to integrate the transfer function over energy \citep[see][for details]{Mastroserio-2018,Mastroserio-2019}.

In this paper we will mostly use the term transfer function to refer to the first case. Specifically, as stated in \S\ref{mathematical}, $\arg{[H(\nu)]}$ gives the phase-lags of light curve $y(t)$ with respect to light curve $x(t)$ at frequency $\nu$. Hence, when we fit Lorentzian functions to the power and cross spectra of a source, as explained above, we will be talking of the frequency-dependent lags of the Lorentzian that represents the variability in the power and cross spectra in one energy band with respect to the Lorentzian that represents the same variability in another energy band. In some passages, when we talk about global transfer or response functions, we refer to the second case.

\section{Application to data}
\label{examples}
In this section we apply the formalism presented in \S\ref{mathematical} to five observations of three black-hole X-ray binaries: an observation of GX 339--4 and two observations of GRS 1915+105 with the Proportional Counter Array \citep[\textit{PCA};][]{Jahoda-2006} on board the \textit{Rossi X-ray Timing Explorer}  \citep[\textit{RXTE};][]{Bradt-1993}, and two observations of MAXI J1820+070 with the \textit{Neutron Star Interior Explorer} \citep[\textit{NICER};][]{Gendreau-2012}. Each of the examples highlights certain aspects of the method or provide useful insight into aspects of the variability of accreting X-ray binaries. (We have already presented results obtained with an initial version of this method in \citealt{Alabarta-2022} and \citealt{Peirano-2022}.) 

In each of the subsections we first give very briefly the necessary information of how we process the data and produce power, cross and lag spectra and coherence function of the source, and we then give the results of fitting those data. In all cases, when we fit simultaneously the power spectrum and the Real and Imaginary parts of the cross spectrum
, we link the frequency and FWHM of each Lorentzian in the model so that these parameters are the same in the PS and CS. In some cases we fix the frequency and FWHM of all the Lorentzians to the values we obtain from an initial fit to the PS alone, while in other cases we leave these parameters free to vary, but we always link them across the PS and the Real and Imaginary parts of the CS. In each case we indicate whether we do one or the other. On the other hand, we always leave free the normalisations of all the Lorentzians and allow them to vary independently in the PS and the CS. If we plot the phase-lag frequency spectrum or the intrinsic coherence function together with the best fitting model, we do not fit the model to the phase-lag frequency spectrum or the intrinsic coherence function, but we derive the model from the best-fitting model to the Real and Imaginary parts of the CS. We call this the ``derived model'' of the phase-lag spectrum/intrinsic coherence function.

To simplify the reading we give the definition of some acronyms and language shortcuts that we use often in the text in Table~\ref{tab:definitions}. For completeness, we also give the acronym in the text the first time we define it.

We use the recommended tools for each mission to extract calibrated clean event files for each observation. When extracting events in certain energy bands, we always use the channel space of the detector, and convert those values into energies taking into account the channel to energy conversion for each instrument. In all cases we give both the channel and the (approximate) energy range that we use to analyse the data.

Once we have obtained the final event files, we use GHATS\footnote{\url{http://astrosat-ssc.iucaa.in/uploads/ghats_home.html}} to compute the Fast Fourier Transform (FFT) in each band of interest. Except in one case (see below), we always compute the FFT of the data from the full observation (we give the exact ObsIDs that we use in each subsection). For this we compute FFTs on segments of a given duration, $T_{\rm FFT} < T_{\rm exp}$, where $T_{\rm exp}$ is the total duration of the observation, with a time resolution, $\Delta t$, that allows us to reach (and exceed) the frequency of interest in the PS and CS, and we average the PS and CS of the individual segments \citep{vanderKlis-1989}. The number of segments that we average for each PS and CS is therefore $M = \mathrm{integer}(T_{\rm exp}/T_{\rm FFT})$. From this, the minimum frequency and the frequency resolution of the PS and CS are $\nu_{\rm min} = \Delta\nu = 1/T_{\rm FFT}$, while the Nyquist frequency is $\nu_{\rm Nyq} = 1/(2 \Delta t)$. The averaging of the PS and CS per observation increases the signal-to-noise ratio by a factor $\sim \sqrt{M}$. We further rebin the average PS and CS of an observation logarithmically in frequency such that the size of a bin increases by a factor $\approx 1.023 (=10^{1/100})$ compared to the size of the previous bin to increase the signal-to-noise ratio further. 

We use \texttt{Xspec v.12.13.0} \citep{Arnaud-1996} to fit the PS and CS. We always start by first fitting the full-band PS with a model with only one Lorentzian, and we add a new Lorentzian until the reduced $\chi^2$ is about one and the fit shows no significant structured residuals. The final model of the PS, consisting of $n$ Lorentzians (eq.~\ref{eq3}), is $\sum_{i=1}^n D_i L(\nu;\nu_{0,i},\Delta_i)$. In \S\ref{339-4}-\ref{1820} we fit simultaneously the full-band PS, $G_{zz}(\nu)$, and the the Real and Imaginary parts of the CS in two bands, $\mathrm{Re}\left[G_{xy}(\nu)\right]$ and $\mathrm{Im}\left[G_{xy}(\nu)\right]$, while in \S\ref{339-4}, \S\ref{1820} and \S\ref{1820_b} we also use the PS in the two bands, $G_{xx}(\nu)$ and $G_{yy}(\nu)$, to be able to compute the coherence function.

When we fit the PS and the CS simultaneously, we need to assume the frequency dependence of the phase lags (see \S\ref{phaselagmodel}). Here we only consider two cases: either the phase lags or the time lags are constant with frequency (constant phase-lags or constant time-lags model, respectively); although it is possible that the phase lags of different variability components depend differently upon frequency, here we always assume that the same functional dependence applies to all the Lorentzian components. We mention in the text and in the Figure and Table captions the functional dependence of the lags that we use when we fit the data. 

Specifically, when we use the constant phase-lags model we fit functions of the form 
$\sum_{i=1}^n C_iL(\nu;\nu_{0,i},\Delta_i) \cos{(2\pi k_i)}$ to the Real part of the CS and
$\sum_{i=1}^n
C_iL(\nu;\nu_{0,i},\Delta_i) \sin{(2\pi k_i)}$ to the Imaginary part,
where $C_i$ is the integral from zero to infinity of the modulus squared of the cross vector (CV) and $2\pi k_i$ are the phase lags for each component (eq.~\ref{eq8} with $\Delta\phi_{xy,i}(\nu) = 2\pi k_i$). The $C_i$ and $k_i$ are free parameters that need to be fitted.

When we use the constant time-lags model we fit the Real part of the CS with $\sum_{i=1}^n C_i L(\nu;\nu_{0,i},\Delta_i) \cos{(2\pi k_i \nu)}$ and the Imaginary part with
$\sum_{i=1}^n
C_i L(\nu;\nu_{0,i},\Delta_i) \sin{(2\pi k_i \nu)},$
where the $k_i$ are now the time lags of each component (eq.~\ref{eq8} with $\Delta\phi_{xy,i}(\nu) = 2\pi k_i \nu$).

Since we normalise the PS and CS to units of fractional rms squared per Hz \citep{Belloni-1990}, the integrated fractional rms amplitude of a Lorentzian is the square root of its normalisation. We consider that a Lorentzian is significantly needed in the model of the PS if the normalisation of that Lorentzian divided by its 1-$\sigma$ error is larger than 3. For the CS we take the modulus squared of the CV and its error. If the errors are asymmetric, we take the negative error to calculate the significance. Unless otherwise indicated, we give the 1-$\sigma$ error for one parameter and, if a Lorentzian is absent, we give the 95\% confidence upper limit to the rms fractional amplitude of that Lorentzian in the power or the cross spectrum.

In the literature, authors call ``the phase lags of a QPO'' to the inverse tangent function of the ratio of the average of the Imaginary part to the average of the Real part of CS; these averages are computed over a fixed frequency range across the QPO profile, usually covering a FWHM, from $\nu_{\rm QPO} - \Delta/2$ to $\nu_{\rm QPO} + \Delta/2$. From this, ``the time lags of the QPO'' are defined as the phase lags of the QPO divided by $2\pi$ times the centroid frequency of the QPO (but see \S\ref{averages} for the problems of this definition). We call these quantities the ``traditional phase/time lags of a QPO''. 

On the contrary, following the explanation in \S\ref{phaselagmodel}, here we call ``the phase lags of a QPO'' to the quantity $2\pi k_i$ for the Lorentzian that fits the QPO in the CS for the constant phase-lags model, and ``the time lags of a QPO'' to the parameter $k_i$ for the Lorentzian that fits the QPO in te CS for the constant time-lags model. As we explained in \S\ref{averages}, this means that we cannot compute the time lags of a QPO when we use the constant phase-lags model, and vice versa. Withal, the phase and time lags of a QPO are not related via the centroid frequency of the QPO as in the case of the traditional definition of the phase and time lags of the QPO (see \S\ref{mathematical}).

Finally, as is customary, we say that lags are hard or positive if the high-energy photons lag the low-energy ones, and soft or negative if the opposite happens.

%
%
\begin{figure*}
\centering
\includegraphics[width=0.45\textwidth]{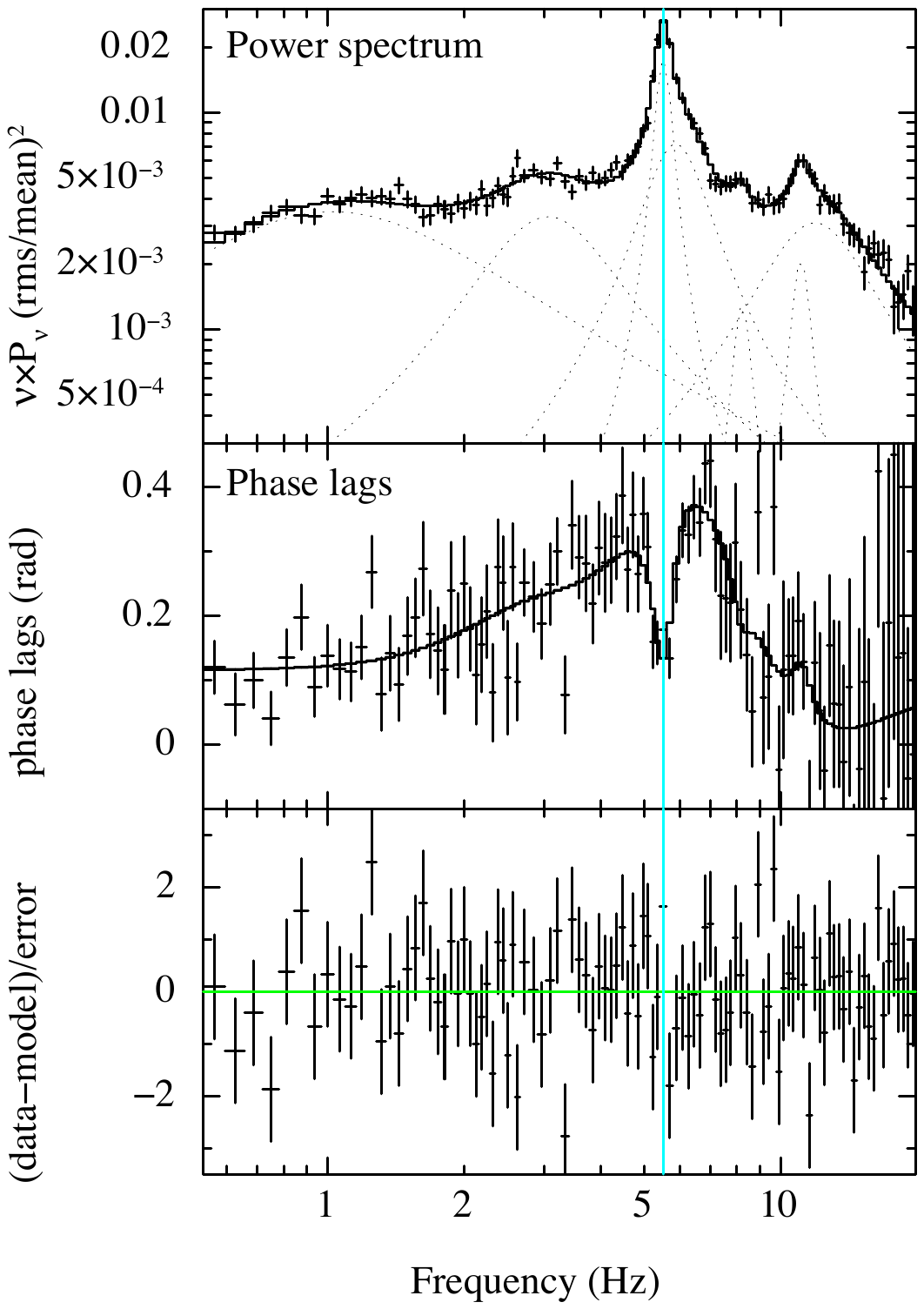}
\includegraphics[width=0.45\textwidth]{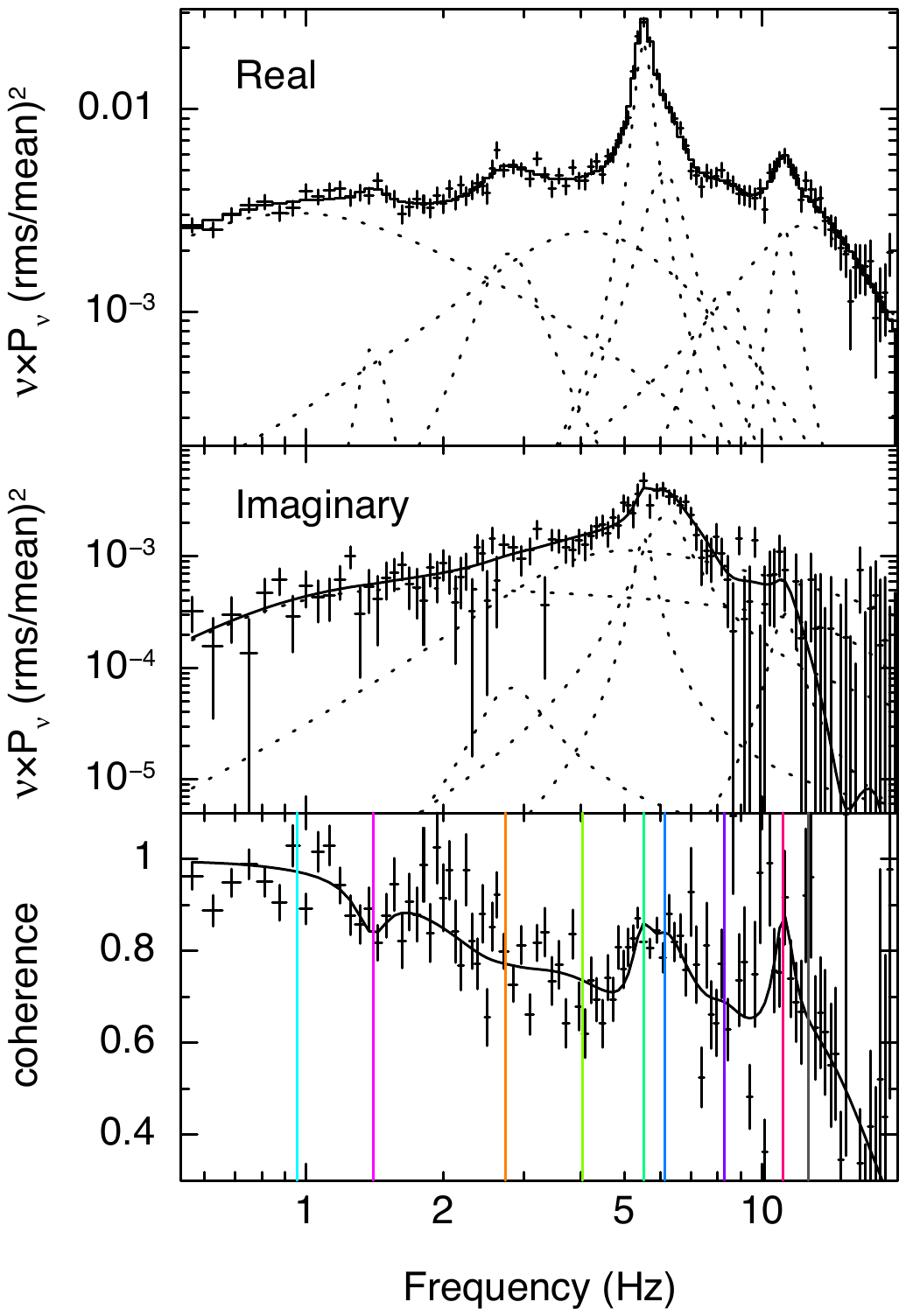}
\caption{Top left panel: Full-band PS of the RXTE observation 92035-01-03-06 of GX 339--4 fitted with a model (solid thick line) consisting of seven Lorentzian functions (thin dotted lines).
Middle left panel: Phase-lag vs. Fourier frequency (middle panel) together with the derived model (thick solid line) from the CS (not shown). During the fit, we let the centroid frequency and FWHM of each component free to vary but we link them to be the same in the PS and the CS, further assuming the constant phase-lags model (see Table~\ref{tab:definitions} and \S\ref{phaselagmodel}). Bottom left panel: Residuals of the phase lags with respect to the derived model. The vertical line indicates the centroid frequency of the narrow QPO.
Right column: Real (top) and Imaginary (middle) parts of the CS with the best-fitting model assuming constant time lags. The bottom panel shows the intrinsic coherence together with the derived model. For each Lorentzian in the model, the vertical lines mark $\nu_{\rm max}$, the frequency at which the Lorentzian function peaks in the $\nu \times P_\nu$ representation \citep{Belloni-2002}.}

\label{fig1}
\end{figure*}

\subsection{Case study 1: The complex lag spectrum of the type-C QPO in GX 339--4}
\label{339-4}

We use the \textit{RXTE} observation 92035-01-03-06 of GX 339--4 that was analysed by \citet[][see their Fig. 1]{Zhang-2017} and \citet[][see their Fig. 3]{Altamirano-2015}. This observation contains a prominent type-C QPO \citep[see][for the definition of the three types of low-frequency QPOs, types A, B and C, in black-hole X-ray binaries]{Wijnands-1999,Remillard-2002,Casella-2005} at $\sim 5.5$ Hz that shows a rather complex phase-lag frequency spectrum: The hard phase lags around the QPO are double-peaked, with the minimum centred on the QPO centroid frequency (see top and middle panels of Fig.~\ref{fig1})

Although no systematic study of this type of lag-frequency spectra exists, this is not an isolated case, and a similar trend is apparent in several other observations of GX 339--4 \cite[see Fig. 1 of][for other examples]{Zhang-2017} and GRS 1915+105 \citep[][see also \S\ref{1915_1} below]{Zhang-2020}. 

We take the Binned and Event Mode files of this observation to compute the full-band (channels 0 to 249) PS and CS of photons in the $5.4-115$ keV band\footnote{\url{https://heasarc.gsfc.nasa.gov/docs/xte/e-c_table.html}} (channels 14 to 249) with respect to those in the $1.95-5.4$ keV band (channels 0 to 13). Both for the PS and the CS we use $T_{\rm FFT}=16$ s, yielding $\nu_{\rm min}=\Delta\nu=1/16$ Hz, at a time resolution $\Delta t= 1/2048$ s such that $\nu_{\rm Nyquist}=1024$ Hz.

We initially fit the PS in the range $0.5-20$ Hz adding one Lorentzian at a time as described in \S\ref{examples}. A fit with six Lorentzians gives $\chi^2=111.8$ for 87 degrees of freedom (dof), while that with seven Lorentzians gives $\chi^2=93.4$ for 84 dof, and all Lorentzians are at least 4-$\sigma$ significant. Adding an eighth Lorentzians does not improve the fit significantly, so we stop at seven.

\textit{Constant phase-lags model:} We next fit simultaneously the full-band PS and the CS with the same number of Lorentzians, fixing the frequency and FWHM of each Lorentzian to the values that we obtain from the fits to the PS; for this fit we assume the constant phase-lags model. The fit gives $\chi^2=305$ for 294 dof, with structured residuals around 5 Hz (not shown). We next let the frequency and FWHM free but linked in the PS and CS for each Lorentzian. This fit, shown in Figure~\ref{fig1}, gives $\chi^2=287.9$ for 280 dof; while it is not statistically better than the one with fixed frequency and FWHM of the Lorentzians, this fit shows no structured residuals around 5 Hz. Because of this, and because later on we find that in other cases the fit with fixed frequencies and FWHM does not work, we adopt this one as the final model.

As it is apparent from Figure~\ref{fig1}, the QPO is actually fitted with two Lorentzians, a relatively broad one that fits the wings, and a relatively narrow one that fits the central part of the QPO profile. The broad QPO component is reminiscent of the shoulder mentioned by 
\citet[][later on called the ``hump'' in \citealt{Belloni-2002}; see also \citealt{vanDoesburgh-2020}]{Belloni-1997}, and we therefore call this the QPO shoulder. Both the QPO and the QPO shoulder are significantly required by the fit, with a significance of 12-$\sigma$ and 20-$\sigma$, respectively.
Since the QPO and the QPO shoulder have slightly, but significantly ($\sim 5.5$-$\sigma$), different frequencies, $\nu_{\rm QPO} = 5.50 \pm 0.01$ Hz and $\nu_{\rm shoulder} = 5.78 \pm 0.05$ Hz, the QPO profile appears to be slightly asymmetric (top panel of Fig.~\ref{fig1}). 

%
%
\begin{table}
    \centering
    \caption{Phase and time lags ($5.4-115$ keV vs. $1.95-5.4$ keV) of the QPO and the QPO shoulder in GX 339--4 with our method}
    \label{tab:data-339-4}
    \footnotesize
\hspace{-0.5cm}    
\begin{tabular}{c|c|c|c}
\hline
\textbf{Component} & $\nu_0$ (Hz) & Phase lags$^{\dagger}$ (rad) & Time lags$^{\ast}$ (ms)\\    
\hline
QPO   &  $5.50 \pm 0.01 $ & $-0.004 \pm 0.024$ & $0.9 \pm 0.7$\\
Shoulder & $5.78 \pm 0.05$ & $0.51 \pm 0.03$ &  $10.8 \pm 0.7$ \\ 
\hline
\multicolumn{4}{l}{\small $^{\dagger}$Using the constant phase-lags model.}\\
\multicolumn{4}{l}{\small$^{\ast}$Using the constant time-lags model.}\\
\end{tabular}
\end{table}

In principle, it is possible that the asymmetric QPO profile is due to slight changes of the QPO frequency during the course of the observation, with the QPO spending most of the time at around $\sim 5.5$ Hz, and a small fraction of the time at $\sim 5.8$ Hz. To check this, we compare the full-band rms amplitude and phase lags between the two bands given above, $\Delta\phi$, of the two components, the QPO and the shoulder; we find rms$_{\rm QPO} = 4.9 \pm 0.2 \%$, rms$_{\rm shoulder} = 5.5 \pm 0.3 \%$, $\Delta\phi_{\rm QPO} = -0.004 \pm 0.024$ rad and $\Delta\phi_{\rm shoulder} = 0.51 \pm 0.03$ rad. While the rms amplitudes of the QPO and the QPO shoulder are consistent with being the same, the phase lags are significantly ($\sim 13.4$-$\sigma$) different. Since the lags of the QPO do not change so rapidly with QPO frequency \citep{Zhang-2017,Zhang-2020}, we conclude that the QPO and the QPO shoulder are two different components. 

\textit{Constant time-lags model}: We also compute the time lags of the QPO and the QPO shoulder by fitting the PS and the CS assuming the constant time-lags model, and we find that time lags of the QPO are significantly different from those of the QPO shoulder. The fit yields $\chi^2 = 284.0$ for $290$ dof. We give the frequencies, phase and time lags of the QPO and QPO shoulder in Table~\ref{tab:data-339-4}. 

As we explained in \S\ref{phaselagmodel}, in our method the phase (time) lags of the Lorentzians are parameters of the constant phase-lags (time-lags) model. Under the assumption that the phase (time) lags are constant, the time (phase) lags change with Fourier frequency across the QPO profile and, therefore, the time lags of a QPO are not exactly equal to the phase lags divided by $2\pi$ times the QPO centroid frequency.  

\textit{Traditional lags}: The case of the shoulder of the QPO highlights the problem of measuring lags in the traditional way, which we explained in \S\ref{introduction}. To show this we compute the traditional phase-lag, $\Delta\phi_{\rm trad}$, of the QPO in the $5.285-5.715$ Hz frequency range (one FWHM), $\Delta\phi_{\rm QPO, trad} = 0.23 \pm 0.01$ rad, which is significantly larger than the lag using our method (see Tab.~\ref{tab:data-339-4}), $\Delta\phi_{\rm QPO} = -0.004 \pm 0.024$ rad. Similarly, the traditional phase lag of the QPO shoulder within one FWHM ($4.795-6.765$ Hz) is $\Delta\phi_{\rm shoulder, trad} = 0.17 \pm 0.02$ rad, significantly smaller than the phase lag using our method, $\Delta\phi_{\rm shoulder} = 0.51 \pm 0.03$ rad. 

Likewise, the time lags computed using the traditional method over the same frequency ranges,  $\Delta\tau_{\rm QPO,trad}= 4.9 \pm 0.5$ ms and $\Delta\tau_{\rm shoulder,trad}= 6.5 \pm 0.3$ ms, are significantly different from those obtained with our method, 
$\Delta\tau_{\rm QPO}=0.9 \pm 0.7$ ms and 
$\Delta\tau_{\rm shoulder}= 10.8 \pm 0.7$ ms. 

%
%
\begin{figure*}
\centering
\hspace{-0.5cm}
\includegraphics[width=0.38\textwidth]{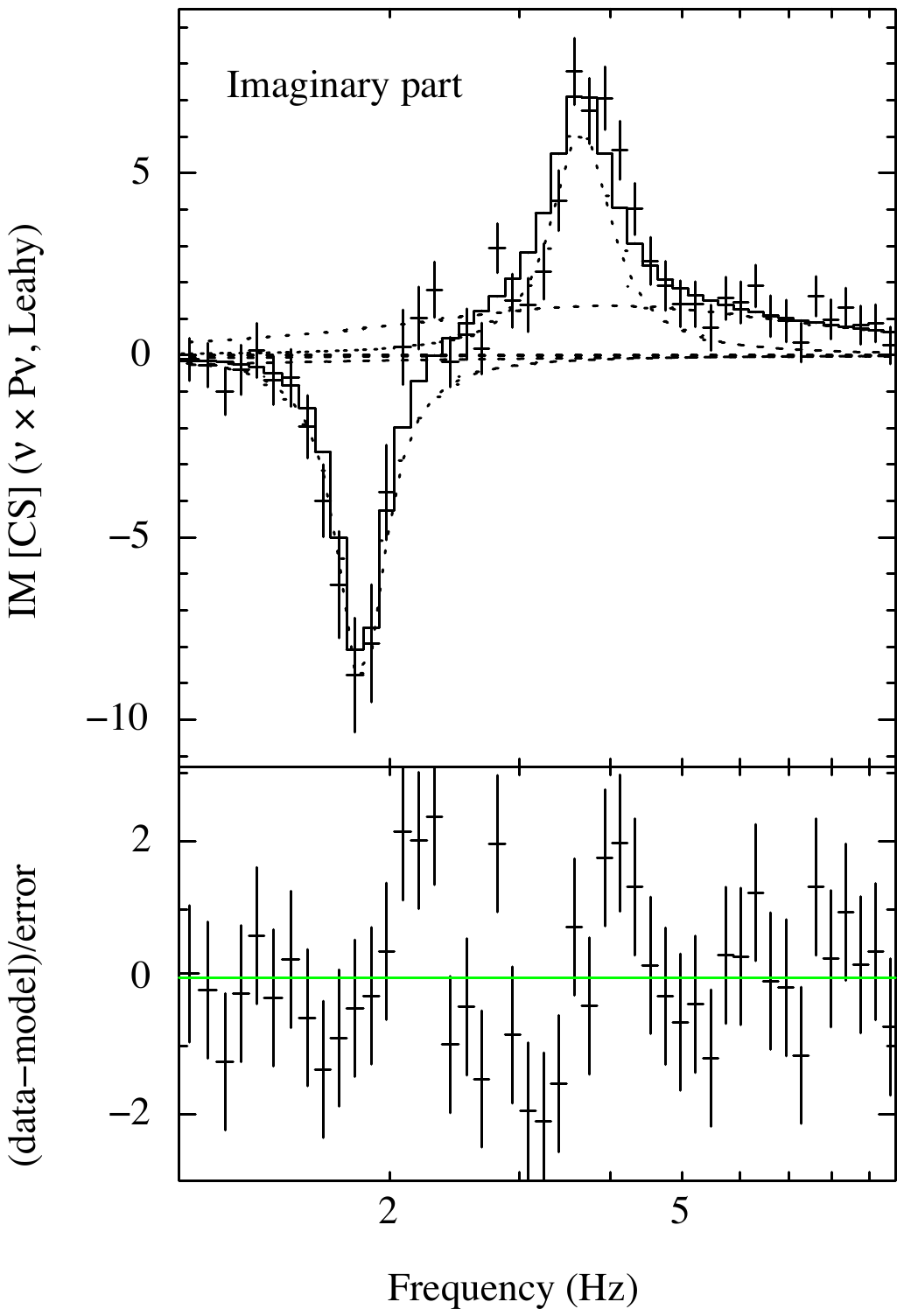}
\hspace{-1.3cm}
\includegraphics[width=0.38\textwidth]{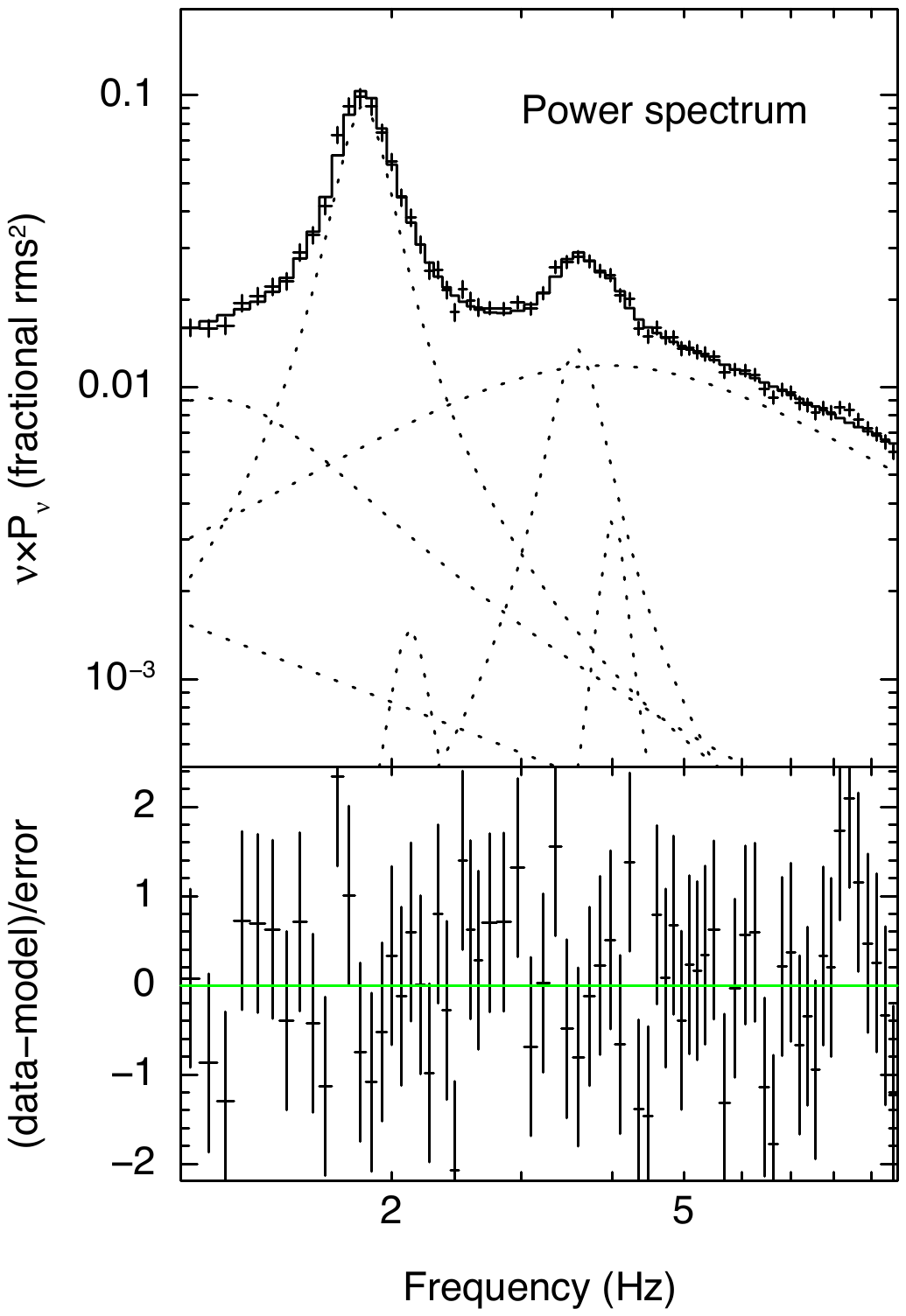}
\hspace{-1.1cm}
\includegraphics[width=0.38\textwidth]{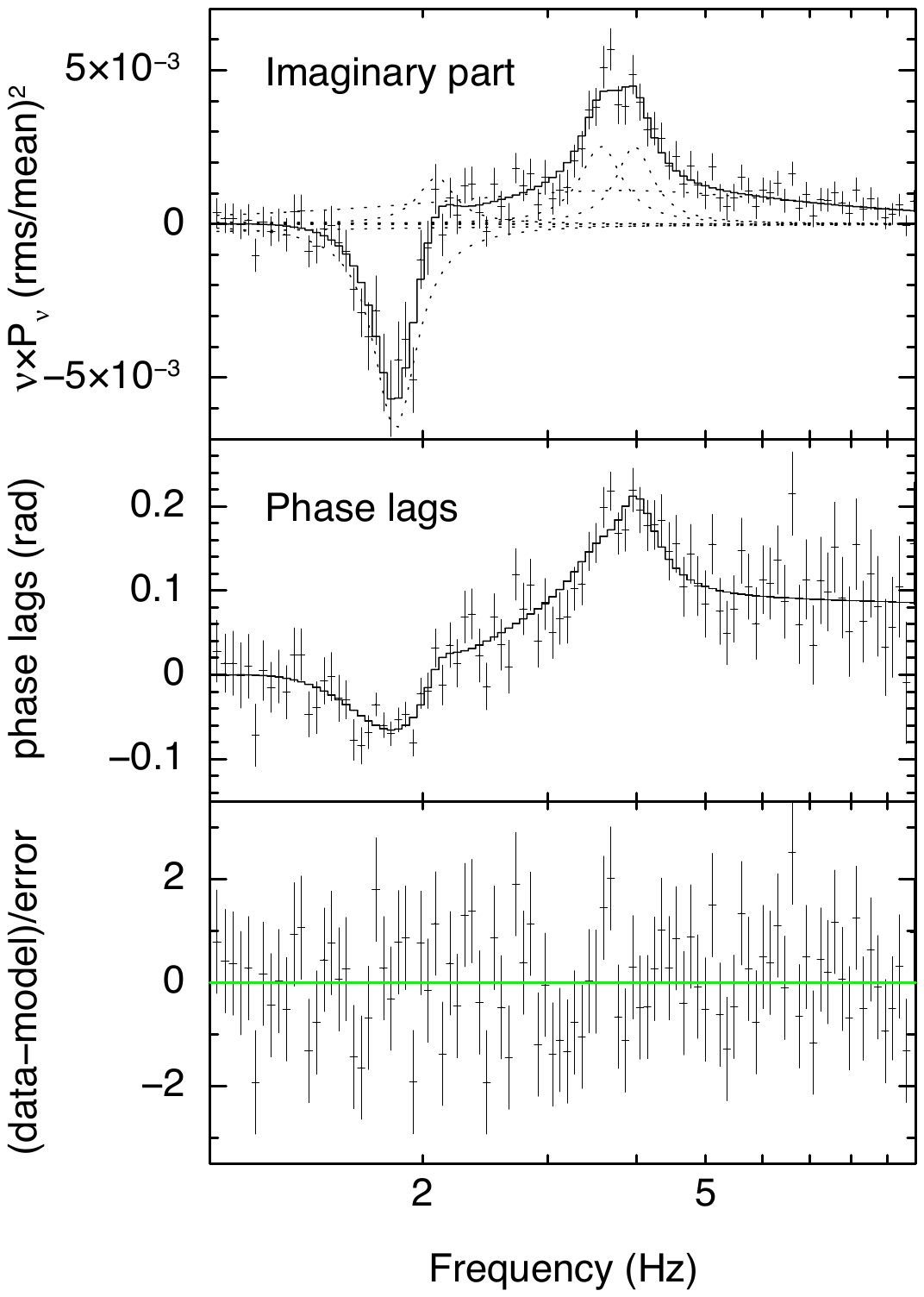}
\caption{Left panel: Imaginary part of the CS of the RXTE observation 30703-01-34-00 of GRS 1915+105 fitted with six Lorentzian functions. The upper panel shows the data and the best-fitting model (thick solid line) and the Lorentzian functions (thin dotted lines), while the lower panel shows the residuals with respect to the best fitting model. Structured residuals are visible at $\sim2$ Hz and $\sim 4$ Hz, just above the frequency of, respectively, the QPO fundamental and the second harmonic (see Table~\ref{tab:data-1915_1}). 
Middle panel: PS of the same observation, but now fitted with eight Lorentzian functions. The upper panel shows the data and the best-fitting model (thick solid line) and the Lorentzian functions (thin dotted lines), while the lower panel shows the residuals with respect to the best fitting model. \citep[The eighth Lorentzian is the high-frequency bump,][which is only visible above 10 Hz.]{Trudolyubov-2001,Zhang-2022}
Right panel: Imaginary part of the CS  fitted with eight Lorentzians (upper panel), phase lags vs. Fourier frequency (middle panel) and residuals of the phase lags with respect to the derived model (bottom panel). In all cases, during the fits we let the centroid frequency and FWHM of each component free to vary but we link them to be the same in the PS and the CS, further assuming the constant phase-lags model  (see Table~\ref{tab:definitions} and \S\ref{phaselagmodel}).}
\label{fig2}
\end{figure*}

Finally, we note that the second harmonic of the QPO is also fitted by two Lorentzian components (see Fig.~\ref{fig1}), a narrow one at $11.07 \pm 0.06$ Hz and a broad one at $11.4 \pm 0.3$ Hz, both of them significant (11.2-$\sigma$ and 3.4-$\sigma$, respectively). The frequencies of both these components are consistent with being in a 2:1 relation with, respectively, the QPO fundamental and the QPO shoulder. The phase lags of these two components are, respectively, $0.25 \pm 0.11$ rad and $-0.07 \pm 0.04$ rad, and the time lags are, respectively, $3 \pm 2$ ms and $-2.3 \pm 0.4$ ms, marginally consistent within errors; we therefore cannot assess whether the QPO second harmonic and its shoulder are different components or the same component that drifts slightly during the observation period. Guided by this, in the next section we explore the presence of a shoulder of the QPO second harmonic in an observation of GRS 1915+105.

We subsequently compute PS of this same observation for the channels $0-13$ and $14-249$, using the same parameters mentioned at the start of this subsection. We fit these two PS together with the CS in the same bands above with a model consisting of nine Lorentzians\footnote{The model requires two extra Lorentzians when we fit the PS in the two bands instead of the PS in the full band as we did above. The two extra Lorentzians are the ones that peak at $\sim 1.5$ Hz and $\sim 4.75$ Hz in the plot of the Real part of the CS in the top-right panel of Figure~\ref{fig1}.}, and assuming the constant time-lags model. In the right panels of Figure~\ref{fig1} we plot the Real (top) and Imaginary (middle) parts of the CS and the intrinsic coherence with the derived model (bottom). The derived model reproduces the data rather well, with the largest deviations appearing above $\sim 10$ Hz, where the signal-to-noise ratio drops very quickly. The vertical lines in the bottom-right panel mark the characteristic frequency of each Lorentzian, $\nu_{\rm max} = \sqrt{\nu^2_0 + \left(\frac{\mathrm{FWHM}}{2}\right)^2}$ \citep{Belloni-2002}.  We note that the changes in the behaviour of the coherence happen at $\nu_{\rm max}$ and not at $\nu_0$ of the corresponding Lorentzians.

The coherence is $\sim 1$ at low frequencies where a single, broad, Lorentzian with $\nu_{\rm max} = 1$ Hz dominates (see top-right panel of Figure~\ref{fig1} for the Lorentzians mentioned in this part of the text); the coherence then drops at $\sim 1.5$ Hz when a new Lorentzian with $\nu_{\rm max} = 1.41$~Hz appears. This Lorentzian is not strong enough to lead to a peak in the coherence, but its presence interferes with the previous Lorentzian and leads to a drop. After that, the coherence drops more or less steadily over a frequency range in which the Lorentzian with $\nu_{\rm max} = 1$~Hz and two other Lorentzians with, respectively, $\nu_{\rm max} = 2.74$~Hz and $\nu_{\rm max} = 4.04$~Hz, alternate dominance. The coherence goes up again at $\sim 5$~Hz, where the strong QPO starts to dominate the variability. As the frequency increases further, in the model the coherence decreases and increases again going from the QPO to the shoulder, but the errors in the data are too large to see that. At higher Fourier frequency the model is consistent with the data, but there, the coherence has large errors.

The results of this part show that the best-fitting model of the PS and the CS correctly describes both the phase-lag spectrum and the coherence function. These two outcomes align with the model's predictions, which are based on the assumptions outlined in Section \S\ref{mathematical}.

\subsection{Case study 2: Shoulders of the fundamental and second harmonic of the type-C QPO in GRS 1915+105}
\label{1915_1}

For the next example we use the \textit{RXTE} observation 30703-01-34-00 of GRS 1915+105 with a strong type-C QPO at $\sim 1.8$ Hz, which was analysed by \cite{Zhang-2020}. We select this observation because the type-C QPO shows a significant second harmonic at $\sim 3.6$ Hz that displays a double peak in the phase-lag frequency spectrum. Motivated by this, and after our findings in \S\ref{339-4}, we decided to study the potential presence of a shoulder to the QPO fundamental and the second harmonic.

We take the Single-Bit and Event Mode data of this observation to compute the full-band (channels 0 to 249) PS and the CS of photons with energies in the $5.7-100$ keV band (channels 14 to 249) relative to those with energies in the $2-5.7$ keV band (channels 0 to 13). Both for the PS and the CS we use $T_{\rm FFT}=256$ s, which yields $\nu_{\rm min}=\Delta\nu=1/256$ Hz, at a time resolution $\Delta t= 1/8192$ s such that $\nu_{\rm Nyquist}=4096$ Hz.

We initially fit the full-band PS in the range $0.1-64$ Hz adding one Lorentzian at a time as described in \S\ref{examples}, and find that we need a model consisting of six Lorentzians. The fit gives $\chi^2=144.7$ for 143 dof, 
and all Lorentzians are at least 3-$\sigma$ significant.

\textit{Constant phase-lags model:} We next fit simultaneously the PS and the CS with the same number of Lorentzians, fixing the frequency and FWHM of each Lorentzian to the values that we obtain from the fit to the PS; for this fit we assume the constant phase-lags model. The fit gives $\chi^2=684.0$ for 619 dof, with structured residuals in the Imaginary part of the CS around 2 Hz and 4 Hz, close to the frequencies of the QPO fundamental and second harmonic. We next let the frequency and FWHM of each Lorentzian free but linked in the PS and CS. This fit gives $\chi^2=675.3$ for 607 dof, but the residuals around 2 Hz and 4 Hz in the Imaginary part of the CS remain. We show the Imaginary part of the CS and the residuals of the fit with six Lorentzians in the left panel of Figure~\ref{fig2}.

%
%
\begin{table}
    \centering
    \caption{Phase and time lags ($5.7-100$ keV vs. $2-5.7$ keV) of the QPO fundamental and second harmonic and the corresponding QPO shoulders in GRS 1915+105 with our method}
    \label{tab:data-1915_1}
    \footnotesize
\hspace{-0.5cm}    
    \begin{tabular}{cccc}
\hline
\textbf{Component} & $\nu_0$ (Hz) & Phase lags$^{\dagger}$ (rad) & Time lags$^{\ast}$ (ms)\\    
\hline
QPO fund  &  $1.822 \pm 0.006$ & $-0.076 \pm 0.006$ & $-7.1 \pm 0.7$\\
Should. fund & $2.11 \pm 0.06$ & $1.9^{\,\,+0.7}_{-1.0}$ &  $114 \pm 20$ \\ 
2nd harm & $3.56 \pm 0.02$ & $0.20 \pm 0.2$ & $11.2 \pm 1$ \\
Should. 2nd &$4.00\pm 0.03 $ & $0.49 \pm 0.08$ & $21.7 \pm 5$ \\
\hline
\multicolumn{4}{l}{\small $^{\dagger}$Using the constant phase-lags model.}\\
\multicolumn{4}{l}{\small$^{\ast}$Using the constant time-lags model.}\\
\end{tabular}
\end{table}

We therefore add two new Lorentzians to the model at, respectively, $\sim 2$ Hz and $\sim 4$ Hz with, as for the other Lorentzians, the centroid frequency and FWHM free and linked in the PS and CS. This fit, shown in the middle and right panels of Figure~\ref{fig2}, gives $\chi^2=627.1$ for 598 dof, significantly better than the one with six Lorentzians (and also better than a model with seven Lorentzians not shown here), and the fit no longer shows the structured residuals at around 2 Hz and 4 Hz. In fact all Lorentzians are at least 3-$\sigma$ significant in either the PS, the CS or both (see below). We therefore adopt this one as the final model. 

%
%
\begin{figure*}
\centering
\hspace{-0.5cm}
\includegraphics[width=0.38\textwidth]{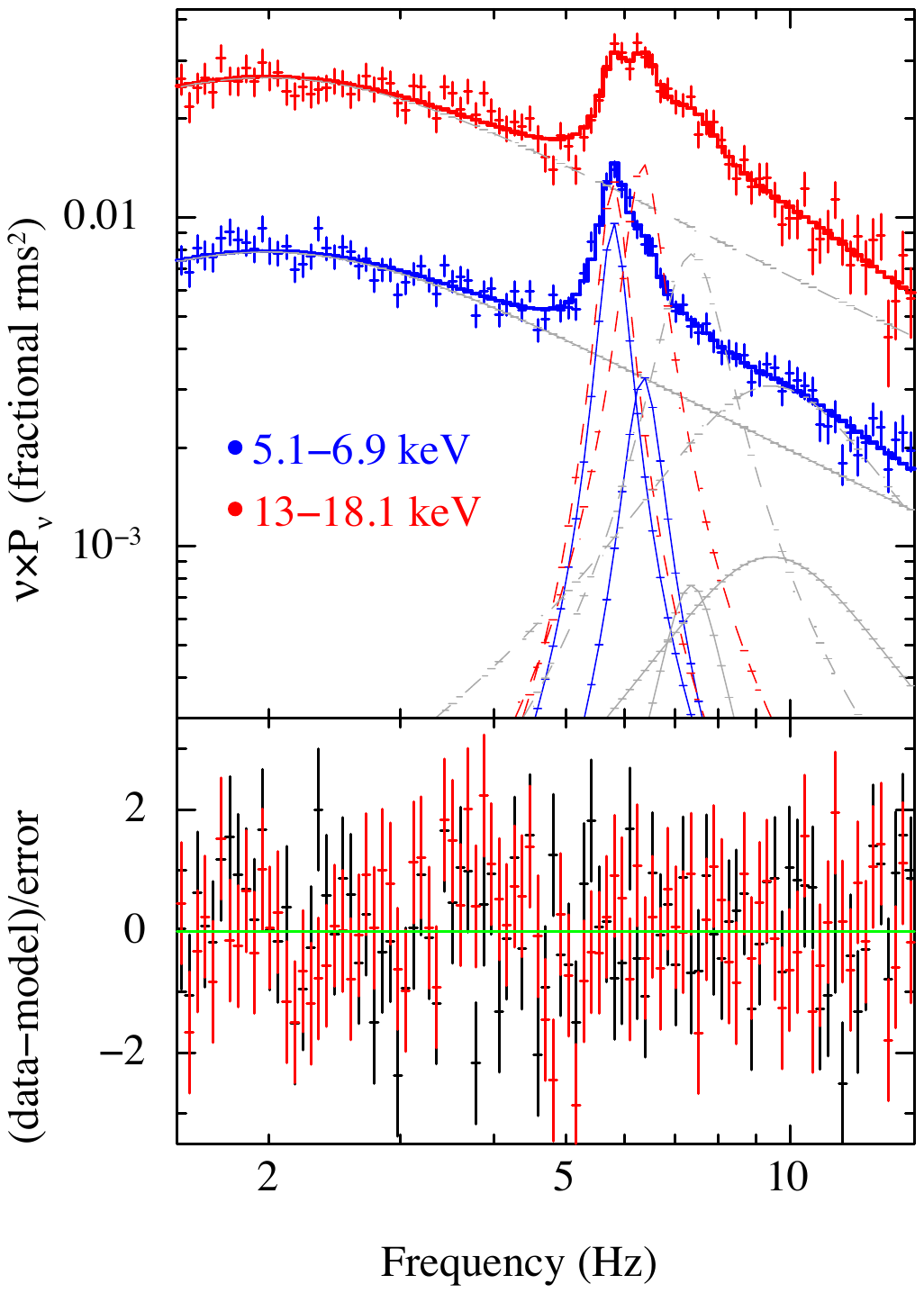}
\hspace{-1.2cm}
\includegraphics[width=0.38\textwidth]{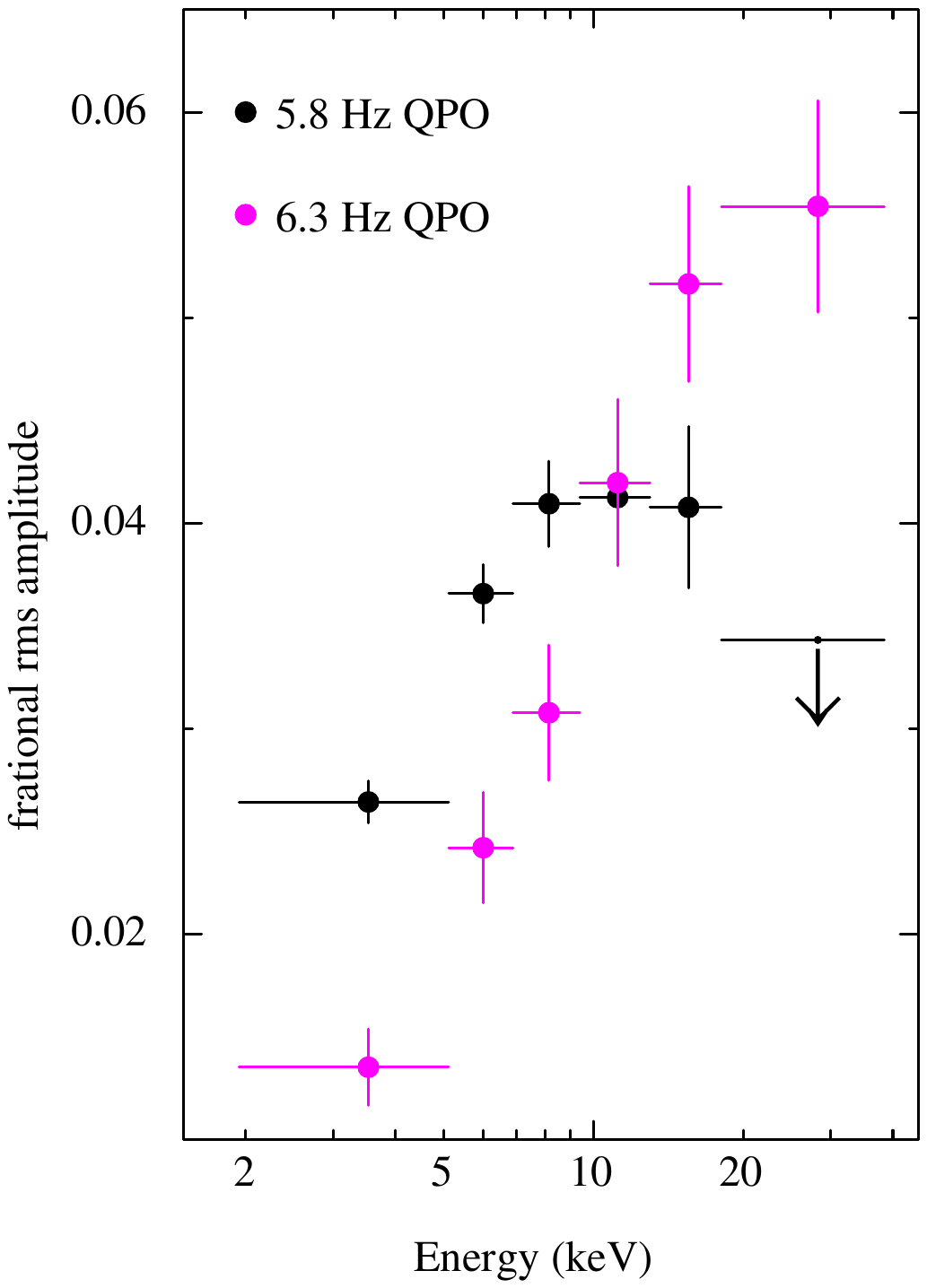}
\hspace{-1.2cm}
\includegraphics[width=0.38\textwidth]{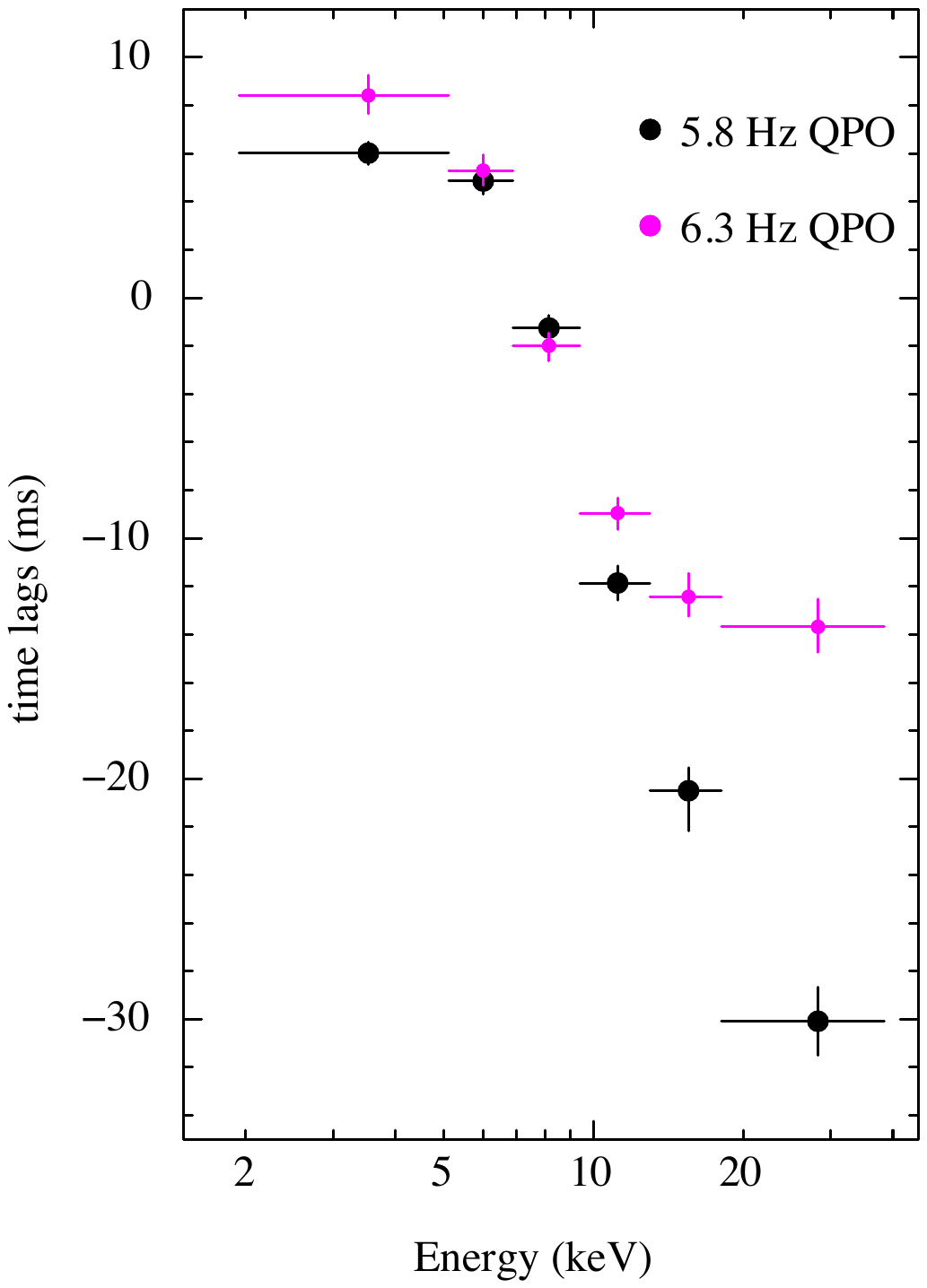}

\caption{Left panel: PS of the second segment of the RXTE observation 10408-01-27-00 of GRS 1915+105 in two energy bands (see legend). The upper panel shows the data and the best-fitting model (thick solid line), consisting of five Lorentzian functions (thin solid and dashed lines). We fitted the two PS simultaneously with the centroid frequency and FWHM of all Lorentzians linked so that they are the same in both bands. The two Lorentzians plotted using thin solid blue lines represent, respectively, the main QPO at 5.8 Hz and the QPO shoulder at 6.3 Hz in the $5.1-6.9$ keV band; similarly, the Lorentzians plotted using the thin dashed red lines represent, respectively, the main QPO at 5.8 Hz and the QPO shoulder at 6.3 Hz in the $13-18.1$ keV band. The bottom panel shows the residuals with respect to the best fitting model. 
Middle panel: Fractional rms amplitude of the main QPO at 5.8 Hz (black) and the QPO shoulder at 6.3 Hz (purple) vs. energy.
Right panel: Time lags of the main QPO and the QPO shoulder vs. energy. The reference band for the lags is the full band (channels 0-249).}
\label{fig3}
\end{figure*}

The middle panel of Figure~\ref{fig2} shows the PS (upper panel) and the residuals (bottom panel) of the fit with the adopted model with eight Lorentzians. As is apparent from the Figure, both the QPO fundamental and the QPO second harmonic are fitted with two Lorentzians, a relatively broad one that fits the bulk of the QPO profile, and a relatively narrow one at slightly higher frequency than the other one. While both the fundamental QPO at $\nu_{\rm QPO} = 1.822 \pm 0.006$ Hz and the second harmonic at $\nu_{\rm 2nd}= 3.56 \pm 0.02$ Hz are significantly required by the fit (24.6-$\sigma$ and 6.8-$\sigma$, respectively), the shoulder of the fundamental QPO at $\nu_{\rm shoulder,1} = 2.11 \pm 0.06$ Hz and that of the second harmonic at $\nu_{\rm shoulder,2} = 4.00 \pm 0.03$ Hz are not detected significantly in the PS ($< 1$-$\sigma$ and $2.6$-$\sigma$, respectively). However, both shoulders are significantly required (4-$\sigma$ and $\sim 5.5$-$\sigma$, respectively) to fit the CS. This is remarkable, since a component that is not required to fit the PS is significantly required to fit the CS. As this occurs again in \S\ref{1820}, we expand on this there. 

The right panel of Figure~\ref{fig2} shows the Imaginary part of the CS (upper panel), the phase-lag spectrum (middle panel) and the residuals of the phase-lag spectrum with respect to the derived model with eight Lorentzians (bottom panel; we do not plot the Real part of the PS because it looks very similar to the PS ). We give the frequencies and phase lags of the fundamental, second harmonic and the shoulders of the QPO fundamental and second harmonic in Table~\ref{tab:data-1915_1}. The phase lags of the shoulder of the second harmonic are significantly (3.5-$\sigma$) larger than those of the second harmonic itself, but the phase lags of the shoulder of the fundamental have very large errors, therefore we cannot conclude whether the phase lags of the shoulder and the fundamental QPO are different.  

Using the traditional method, the lags of the QPO, the QPO shoulder, the second harmonic and the shoulder of the second harmonic are, respectively, $-0.056 \pm 0.005$ rad, $-0.001 \pm 0.009$ rad, $0.123 \pm 0.008$ rad and $0.16 \pm 0.01$ rad. Given that the error bars in our method are relatively large compared to those in the traditional method (see Appendix \ref{simulations} for a discussion of this), the lags of the QPO fundamental, the QPO shoulder and the second harmonic are all consistent with being the same using either the traditional or our method (see Table~\ref{tab:data-1915_1}). On the contrary, the lags of the shoulder of the second harmonic using our and the traditional method are $4\sigma$ different. In Appendix \ref{simulations} we use simulations to explore the difference in the results of our and the traditional method.

\textit{Constant time-lags model:} When we fit the data assuming the constant time-lags model (this fit is marginally worse, $\chi^2=633.7$ for 598 dof, than the one with the constant phase-lags model), the time lags of the fundamental QPO and those of its shoulder are 6.1-$\sigma$ different, whereas the time lags of the second harmonic and its shoulder are only 2.1-$\sigma$ different. We give the time lags from this fit in Table~\ref{tab:data-1915_1}.

To summarise, we find that in this observation of GRS 1915+105 either the phase or the time lags of the shoulders are significantly different from those of the corresponding QPO, and hence the shoulders are not due to small drifts of the QPO frequency during the observation, but are separate components in the PS or the CS of the source.

\subsection{Case study 3: No energy dependence of the QPO frequency in GRS 1915+105}
\label{1915_2}

%
%
\begin{table*}
    \centering
    \caption{Phase and time lags (given band vs. the full band) of the QPO and QPO shoulder in GRS 1915+105 with our method}
    \label{tab:data-1915_2}
    \footnotesize
\hspace{-0.5cm}    
    \begin{tabular}{cccccc}
\hline
\textbf{Component} & $\nu_0$ (Hz) & Phase lags$^{\dagger}$ (rad) & Time lags$^{\ast}$ (ms) & Phase lags $^{\dagger}$ (rad) & Time lags$^{\ast}$ (ms) \\    
\multicolumn{2}{l}{} & \multicolumn{2}{c}{($5.1-6.9$ keV)} & \multicolumn{2}{c}{($13-18.1$ keV)}\\
\hline
QPO &  $5.78 \pm 0.02$ & $0.16 \pm 0.03$ & $4.9 \pm 0.5$ & $-0.61 \pm 0.06$ & $-20 \pm 1$\\
QPO shoulder & $6.34 \pm 0.06$ & $0.24 \pm 0.08$ &  $5.5 \pm 0.6$ & $-0.36 \pm 0.08$ &  $-12.5 \pm 0.8$\\ 
\hline
\multicolumn{6}{l}{\small $^{\dagger}$Using the constant phase-lags model.}\\
\multicolumn{6}{l}{\small$^{\ast}$Using the constant time-lags model.}\\
\end{tabular}
\end{table*}

Now that we have established that the shoulders of the QPOs are separate components, we explore whether previous claims that the QPO frequency changes with energy \citep{Qu-2010,Li-2013a,Li-2013b,Yan-2018} are connected to the presence of a shoulder of the QPO. For this we analyse the \textit{RXTE} observation 10408-01-27-00 of GRS 1915+105 for which \cite{Qu-2010} reported a significant change of the centroid frequency of the QPO with energy, from $\nu_{\rm QPO}=5.9$ Hz for energies below 5 keV to $\nu_{\rm QPO}=6.7$ Hz for energies above 20 keV (see their Figure 2). We speculate that this apparent change of the QPO frequency with energy could be due to the presence of a shoulder if the rms spectrum of the QPO and the shoulder are different. We will therefore call the QPO reported by \cite{Qu-2010} the ``QPO feature'', since it could be a combination of a QPO and a QPO shoulder.

We take the Binned and Event Mode data of this observation to compute the full-band (channels 0 to 249) spectrogram \citep[also known as dynamical PS; e.g.,][]{Markwardt-1999} of the observation, and find that the frequency of the QPO feature changes with time (not shown), from above $\sim 6.5$ Hz at the start to $\sim 5$ Hz at the end of the observation; the frequency of the QPO feature appears to be more or less constant at $\sim 6$ Hz in the second segment of the observation, therefore we use only that segment for the rest of the analysis. We then compute PS in the same six energy bands used by \citet[][channels $0-13$, $14-18$, $19-25$, $26-35$, $36-49$ and $50-103$; see their Table 2 for the energy ranges]{Qu-2010} to study the rms spectrum of the QPO feature, plus the CS of the photons in the individual bands with respect to those in the full band (channels 0 to 249). Both for the PS and the CS we use $T_{\rm FFT}=128$ s, yielding $\nu_{\rm min}=\Delta\nu=1/128$ Hz, at a time resolution $\Delta t= 1/512$ s that yields $\nu_{\rm Nyquist}=256$ Hz. 

The left panel of Figure~\ref{fig3} shows the PS of the data during the second segment of the observation in two energy bands, $5.1-6.9$ keV and $13-18.1$ keV. We fit the two PS simultaneously in the range $0.5-15$ Hz with five Lorentzians, with the frequency and FWHM of each Lorentzian free but linked to be the same in both energy bands. As it is apparent from the Figure, the QPO feature is fitted with two Lorentzians, one at $\sim 5.8$ Hz and the other at $\sim 6.3$ Hz. Notably, the Lorentzian at 5.8 Hz is more or less equally strong in the two bands whereas the Lorentzian at 6.3 Hz is stronger in the $13-18.1$ keV band than in the $5.1-6.9$ keV band. (It is also true that all the other Lorentzian components in the PS are stronger in the high- than in the low-energy band.) As before, we call these two Lorentzian components the QPO and the shoulder of the QPO, respectively. 

%
%
\begin{figure*}
\centering
\includegraphics[width=0.5\textwidth]{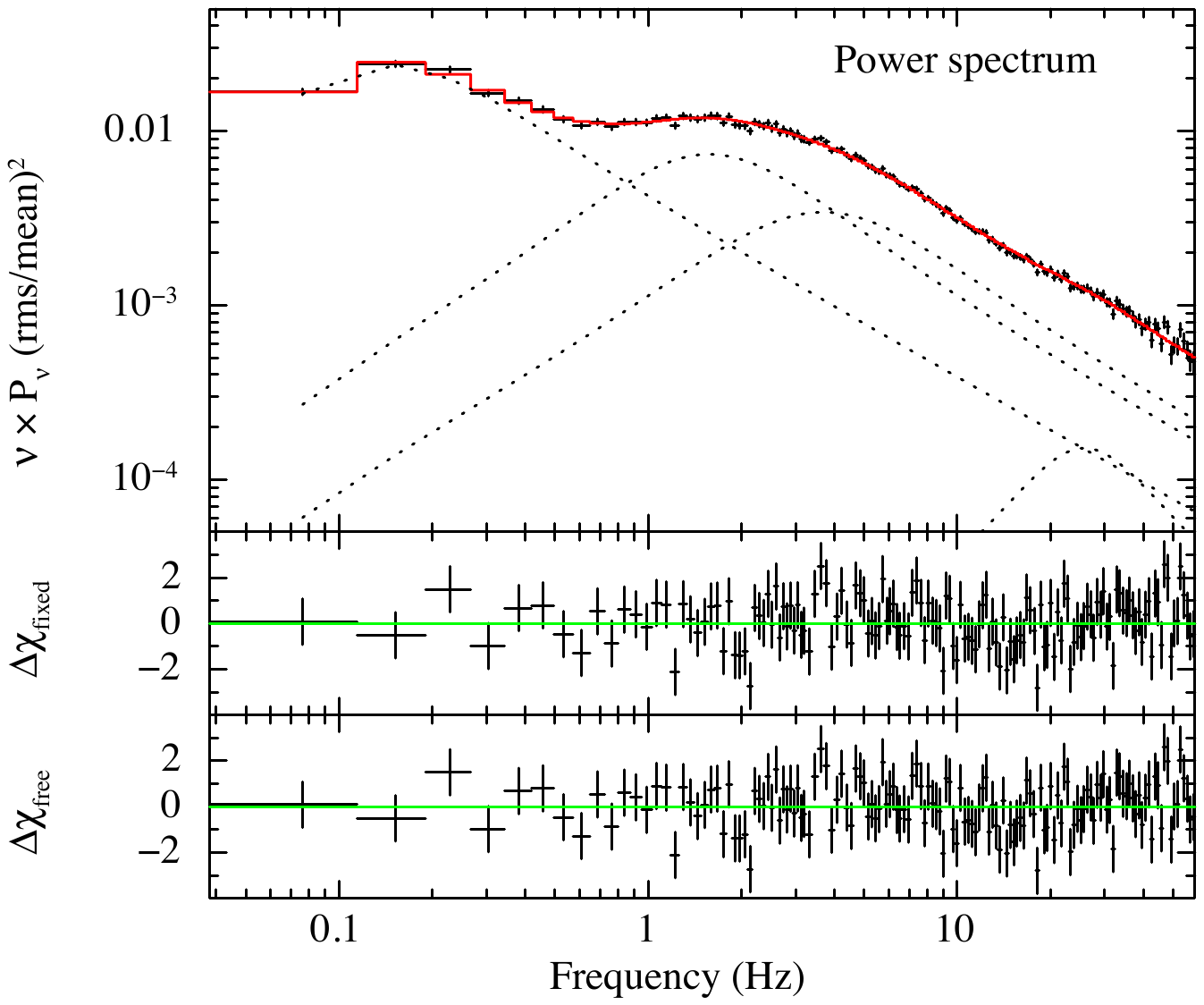}
\hspace{-1cm}
\includegraphics[width=0.5\textwidth]{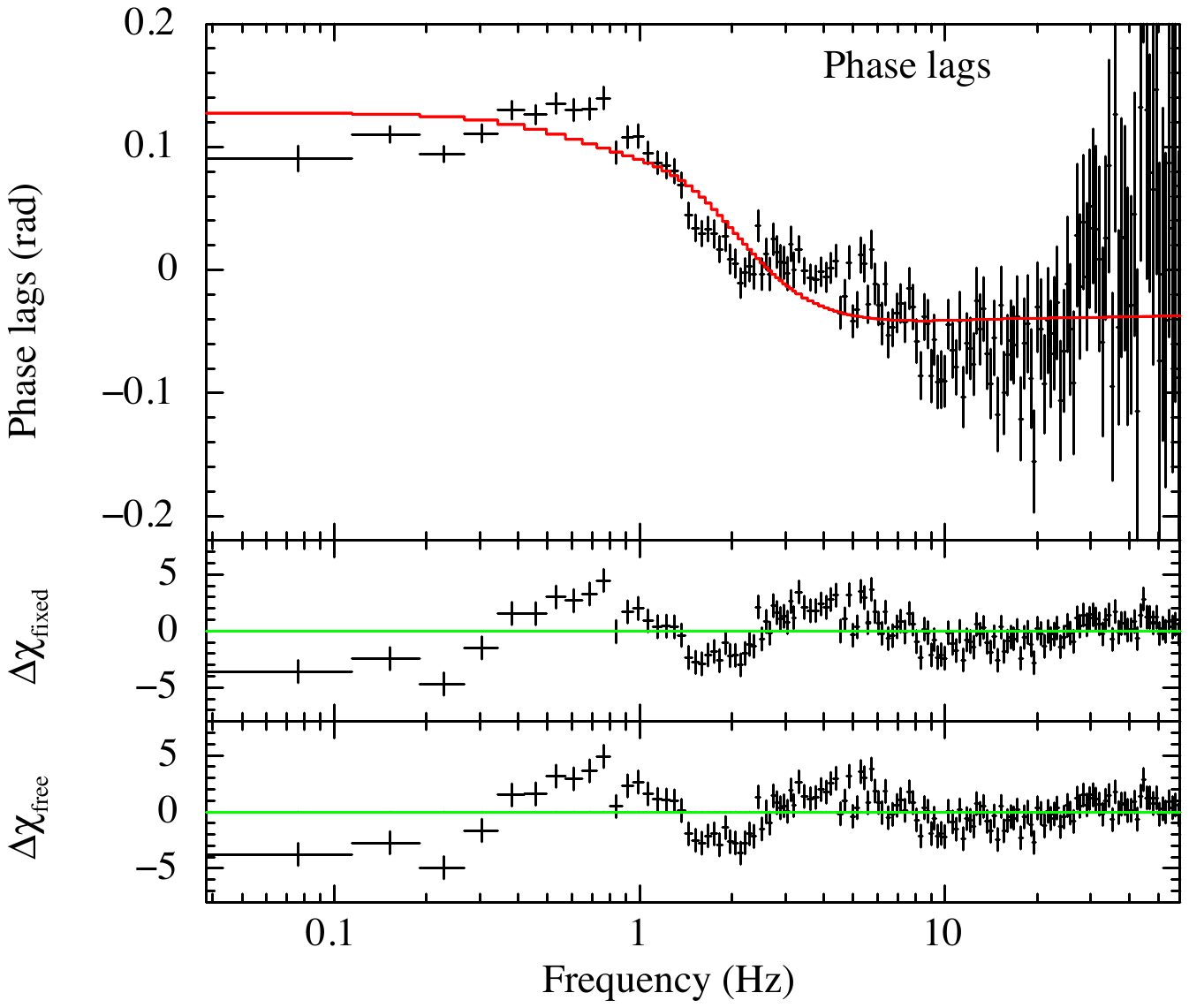}\\
\includegraphics[width=0.5\textwidth]{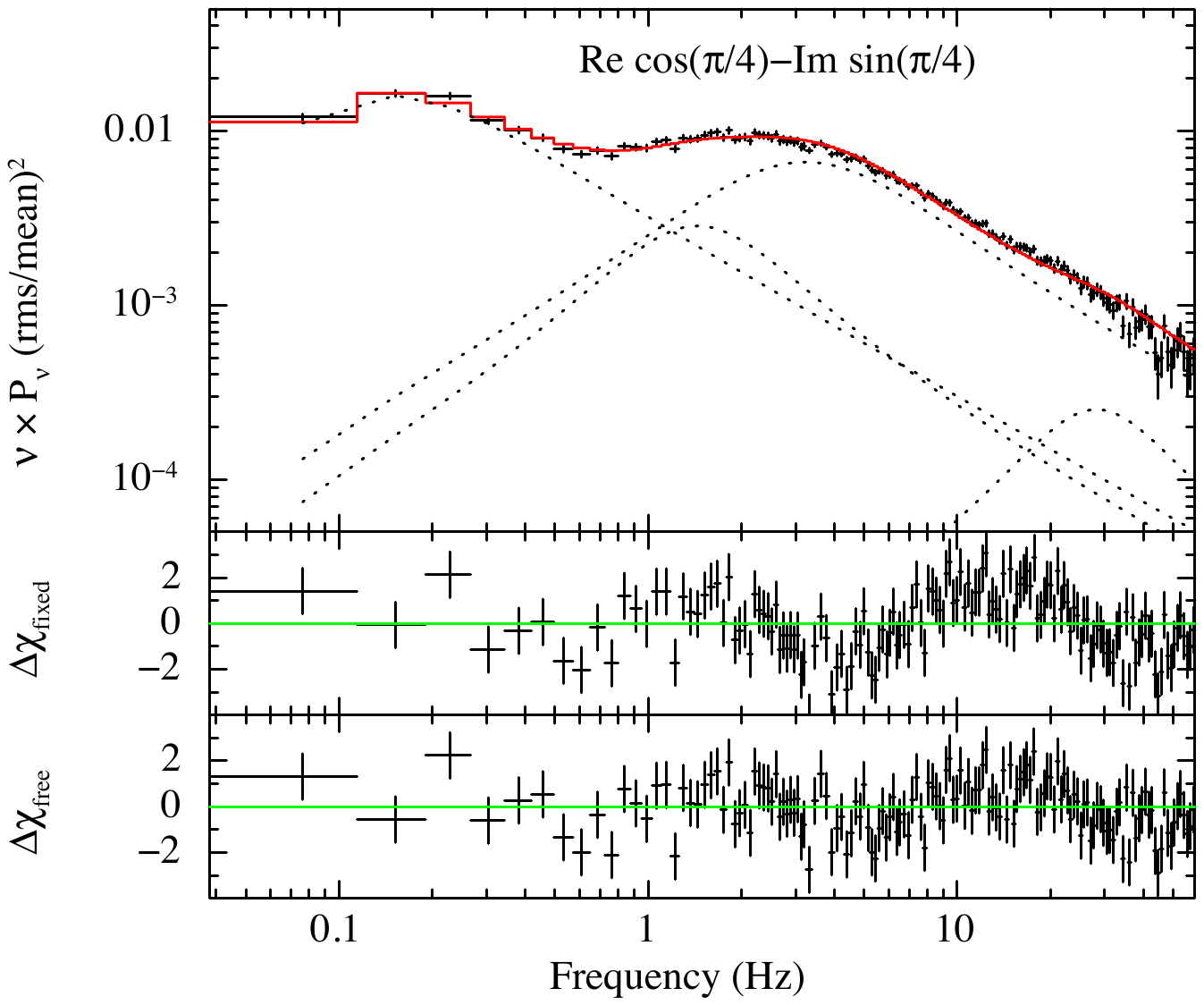}\hspace{-1cm}
\includegraphics[width=0.5\textwidth]{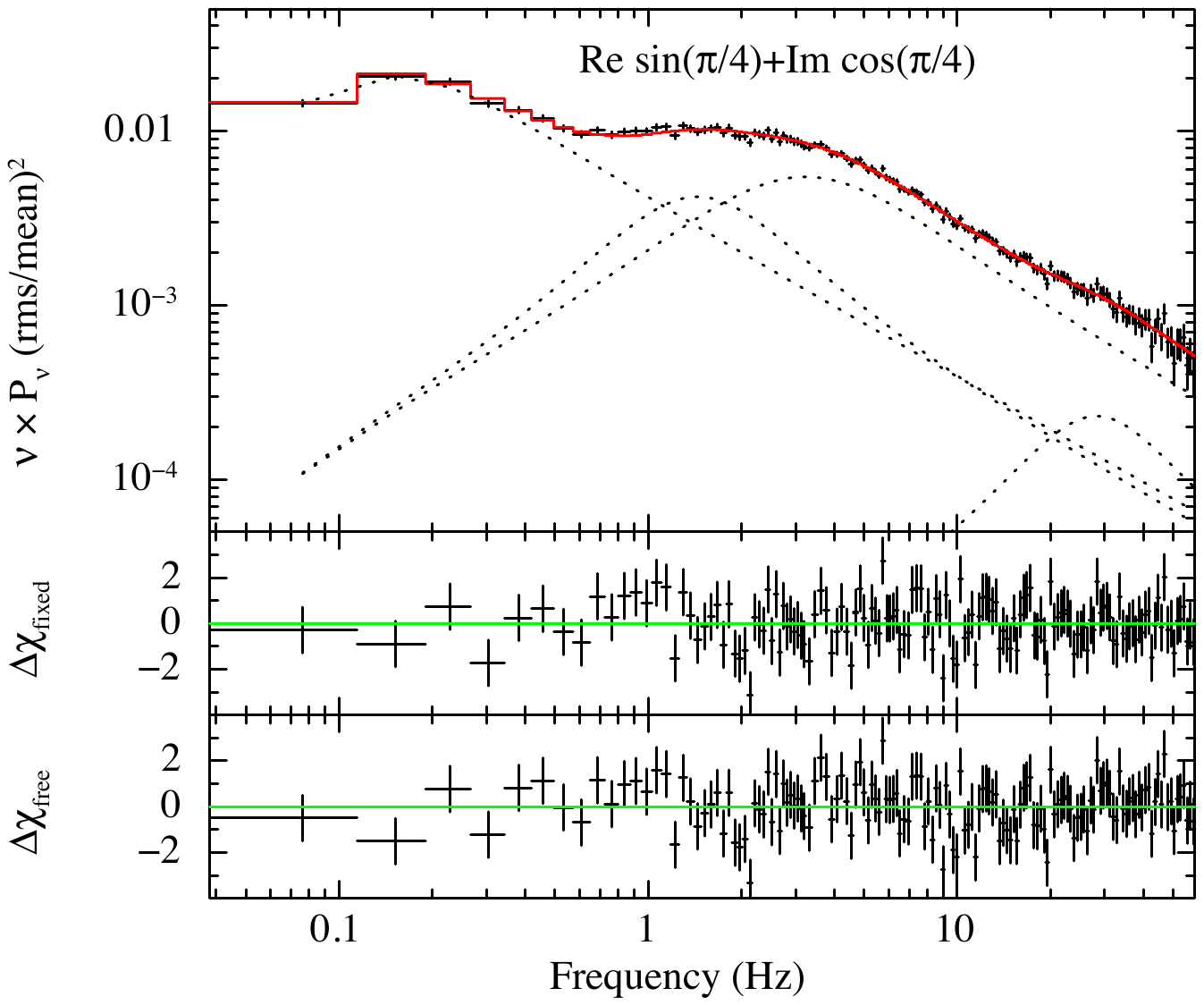}\caption{PS in the $0.3-12$ keV energy band (top left), Real (bottom left) and Imaginary (bottom right) part of the CS for the $2-12$ keV band with respect to the $0.3-2$ keV band  of the \textit{NICER} observation 1200120120 of MAXI J1820+070 fitted with a model (red solid line) consisting of four Lorentzians (dotted lines). For the fitting and the plot we rotate the cross vector by $45^\circ$ (see text for details), we assume the constant phase-lags model (see Table~\ref{tab:definitions} and \S\ref{phaselagmodel}) and we fix the frequency and FWHM of the Lorentzians to the values we obtain from the best-fitting model of the PS. The top right panel shows the phase-lag spectrum with the derived model. In each case the top sub-panels show the data and the model and the middle sub-panels show the residuals. The bottom sub-panels show the residuals when we fit the data letting the frequency and FWHM of the Lorentzians free, but linked to be the same in the PS and the CS.
}
\label{fig4}
\end{figure*}

In the $5.1-6.9$ keV band the rms amplitude of the QPO and the shoulder are rms$_{\rm QPO} = 3.6 \pm 0.3 \%$ (5-$\sigma$ significant) and rms$_{\rm shoulder} = 2.4 \pm 0.3 \%$ (3.7-$\sigma$ significant). In the $13-18.1$ keV band the rms amplitude of the QPO and the shoulder are rms$_{\rm QPO} = 4.2 \pm 0.4 \%$ (4.4-$\sigma$ significant) and rms$_{\rm shoulder} = 5.1 \pm 0.7 \%$ (4.2-$\sigma$ significant). We give the frequencies, phase lags (obtained using the constant phase-lags models) and time lags (using the constant time-lags model) of the QPO and the shoulder in Table~\ref{tab:data-1915_2}. While the phase lags ($13-18.1$ keV with respect to $5.1-6.9$ keV) of the QPO and the QPO shoulder are consistent with being the same within 3-$\sigma$, their time lags are 4.9-$\sigma$ different. (See the discussion about the difference between phase and time lags at the end of \S\ref{examples}.) 

%
%
\begin{figure*}
\centering
\includegraphics[width=0.5\textwidth]{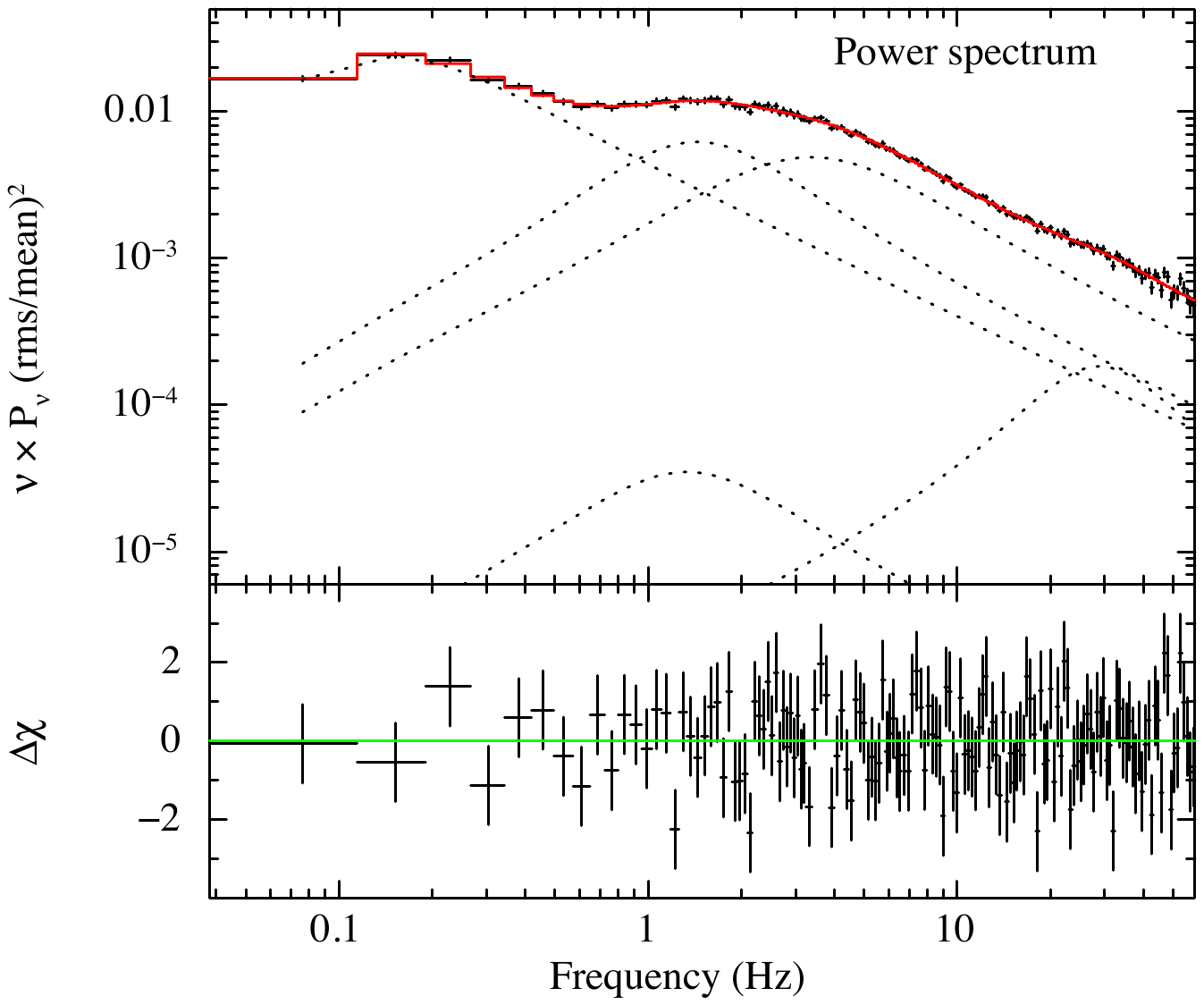}\hspace{-1cm}
\includegraphics[width=0.5\textwidth]{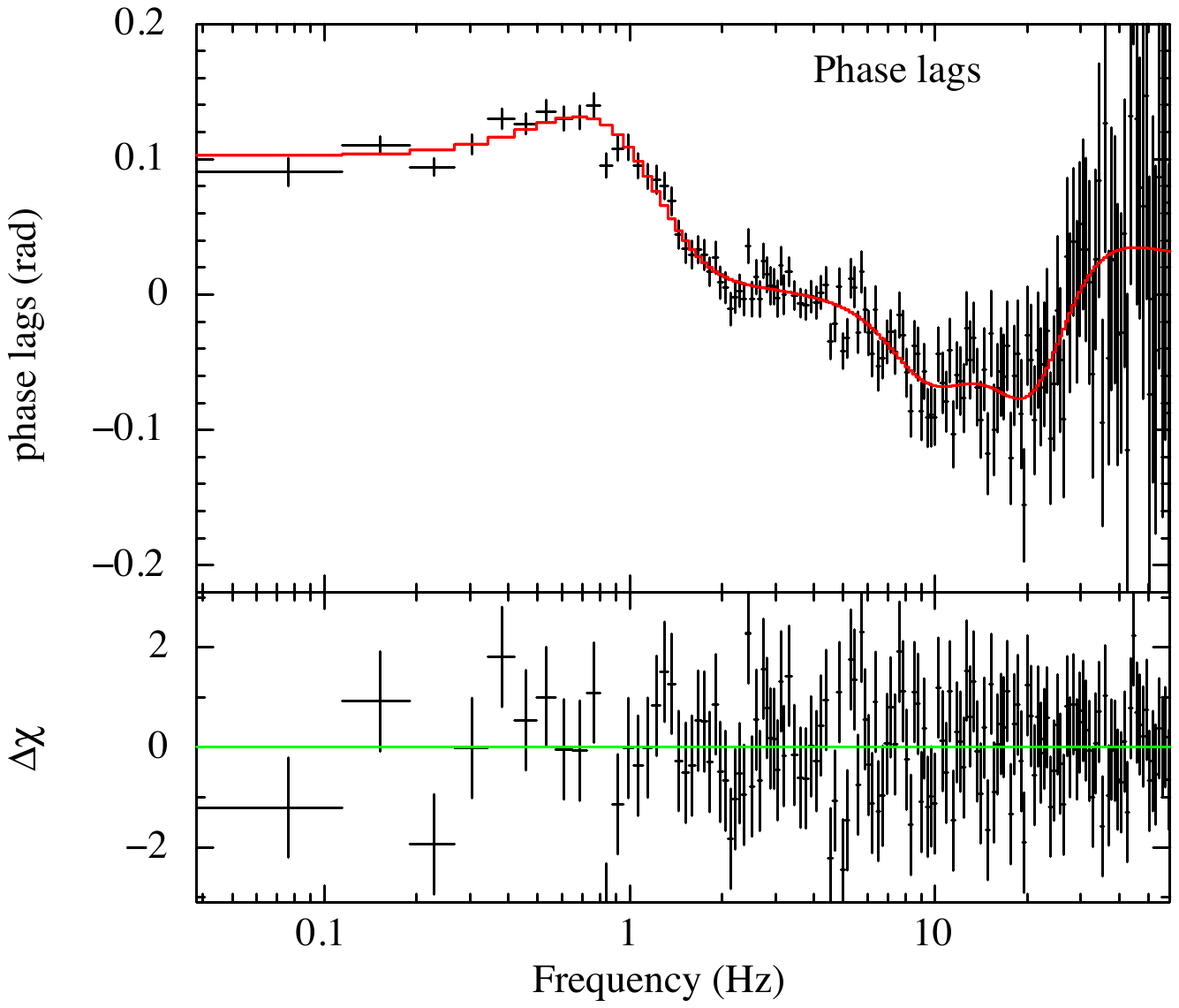}\\
\includegraphics[width=0.5\textwidth]{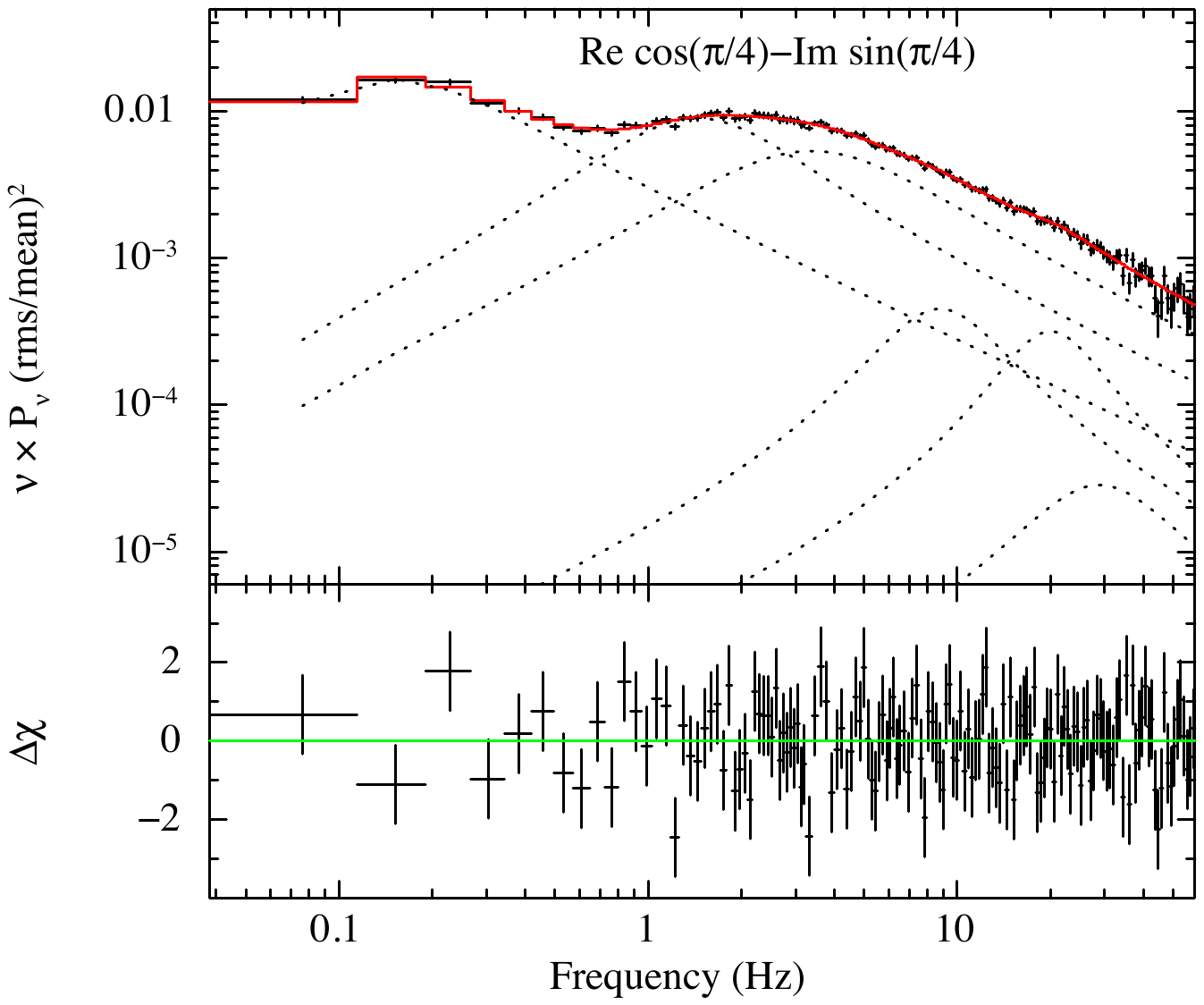}\hspace{-1cm}
\includegraphics[width=0.5\textwidth]{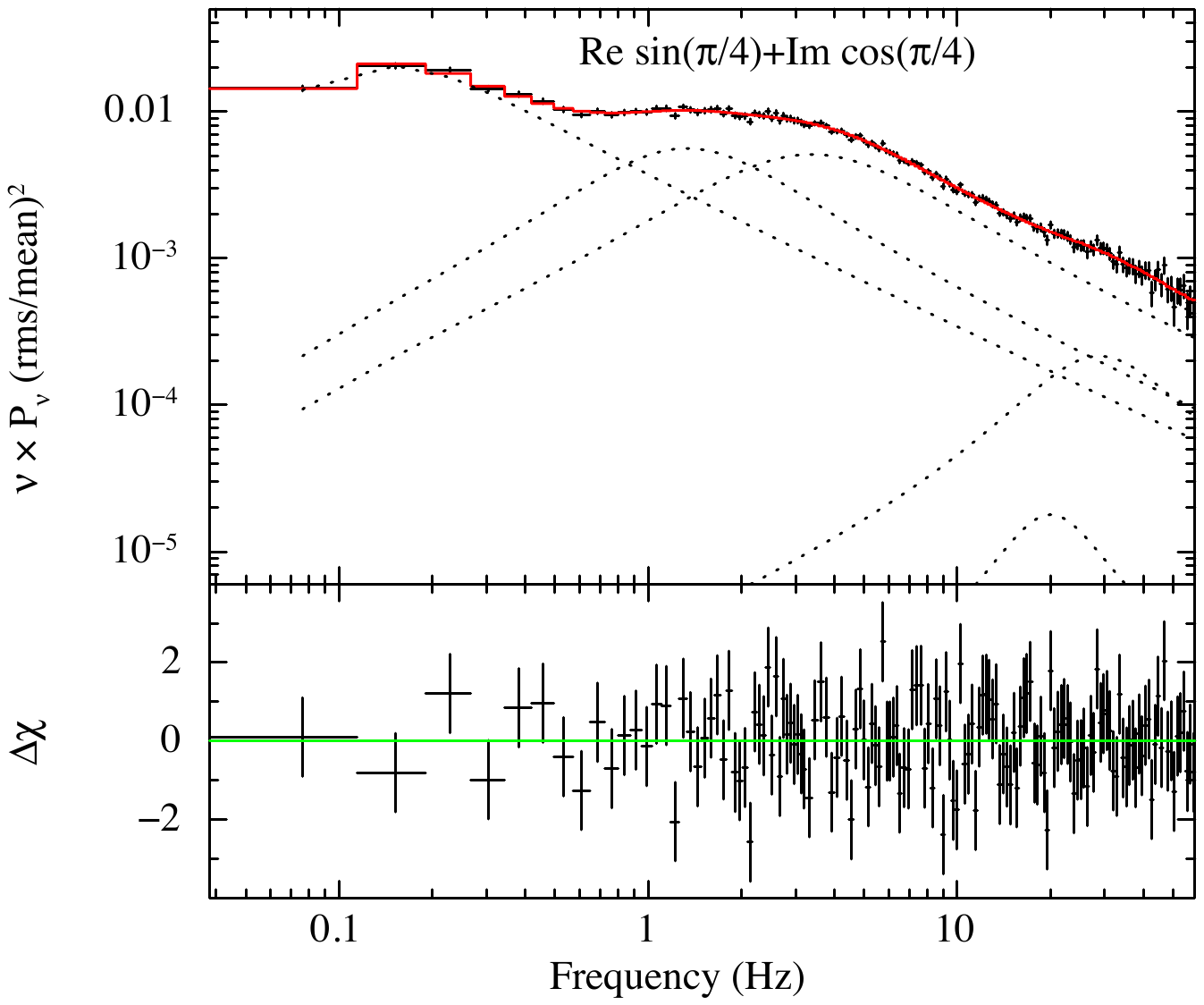}
\caption{Same as Figure~\ref{fig4} but now fitted with seven Lorentzians instead of four, assuming the constant phase-lags model. The bottom sub-panels show the residuals with respect to the best-fitting model.
}
\label{fig5}
\end{figure*}

The middle panel of Figure~\ref{fig3} shows the rms spectrum of the QPO and the QPO shoulder. From this Figure it is apparent that the rms amplitude of the QPO increases with energy up to about 10 keV and at that energy it levels off or decreases slightly, whereas the rms amplitude of the QPO shoulder increases steadily with energy, also above $\sim 10$ keV. Because of this, at energies below $\sim 8$ keV the QPO is stronger than the QPO shoulder, whereas above that energy the opposite is true. The right panel of Figure~\ref{fig3} shows the time-lag spectrum of the QPO and the QPO shoulder obtained using the constant time-lags model and taking the full band as reference. (The phase-lag spectrum, from the fit with the constant phase-lags model, looks almost exactly the same as this one, with slightly larger error bars.) The time lags of both the QPO and the QPO shoulder become more negative as the energy increases \citep[compare with Fig. 2b in][in which they plot the phase lags with respect to the lowest energy band]{Qu-2010}; both time-lag spectra are consistent with being the same up to $\sim 10$ keV, at which point the time lags of the QPO continue decreasing while the time lags of the QPO shoulder level off. 

The fits of the PS in two energy bands (Fig.~\ref{fig3}, left panel), the rms spectrum (Fig.~\ref{fig3}, middle panel) and the time-lag spectrum (Fig.~\ref{fig3}, right panel) of the QPO and the QPO shoulder show that the data are consistent with the presence of two separate components, the QPO and the QPO shoulder, with different rms and time-lag spectra. In our model, the centroid frequency of both the QPO and the QPO shoulder are linked to be the same in the different energy bands, it is only that the change of the relative strength of the two mimics a frequency dependence of the QPO feature with energy.  

In fact, our model fits the data better and with less free parameter than the model of a QPO with frequency that changes with energy. Our model of the QPO feature in $N$ energy bands with two Lorentzians that have each the same frequency and FWHM in all the bands has $4+2N$ parameters: Two centroid frequencies, two FWHM and $2N$ normalisations. The model with two separate QPOs in $N$ energy bands has $3N$ parameters: $N$ centroid frequencies, $N$ FWHM and $N$ normalisations. 
When we fit the PS of the six bands simultaneously with a model in which the QPO feature is a single Lorentzian with the frequency and FWHM of the Lorentzian free to change with energy, like in \cite{Qu-2010}, we get $\chi^2 = 601.9$ for 528 dof. The same fit with a model that has a Lorentzian for the QPO and another one for the QPO shoulder with the frequency and FWHM of each Lorentzian linked to be the same in all bands gives a better fit, $\chi^2 = 566.1$ for 530 dof.

In conclusion, in this observation of GRS 1915+105 the data are better fit with a model in which the QPO feature consists of two components with frequencies that do not change with energy rather than with a single Lorentzian with centroid frequency and FWHM that do change with energy. The rms spectra of the two Lorentzian components combine in such a way that the frequency of the QPO feature appears to depend upon energy. 

\subsection{Case study 4: The broadband variability in MAXI J1820+070}
\label{1820}

As a final example, in this and the next section we study the PS and CS of the transient black-hole X-ray binary MAXI J1820+070. We select this source because it was very bright and highly variable during outburst \citep[][]{Wang-2022} and because, as we explain below, it provides a stringent test to the hypothesis that we put forward here.

From 2018 March 6 to 2018 November 21 MAXI J1820+070 was observed almost daily with \textit{NICER}. During the rising part of the outburst, in the low-hard and hard-intermediate states, the source showed strong broadband X-ray variability \cite[e.g.][]{deMarco-2021}. Notably, except for a few cases \citep[e.g.,][]{Homan-2020,Ma-2023b} the power spectra of most of these observations were relatively featureless \citep[e.g.,][]{Kawamura-2022} and showed no strong and narrow QPOs. Furthermore, the magnitude of the time lags was rather small and the time-lag spectra were smooth \citep[e.g.,][]{Kara-2019}. All these characteristics can potentially challenge our proposal that, as for the PS, the CS consists of a combination of Lorentzian components.

Here we use ObsID 1200120120, which was previously analysed by \cite{Kara-2019} and \cite{Wang-2022}. We process the data with the tool \texttt{nicerl2} to produce clean event files;  following the recommendations on the \textit{NICER} website\footnote{\url{https://heasarc.gsfc.nasa.gov/docs/nicer/analysis_threads/}}, we discard the data of detectors 14 and 34 that show episodes of increased detector noise. We compute a PS in the $0.3-12$ keV band and a CS of photons in the $2-12$ keV band with respect to those in the $0.3-2$ keV band. (In this case we do not give the channels because for NICER the channel number can be calculated directly as the energy in keV multiplied by 100.) Both for the PS and the CS we use $T_{\rm FFT}=13.1072$ s, yielding $\nu_{\rm min}=\Delta\nu=0.07629$ Hz, at a time resolution $\Delta t= 1/2500$ s such that $\nu_{\rm Nyquist}=1250$ Hz.

We initially fit the PS in the range $0.07629-60$ Hz following the procedure described in \S\ref{examples}, and find that we need a model consisting of four Lorentzians. The fit gives $\chi^2=153.9$ for 146 dof, and all Lorentzians are at least 3-$\sigma$ significant.

Before we discuss the joint fits of the PS and CS we note that, because the lags of MAXI J1820+070 are close to zero over a broad frequency range \cite[e.g.,][]{Kara-2019}, at all Fourier frequencies the Real part of the CS is much larger than the Imaginary part. Since all fitting routines are more stable when the free parameters are of the same order \citep[see, for instance, the Levenberg–Marquardt algorithm in the Xspec code,][]{Arnaud-1996}, we rotated all the cross vectors by $45^\circ$ such that those with a zero phase lag would end up having more or less equal Real and Imaginary parts. Since the rotation does not change the modulus of the cross vector, and the rotation angle is known precisely, the rotation has no effect on the parameters of the fit, while it makes the fit more stable. Because of this, in the Figures in this section we plot the components of the rotated cross vector, $\mathrm{Re}\,\cos{(\pi/4)} - \mathrm{Im}\,\sin{(\pi/4)} \propto (\mathrm{Re} - \mathrm{Im})$ and $\mathrm{Re}\,\sin{(\pi/4)} + \mathrm{Im}\,\cos{(\pi/4)} \propto (\mathrm{Re} + \mathrm{Im})$, which also allows us to plot the components of the (rotated) cross vector vs. Fourier frequency  using logarithmic axes, because both rotated quantities are always positive. Regardless of this, we always report the best-fitting phase lags minus $\pi/4$ so that the lags are given again with respect to the chosen energy band for the non-rotated cross vector.

\textit{Constant phase-lags model\footnote{We describe the results and show the plots of the same analysis using the constant time-lags model in Appendix \ref{Appendix-A}.}:} The top-left and the two bottom panels of Figure~\ref{fig4} show the PS and the CS of MAXI J1820+070 in the range $0.07629-60$ Hz with the same four Lorentzians assuming the constant phase-lags model. During the fit of the CS, we initially fix the frequency and FWHM of the four Lorentzian components to the values obtained from the fit to the PS. The top-right panel shows the phase-lag spectrum with the derived model. In each case the top sub-panels show the data and the model and the middle sub-panels show the residuals.

It is apparent from this Figure that, while the PS is well fitted with four Lorentzians, the Real and Imaginary parts of the CS are not, with structured residuals in the full frequency range. The joint fit of the PS and the CS yields $\chi^2 =593.7$ for 438 dof, while the derived model of the lags gives $\chi^2=412.1$ for 150 dof.

We next fit the model letting the frequency and the FWHM of each of the four Lorentzian components free but linked across the three spectra. The bottom sub-panels of Figure~\ref{fig4} show the residuals in this case. While the fits of the CS and of the derived model of the phase lags improve slightly, significant residuals remain, and this time there are also structured residuals in the fit of the PS because in the model the frequency and FWHM of the Lorentzians change. The joint fit in this case gives $\chi^2 =529.6$ for 430 dof, while the derived model of the lags gives $\chi^2=417.4$ for 150 dof.

%
%
\begin{table*}
\centering
\caption{Parameters of the best-fitting model to the PS and CS of NICER observation 1200120120 of MAXI J1820+070 with seven Lorentzians assuming the constant phase-lags model}
\label{tab:phaselags}
\begin{tabular}{cccccccc}
\hline
\textbf{Component} & \textbf{$\nu_0$ (Hz)} & \textbf{FWHM (Hz)} & \textbf{rms$_{\rm PS}$ (\%)} & \textbf{significance$^{*}$} & \textbf{phase lags (rad)} & \textbf{rms$_{\rm CS}$ (\%)} & \textbf{significance$^{*}$}\\ 
\hline
Lorentzian 1 & $  0.054  \pm   0.002 $	& $  0.245  \pm   0.002 $	& $  25.4  \pm   0.1 $	& $   97.3 $	& $   0.10  \pm    0.01 $	& $  26.6  \pm   0.1 $	& $  124.7 $ \\
Lorentzian 2 & $  0.85  \pm   0.06 $	& $  2.0  \pm   0.1 $	& $  < 9.5^\dagger             $	& $   < 1  $	& $   1.5  \pm    0.2 $	& $  12.5  \pm   1.4 $	& $    5.3 $ \\
Lorentzian 3 & $  0.99  \pm   0.04 $	& $  2.1  \pm   0.09 $	& $  11.2  \pm   1.2 $	& $    2.4 $	& $  -0.9  \pm    0.2 $	& $  13.5  \pm   1.3 $	& $    4.0 $ \\
Lorentzian 4 & $  1.83  \pm   0.08 $	& $  5.68  \pm   0.08 $	& $  10.7  \pm   0.3 $	& $   13.8 $	& $  -0.03  \pm    0.02 $	& $  13.2  \pm   0.3 $	& $   16.0 $ \\
Lorentzian 5 & $  7.8  \pm   0.4 $	& $  8.3  \pm   1.3 $	& $  < 1.2^\dagger             $	& $   < 1  $	& $  -0.91  \pm    0.07 $	& $   2.5  \pm   0.3 $	& $    5.2 $ \\
Lorentzian 6 & $ 18.5\pm   0.8 $	& $ 15.7  \pm   3.2 $	& $  < 11.2^\dagger            $	& $   < 1  $	& $  -0.73  \pm    0.07 $	& $   1.9  \pm   0.3 $	& $    3.9 $ \\
Lorentzian 7 & $ 24.1  \pm   1.2 $	& $ 31.7  \pm   1.6 $	& $   1.68  \pm   0.09 $	& $    8.2 $	& $   0.7  \pm    0.2 $	& $   1.8  \pm   0.1 $	& $    6.4 $ \\
\hline
$\chi^2/\mathrm{dof}$ & &&&& $424.9/415$\\
\hline
\multicolumn{8}{l}{\small$^{*}$ In units of $\sigma$.}\\
\multicolumn{8}{l}{\small$^{\dagger}$ 95\% upper limit.}\\
\multicolumn{8}{l}{\small The rms amplitude of the power and cross spectra of each Lorentzian are integrated from zero to infinity.}\\
\end{tabular}
\end{table*}

The previous result shows that, even if the fit to the PS with a model consisting of four Lorentzians is statistically acceptable, the fits of the PS and the CS with that model are not. We therefore add extra Lorentzians to the PS and the CS until the simultaneous fit to the three spectra, and the derived model of the phase lags are statistically acceptable, even if some of the Lorentzian components are not significant in either of the three spectra, as long as they are significant in at least one of them. In doing the fits we let the centroid frequency and FWHM of the Lorentzian components free but link them so that they are the same for each component in the PS and the CS.

%
%
\begin{figure}
\centering
\includegraphics[width=0.5\textwidth]{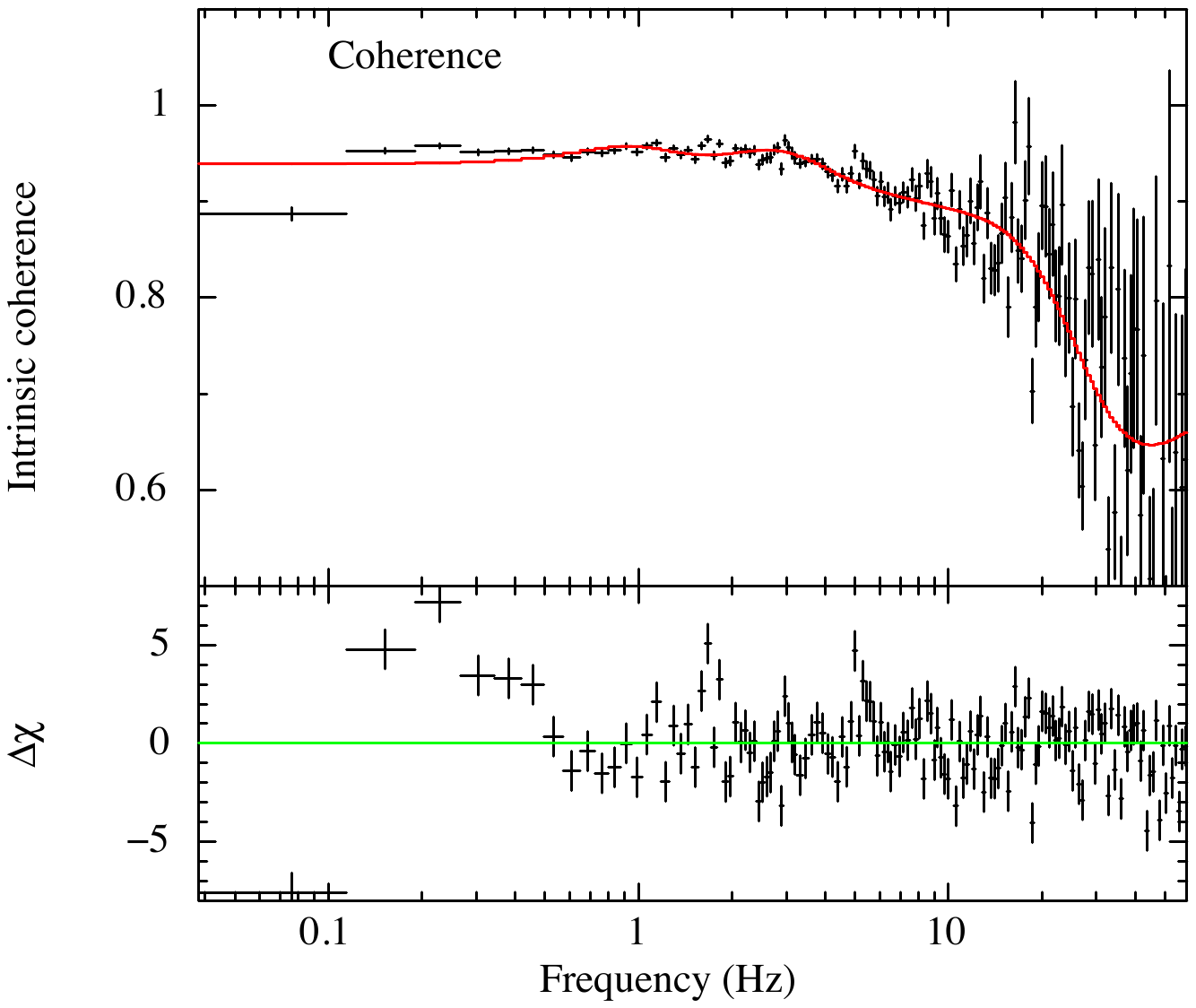}
\caption{Intrinsic coherence function of the same observation of MAXI J1820+070 shown in Figures~\ref{fig4} and \ref{fig5} with the derived model obtained from the fit to the PS and CS assuming the constant phase-lags model.
\label{fig6}
}
\end{figure}

Figure~\ref{fig5} shows the data fitted with a model consisting of seven Lorentzians assuming the constant phase-lags model. As usual, the upper panels show the data and the model, and the lower panels show the residuals. The top-right panel shows the same for the phase-lags frequency spectrum with the derived model. The joint fit of the PS and CS gives $\chi^2 = 424.9$ for 415 dof, while the derived model of the lags gives $\chi^2 = 148.5$ for 150 dof. We give the best-fitting parameters in Table~\ref{tab:phaselags}. Lorentzians number 2, 5 and 6 in this model are less than 3-$\sigma$ significant in the PS, but the three of them are significant in the CS.

We subsequently compute the PS of this observation of MAXI J1802+070 in the $0.3-2$ keV and $2-12$ keV bands using the same parameters mentioned at the start of this subsection, and fit them, together with the CS of those two bands, with the model consisting of seven Lorentzians over the same frequency range as before. In Figure~\ref{fig6} we plot the observed intrinsic coherence with the derived model for the constant phase-lags model (see Appendix~\ref{Appendix-A} for the fit using the constant time-lags model.) The fit gives $\chi^2 = 586.9$ for 558 dof. The derived model reproduces the data rather well, with the largest deviations appearing at $0.2$ Hz, where the PS shows an unresolved QPO (see, e.g., Fig.~\ref{fig5}) that, when studied at a higher frequency resolution (not shown), breaks into two separate QPO peaks.

Once more, similar to what we discuss in \S\ref{339-4}, the most appropriate model for the PS and the CS in the case of this MAXI J1820+070 observation accurately characterises both the phase-lag spectrum and the coherence function. This alignment with the two predictions we detail in \S\ref{mathematical} remains consistent.

%
%
\begin{figure*}
\centering
\includegraphics[width=0.5\textwidth]{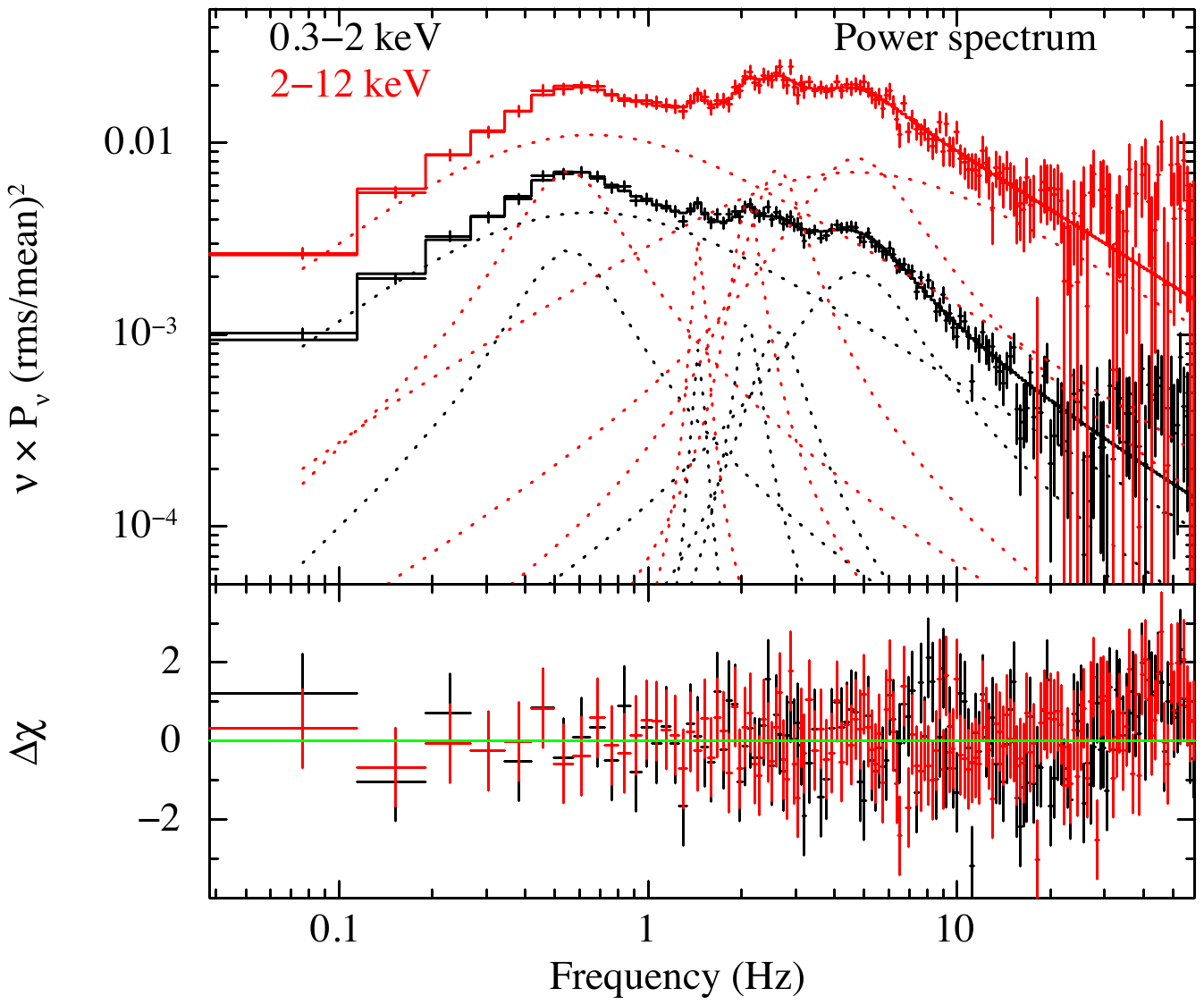}\hspace{-1cm}
\includegraphics[width=0.5\textwidth]{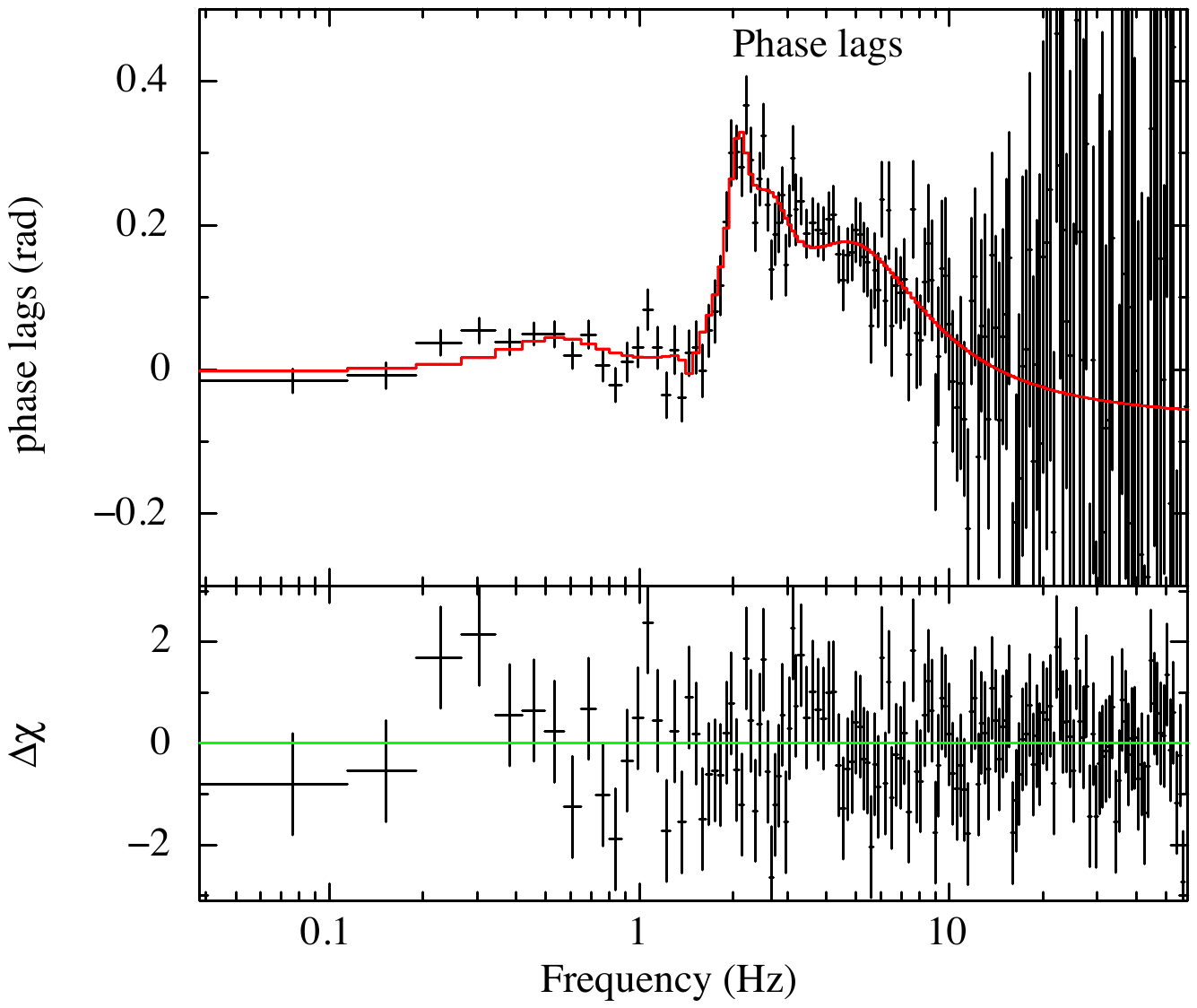}\\
\includegraphics[width=0.5\textwidth]{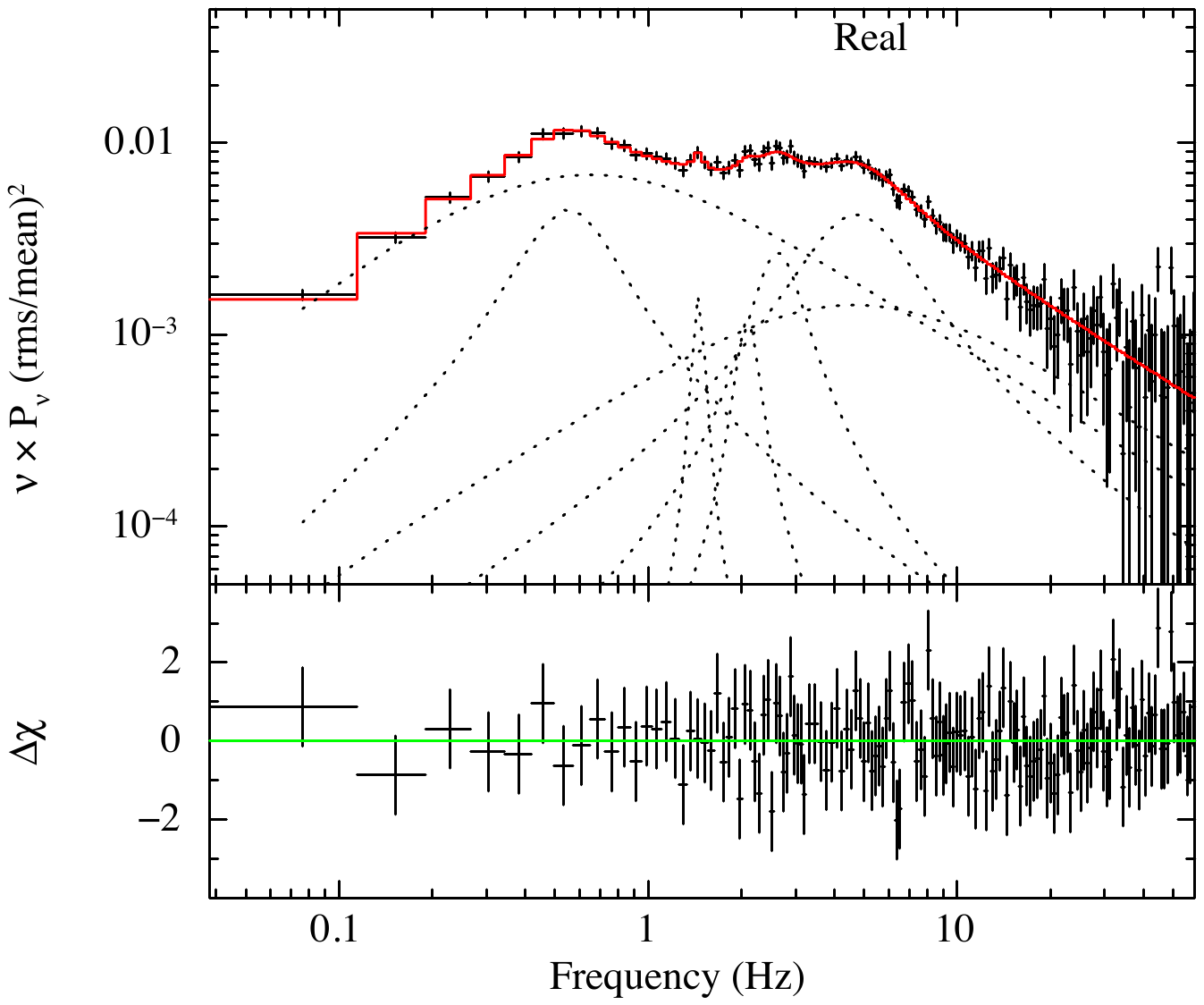}\hspace{-1cm}
\includegraphics[width=0.5\textwidth]{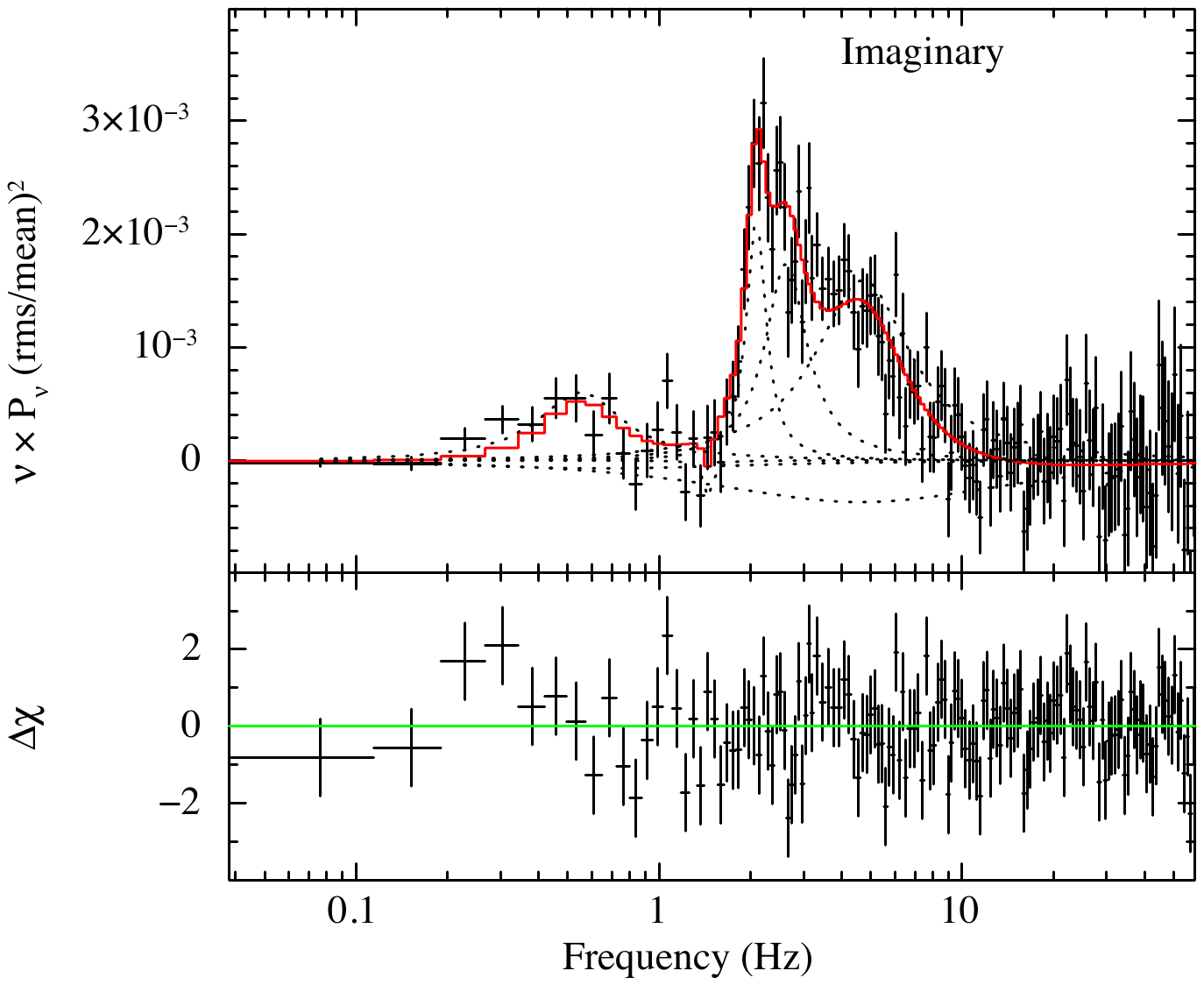}
\caption{PS in two energy bands (top left; see legend for the energies of each band), Real (bottom left) and Imaginary part (bottom right) of the CS for those two same bands of the \textit{NICER} observation 1200120268 of MAXI J1820+070 fitted with a model (thick solid line) consisting of seven Lorentzians (dotted lines). During the fitting we let the frequency and FWHM of each component free to vary but keep them linked to be the same in the PS and the CS. In the model of the CS we assume the constant phase-lags model. The top-right panel shows the phase lags vs. Fourier frequency together with the derived model. The bottom sub-panels show the residuals with respect to the best-fitting model.}
\label{fig7}
\end{figure*}

\subsection{Case study 5: A QPO in the Imaginary part of the cross spectrum of MAXI J1820+070}
\label{1820_b}

We searched the \textit{NICER} archive for an observation in which the coherence function showed a significant drop at some Fourier frequency, and came across ObsID 1200120268 (O. K\"onig, priv. comm.), which was previously analysed by \cite{deMarco-2021}. In this observation MAXI J1820+070 was in the decline of the outburst, in the lower branch of the `q' traced by the source in the hardness-intensity diagram \citep[see, e.g., Fig. 1 of ][or the right panel of Fig. 1 of \citealt{Ma-2023b}]{deMarco-2021}. We follow the procedures in \S\ref{1820} to process the \textit{NICER} data, and compute two PS in the $0.3-2$ keV and $2-12$ keV bands and a CS of photons between those same two bands using the same $\Delta t$ and $T_{\rm FFT}$ of \S\ref{1820} to compute the FFT

\textit{Constant phase-lags model}\footnote{We describe the results and show the plots of the same analysis using the constant time-lags model in Appendix \ref{Appendix-B}.}: We fit the two PS and the CS in the range $0.07629-60$ Hz with seven Lorentzians using the constant phase--lags models, which gives $\chi^2 = 505.1$ for 558 dof. All Lorentzians are at least 3-$\sigma$ significant either in one of the PS or the CS. Figure~\ref{fig7} shows the data and the best fit. While we again rotate the cross vector by $45^\circ$ during the fits, this time we plot the Real and Imaginary parts of the CS without rotation, to highlight the presence of a strong QPO at $\sim 2.1$ Hz in the Imaginary part of the CS. The phase lag of this QPO is $\Delta\phi = 1.06 \pm 0.12$ rad, such that the Imaginary part of the cross vector is about twice larger than the Real part. The 2.1-Hz QPO is not significant in either of the two power spectra (see below). Close to the 2.1-Hz QPO, the Imaginary part of the CS shows also QPOs at $\sim 2.6$ Hz and $\sim 4.3$ Hz, but these QPOs have smaller phase lags ($\sim 0.58$ rad and $\sim 0.36$ rad, respectively) and both are significantly detected in the PS of the two bands.

\begin{figure}
\centering
\includegraphics[width=0.5\textwidth]{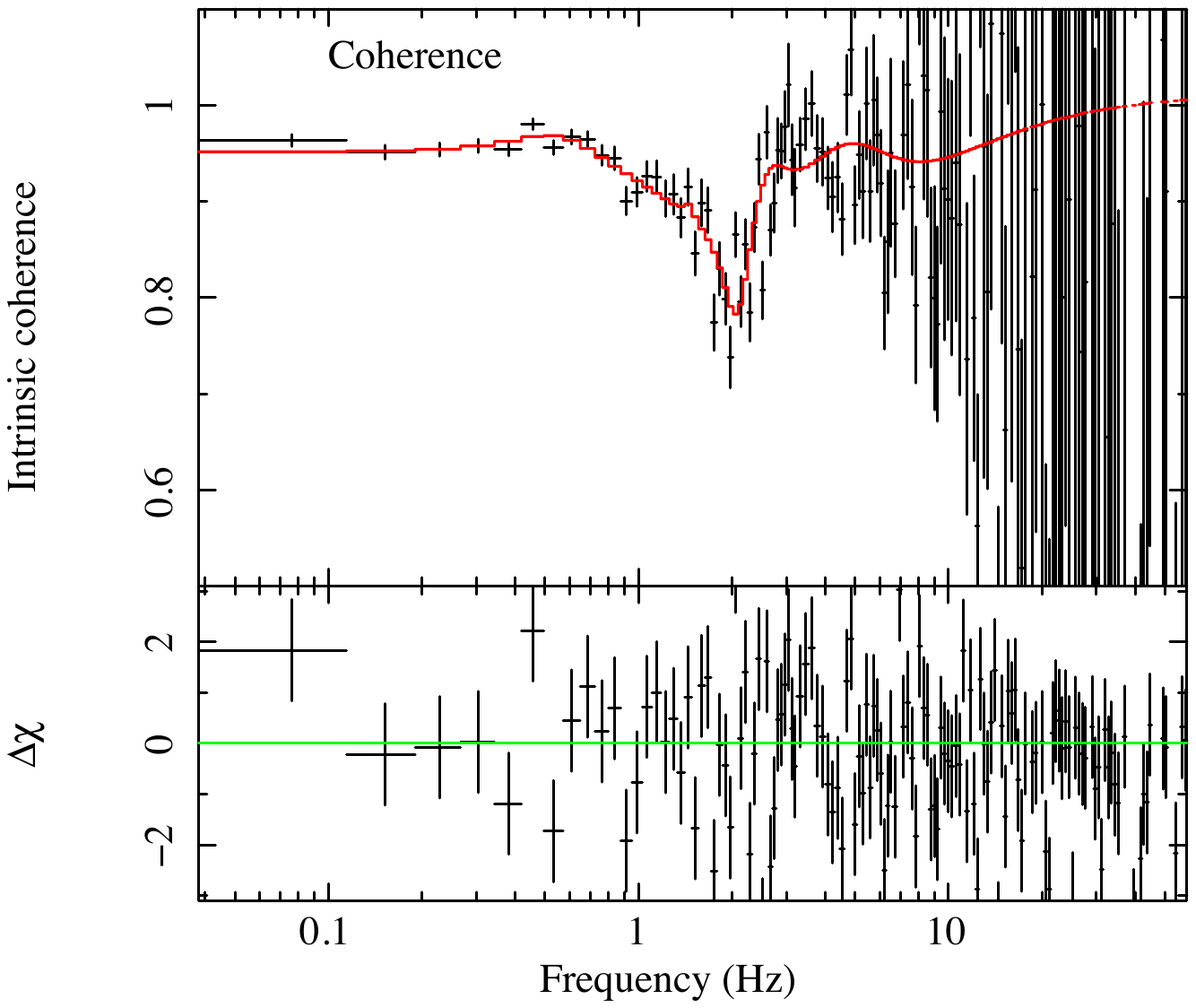}
\caption{Intrinsic coherence function of the same observation of MAXI J1820+070 shown in Figure~\ref{fig7} with the derived model obtained from the fit to the PS and CS assuming the constant phase-lags model.
}
\label{fig8}
\end{figure}

In Figure~\ref{fig8} we plot the observed intrinsic coherence together with the model derived from the simultaneous fit of the two PS and the CS. The observed intrinsic coherence shows a significant drop at $\sim 2.1$ Hz which, as shown in the plot, is perfectly described by the derived model. It is noteworthy that this drop occurs precisely at the frequency of a QPO that is present only in the Imaginary part of the CS (bottom-right panel of Figure~\ref{fig7}). This ``imaginary'' QPO, at $2.09 \pm 0.02$ Hz, is marginally present in the Real part of the CS (bottom left), where it is overshadowed by several other components, and is not significantly detected in any of the two PS (top left; the QPO is $\lesssim 2.9 \sigma$ significant in the $0.3-2$ keV band and $\lesssim 2.5 \sigma$ in the $2-12$ keV band). This underscores the fact that this QPO is only detected significantly in the Imaginary part of the CS, and that the Lorentzian component used to fit this imaginary QPO is incoherent with the other Lorentzians in that frequency range. This aligns with the assumptions made in Section \S\ref{mathematical}, further bolstering our assertion that the variability is comprised of multiple incoherent components.

%
%

\section{Discussion}
\label{discussion}

We show, for the first time, that if the power spectrum (PS) of accreting neutron-stars and black-holes can be fitted with a combination of Lorentzian functions that are coherent in different energy bands but incoherent with each other, the same is true for the Real and Imaginary parts of the cross spectrum (CS). Based on this, we propose a novel method to measure the lags of variability components that is especially useful when those components are weak and overlap with other (stronger) components in the PS and CS. 

Surprisingly, using this method we discover new variability components that are detected significantly only in the CS and not in the PS of these sources. This happens because the PS is insensitive to signals with a large Imaginary part and a small Real part in the CS when such a signal overlaps in frequency with other variability components that have a large Real part in the CS. Because in the last 40 years we have used exclusively the PS to identify variability components in these sources, we have so far missed signals with large positive or negative lags, such that the cross vector has a significant component along the Imaginary axis.

We also show that, contrary to what has been previously claimed, the frequency of a type-C quasi-periodic oscillation (QPO) in the black-hole binary GRS 1915+105 does not depend upon energy. The apparent energy dependence of the QPO frequency can be explained by the presence of a second, significant, component in the PS and CS of the source at a frequency very close to that of the QPO, but with a different rms and lag spectra. This alternative interpretation requires fewer model parameters and is statistically favoured. We propose that the same applies to other QPOs in which a similar energy dependence of the QPO frequency was observed. 

Finally, we demonstrate that, in accordance with the predictions derived from the assumptions underpinning our method, the model that fits both the PS and CS reproduces both the phase-lag spectrum and the coherence function.

\subsection{Weak variability components in the presence of other, stronger, components}
\label{weak}

In \S\ref{mathematical} we show that if the PS of X-ray binaries can be fitted with a linear combination of Lorentzian functions that are coherent in different energy bands and incoherent with each other, the same is true for the CS of linearly correlated light curves of the source in two energy bands. The centroid frequency and FWHM of each Lorentzian are the same in the PS and CS, with the Lorentzians in the Real (Imaginary) part of the CS being multiplied by the cosine (sine) of a function of Fourier frequency, $\Delta\phi_{xy}(\nu)=g(\nu;p_j)$ with parameters $p_j$. These functions are the frequency-dependent phase lags of each Lorentzian. If $g(\nu)$ is constant, the phase lags of that component are independent of Fourier frequency, while if $g(\nu) \propto \nu$ the time lags are independent of Fourier frequency. 

While we show that, under the assumptions discussed in \S\ref{mathematical}, the same Lorentzians should appear both in the PS and the CS, there is no guarantee that all the Lorentzians that are significantly detected in one will be significantly detected in the other. To test this, and to explore the potential of our method, in \S\ref{examples} we fit the PS and CS of five observations of three black-hole binaries. We unveil weak QPO signals at frequencies very close to those of previously detected QPOs in the black-hole binaries GX 339--4 \citep{Zhang-2017,Altamirano-2015} and GRS 1915+105 \citep{Zhang-2020}. The weak signals, which we call QPO shoulders, appear at a slightly higher frequency than that of the QPO fundamental or second harmonic. QPO shoulders have been previously reported \citep{Belloni-1997, Belloni-2002,vanDoesburgh-2020}, but until now it remained unclear if those were really separate QPOs or whether those shoulders were due to the QPO frequency drifting slightly during the time over which one calculated the PS \citep[see, e.g.,][]{Belloni-1997}. Our method allows us, for the first time, to measure a significant difference of the lags and rms spectra of the QPO and the QPO shoulder, which strongly suggests that the QPO and the shoulder are truly different components. In GRS 1915+105 the shoulder of the QPO fundamental and the second harmonic are also consistent with having a harmonic relation. It is therefore possible that all harmonics of a QPO, if studied using this method, will eventually show shoulders that, as the QPOs, are also harmonically related.

An advantage of the method that we propose here is that it allows to measure the lags of components that appear in a frequency region of the PS or CS where other equally strong or stronger components are present. One cannot use the traditional method to measure the lags of such a component in those cases (see \S\ref{339-4}). A way used in the literature to overcome this limitation is to measure the lags in the traditional way within one FWHM of the component of interest and subtract the lags measured outside that range \citep{Ma-2021,Ma-2023}. This method is mathematically incorrect because it ignores the modulus of the cross vector (CV) of the two components. Alternatively, one could try and subtract the Real and Imaginary parts of the CS in a frequency range range outside the QPO from the same quantities within a FWHM of the QPO, however this is rarely possible \citep[see for instance Fig.~\ref{fig1} above or Fig.~1 in][]{Nowak-2000}. 

\subsection{Hidden variability components}
\label{hidden}

Initially, we assumed that we had to fit the PS with a number of Lorentzians first, and subsequently fit the CS with the same Lorentzians with centroid frequencies and FHWM fixed at the values obtained from the fit to the PS. While we find a good fit to the PS with a number of Lorentzians, when we fit the CS with those same Lorentzians with only the normalisation of the Lorentzians  free the fit is bad, with significant structured residuals over the full frequency range. This is most clearly seen in the case of the observation of the black-hole binary MAXI J1820+070 that we present in \S\ref{1820}. As shown in Figure~\ref{fig4} for the constant phase-lags model and Figure~\ref{figA1} for the constant time-lags model, while the PS is well fitted with four Lorentzians, the CS is not; because of this, the phase-lag frequency spectrum cannot be fitted with the derived model either (cf. the residuals in the middle panels of those two Figures). If we let the frequency and FWHM of the Lorentzians free but link each of them to be the same in the PS and CS, the fit improves but remains statistically unacceptable, with significant structured residuals in the CS and the lags. Not only that, but now there are significant residuals also in the PS. The reason for this is easy to understand: In the joint fits the frequency and FWHM of the Lorentzians change to reduce the residuals in the CS, but this degrades the fit of the PS. 

In the case of MAXI J1820+070, in order to get a good fit we need to add three extra Lorentzians to the model. While some of these Lorentzians are not significant in the PS, they are (in some cases very) significant in the CS. This can be seen in Figures~\ref{fig5} and \ref{figA2} and Tables~\ref{tab:phaselags} and \ref{tab:timelags}. For instance, Lorentzians 2, 3 and 6 in the constant phase-lags model (Figure~\ref{fig5} and Table~\ref{tab:phaselags}) and Lorentzian 5 in the constant time-lags model (Figure~\ref{figA2} and Table~\ref{tab:timelags}) are less than 1-$\sigma$ significant in the PS, but they are $3.9-6.5 \sigma$ significant in the CS.

To understand why we do not see these components in the PS, whereas they are significant in the CS, we first need to note that the majority of the variability components in LMXBs have rather small lags \citep[see, for instance, Fig. 2 in][the same is true in other sources]{Zhang-2020}. This means that in the CS the Real part is much larger than Imaginary part (factors $\gtrsim 10$ are usual; see for instance Figs.~\ref{fig1} and \ref{fig7}). Consider two variability components, $X_1$ and $X_2$, that overlap in frequency, each of them perfectly coherent in two energy bands (c.f., assumption (i) in \S\ref{mathematical}). Furthermore, consider that $X_1$ is weak with phase lags close to $\pm \pi/2$ (Real part in the CS close to zero and small but significantly different from zero Imaginary part) and $X_2$ is strong with phase lags close to zero (in the CS large Real part, Imaginary part close to zero). It is easy to show that if the two components have similar rms spectra, the contribution of component $X_1$ in the PS is negligible compared to that of component $X_2$. On the other hand, the statistical errors in the power spectrum will be of the order of the Real part of CS of $X_2$ divided by the square root of the product of the number of segments and the number of frequency bins \citep{Bendat-2010, Ingram-2019b}. As long as this error is larger than the Imaginary part of component $X_1$, the latter will not be detected in the power spectrum, even if it is very significant in the Imaginary part of the CS\footnote{This is a more general issue, which applies in the case of a variability component that overlaps in frequency with another, stronger, variability component, if the two components have a phase lag that differs by $\pm\pi/2$.}. We note that the advantage of detecting such a weak component in the cross spectrum, and not in the power spectrum, will not be available when, instead of the ``contaminating'' component(s),  the Poisson noise dominates the variability.

An example of this is the $2.1$-Hz QPO in MAXI J1820+070 that has a phase lag of $\sim 1.1$ rad, appears close to a strong broad QPO peaking at $\sim 0.6$ Hz (Fig.~\ref{fig7}) with a phase lag of $\sim -0.01$ rad, and is only detected in the Imaginary part of the CS. (A similar effect is likely the cause of the drop of the coherence function at $\sim 0.2-0.3$ Hz, in the observation of GRS 1915+105 shown in Fig. 3 of \citealt{Ji-2003}, where no significant QPO is apparent in the PS.)

All the above demonstrates that by searching for QPOs exclusively in the PS we tend to find only components with relatively small lags, and we are insensitive to signals with lags that approach $\pm \pi/2$. In fact, we would never find lags of $\pm \pi/2$ rad in the CS of any of these sources if we measure the lags over a broad frequency band that contains several components with small lags, because the Real part of the CS in that frequency range will never be zero. By using only the PS to identify and characterise the variability of these sources, and by computing the average lags over broad frequency ranges, all these years we may have missed these components altogether. 

\subsection{No energy dependence of the frequency of QPOs}
\label{noenergy}

It has been argued that the frequency of some low-frequency QPOs in black-hole X-ray binaries, especially the type-C QPO in GRS 1915+105, depends on energy \citep{Qu-2010,Li-2013a,Li-2013b,Yan-2018}; in some cases the QPO frequency appears to increase, and in others to decrease, with energy \citep[see, e.g., Fig.2 of][and Fig. 3 of \citealt{Li-2013a}]{Qu-2010}. This has been explained by \cite{vandenEijnden-2016} in terms of differential Lense-Thirring precession \citep{Ingram-2016,Nathan-2022}, with parts of the inner accretion flow precessing at different rates. An issue with this interpretation is that sometimes the QPO frequency appears to increase whereas other times it appears to decrease with energy \citep{Qu-2010}. Since the characteristic frequencies in the vicinity of a compact object as well as the temperature of the material in the accretion flow generally decrease with increasing distance from the central object, this requires that the outer and cooler parts of the accretion flow precess faster than the inner and hotter parts. 

We find that in one of the observations of GRS 1915+105 in which \cite{Qu-2010} measure an increase of the QPO frequency with energy, the QPO feature can actually be fitted better with a model that consists of two components, a QPO at $\sim 5.8$ Hz and a QPO shoulder at $6.3$ Hz. Because the rms amplitude of the shoulder increases faster with energy than that of the QPO, the QPO dominates at low energies whereas the QPO shoulder dominates at high energies, mimicking a shift of the frequency of the QPO feature with energy. The fit with our model with two Lorentzian components with energy-independent centroid frequency and FWHM is statistically better and requires less parameters than the one with a single Lorentzian with energy-dependent centroid frequency and FWHM. On the other hand, the rms and time-lag spectra of the QPO and the shoulder are significantly different, showing that the shoulder is not due to a drift of the QPO frequency during the observation, but that it is a separate component, and is the reason for the apparent energy-dependence of the QPO frequency. We propose that this holds for other cases in which the QPO frequency appears to change with energy \citep[e.g.,][]{Li-2013a,Li-2013b,Yan-2018}, significantly challenging the ideas that explain the QPO as a single component with its frequency in different bands originating in different parts of the accretion flow. 

The shoulders could be due to amplitude modulation of the QPO signal \citep[see Fig. 3 in][]{vandenEijnden-2016}. Amplitude modulation, however, should produce equally strong sidebands at frequencies below and above that of the QPO, whereas there is no apparent shoulder at frequencies below those of the QPO in the data that we present here. (The 95\% confidence upper limit for the full-band rms amplitude of a possible shoulder at $\sim 0.5$ Hz below the frequency of the QPO is $\sim 1$\%.) The shoulders appearing only at higher frequencies than those of the corresponding QPOs resemble the positive sidebands of the kHz QPOs observed in some neutron-star systems \citep{Jonker-2000,Jonker-2005} although, different from the case of the kHz QPOs, when the frequency of type-C QPOs appears to decrease with energy \citep[e.g.,][]{Qu-2010} the shoulder would be at lower frequencies than those of the QPOs.  

In this observation of GRS 1915+105, and in the observation of GX 339--4 that we present in \S\ref{339-4}, the shoulder appears at a slightly higher frequency than that of the corresponding QPO. In both sources we find a shoulder of the QPO fundamental, while in the observation of GRS 1915+105 in \S\ref{1915_1} we also find a shoulder of the second harmonic of the QPO, with a centroid frequency that is consistent with being twice that of the shoulder of the fundamental QPO. This suggests that the shoulders of the QPO fundamental and second harmonic are physically connected.

\subsection{Broadband noise or QPOs?}
\label{bbn-qpo}

We show that in the black-hole binary MAXI J1820+070, while the PS can be fitted with four broad Lorentzians, the joint fits of the PS and the CS require at least seven, narrower, Lorentzian functions. This result evinces that, as it was already shown for the PS of GX339--4 and Cyg X--1 \citep{Nowak-2000} and several other black-hole and neutron-star X-ray binaries \citep{Belloni-2002}, in MAXI J1820+070 the broadband noise (BBN) in the CS is consistent with a combination of several Lorentzians. 

That the CS, like the PS, comprises multiple Lorentzian components, naturally explains the findings of \cite{Nowak-1999a,Nowak-1999b}. They observe that in Cyg X--1 and GX 339--4, in the frequency range in which one Lorentzian dominates the PS, the phase lags of light curves in two energy bands display a more or less flat shelf. Conversely, when two Lorentzians intersect in the PS, a transition occurs from one characteristic phase-lags shelf to another \citep[compare Figs. 1 and 10 in][]{Nowak-1999a}. \cite{Nowak-1999b} show that, at each Fourier frequency, the total measured phase lags are the average of the phase lags of the individual Lorentzians, weighted by the product of the amplitudes of the Fourier transforms in those two energy bands\footnote{This result is the small-angle approximation of eq.~\ref{eq9}.}. Because over a given frequency range the Fourier amplitudes of one Lorentzian dominate, the source shows a more or less constant shelf if the phase lags of that component are constant with Fourier frequency.

All the above suggests that in MAXI J1820+070 the lags of the BBN component in the CS arise from a transfer function that can be described as the combination of several individual, narrower, responses. This idea is reinforced by the fact that the same linear combination of Lorentzian functions that fits the PS and CS predicts correctly the coherence function, in particular that the coherence is one when a single Lorentzian component dominates in both the PS and CS, and drops when two or more Lorentzians with different cross amplitudes and phase lags overlap in frequency (see Fig.~\ref{fig8} for a remarkable case of this effect). This, in turn, reflects in the fact that  each Lorentzian has its own time or phase lags, unrelated to the lags of all other Lorentzians, possibly indicating separate resonances in the variability properties of the accretion flow \citep[e.g.,][]{Nowak-1997,Mendez-2013,Mendez-2022,Zhang-2022}. Our findings challenge the idea that there is a smooth global transfer function of the accreting system, like the one used to explain the broadband lags in LMXBs as reverberation of corona photons that reflect off the accretion disc.

Several papers in the literature measure the time lags  in MAXI J1820+070 in various frequency bands in the $\sim 0.1-80$ Hz range \citep[e.g.,][]{Kara-2019,deMarco-2015,deMarco-2017,Wang-2022}, which they interpret as the time lags of the BBN produced by the (global) transfer function of the accretion disc. This is generally justified by the fact that the PS of this source appears to be smooth, without very strong and narrow QPO peaks, and in that frequency range it can be fitted by two or three broad Lorentzian functions \citep[e.g.,][]{Kawamura-2022,Kawamura-2023a}. Our results above, however, bring this procedure into question. We show that to fit simultaneously the PS and CS one needs seven Lorentzians. Figures~\ref{fig5} and \ref{figA2} show that there is no unique component that dominates the PS and CS in the full $0.1-80$ Hz range, but different Lorentzians dominate the variability over different parts of that frequency range. Moreover, similar to what we show in \S\ref{339-4} and Appendix \ref{simulations}, in this case the measurements of the lags using the traditional method will clearly misrepresent the actual lags of any of these components. Figures~\ref{fig5} and \ref{figA2} manifestly show that there is no frequency interval over which one can measure the lags of the putative BBN, without being affected by other components.

In short, our findings challenge the validity of conclusions about the geometry of these systems deduced from converting the time lags of the BBN into light travel distances of photons in the accretion flow \citep[e.g.,][]{Kara-2019,Wang-2022} without a model of the frequency-dependent part of the transfer function of the system that properly accounts for the response of the individual components that comprise the BBN \citep[e.g.,][]{,Mastroserio-2018,Ingram-2019c}.

\subsection{Constant phase lags or constant time lags?}
\label{constant}

As we explain in \S\ref{phaselagmodel}, to be able to fit the CS we need to assume the form of the functions $\Delta\phi_{xy,i}(\nu)=g_i(\nu;p_{j,i})$ that represent the phase-lags vs frequency of each Lorentzian component, $i$, and fit the model to get the parameters $p_{j,i}$. In this paper we assume that either the phase lags or the time lags of each Lorentzian component are constant with Fourier frequency, $g_i(\nu;k_i) = 2\pi k_i$, or $g_i(\nu;k_i) = 2\pi k_i \nu$, respectively, with $k_i$ constants to be determined from the fits. Also, while in principle the phase lags of different Lorentzian components could depend differently upon Fourier frequency, here we assume that the same function applies to all the components in the model. 

In \S\ref{examples} we fit both types of models to the data and find that they are statistically equivalent. We note, however, that the parameters of some of the Lorentzians are different in one model and the other. This can be seen from a comparison of Figures~\ref{fig5} and \ref{figA2} and Tables~\ref{tab:phaselags} and \ref{tab:timelags}.
This is not surprising, given that the form of $g(\nu;p_j)$ affects the shape of the profile of the Lorentzian in the CS. In fact, in the constant phase-lags model the components in the CS are also Lorentzian functions, but for the constant time-lags model the components in the Real and Imaginary parts of the CS are Lorentzian functions multiplied by, respectively, $\cos{(2\pi k_i \nu)}$ and $\sin{(2\pi k_i \nu)}$. That in this case these are not Lorentzian functions is apparent in Figure~\ref{figA2}, where some of the individual components, and the total model, show oscillations as a function of Fourier frequency with a period of $1/k_i$ Hz.

We initially hoped that from these data, especially those of MAXI J1820+070, we would be able to conclude whether the constant phase-lags or the constant time-lags model was most likely. Since in the constant time-lags model the CV of each component rotates in the Fourier plane as a function of Fourier frequency, each time the CV rotates by $2\pi$ radians the phase lags will wrap back to the interval $\left[-\pi,\pi\right)$ and the Real and Imaginary parts of the CS will start a new cycle.  We expected that, if this was the case, we could observe in the data the oscillations predicted by the model, or that we could discard that model if the residuals showed significant oscillations with Fourier frequency. Unfortunately the current data do not allow us to choose between the two alternatives.

We stress that, while we can fit the data equally well using either the constant phase- or time-lags model for each component, these two are the simplest of an infinite choice of lag models. It remains to be seen whether a more sophisticated analysis of the data, e.g., combining several observations, would allow us to discard either of the two models, or whether a more complex model is needed to describe the data. 

The models that have been proposed so far to explain the lags (and in some cases also the rms amplitude) of the X-ray variability can be roughly divided into two classes: (i) Models of the broadband variability consider that the lags come from the travel time of photons or accretion-rate fluctuations, and hence the time lags are constant with Fourier frequency. (ii) Models of the QPOs consider a sinusoidal signal (sometimes with harmonics), and hence the distinction cannot be made because, for a given frequency of the sinusoidal function, the phase and time lags are simply related to each other. The first class of models includes reverberation, propagating mass accretion rate fluctuations and Comptonisation in a relativistic jet or outflow. The second class includes Lense-Thirring precession, time-dependent Comptonisation with feedback and a precessing jet.

In the \texttt{RELTRANS} model \citep{Mastroserio-2018,Mastroserio-2019,Mastroserio-2021,Ingram-2019c}, hard photons emitted from the corona \citep[assumed to be a point source along the spin axis of the black hole; see][for the case of a corona consisting of two point sources]{Lucchini-2023} reach the observer first. Hard photons that illuminate and reflect off the accretion disc are reprocessed and re-emitted at lower energies than the corona photons, and reach the observer at later times. In this model the \textit{soft} lags of the BBN component in the PS of these sources reflect the difference of the travel times between the hard and soft photons. (The observed time lags are not simply the difference of those travel times --see above for the shortcomings of that--, but the model computes the lags from the energy- and frequency-dependent transfer function of the disc illuminated by a point source, the so-called lamppost, at a certain height above the disc; see, e.g., \citealt{Reynolds-1999}.)

In the several variants of the model of propagating mass accretion rate fluctuations, \texttt{PROPFLUC} \citep[][see also \citealt{Ingram-2011,Mahmoud-2018,Mahmoud-2019,Kawamura-2023a,Mummery-2023}]{Rapisarda-2014, Rapisarda-2017a, Rapisarda-2017b}, the \textit{hard} lags of the BBN reflect the speed at which those fluctuations propagate in the accretion disc on the viscous time scale. Similarly, in the \textsc{JED-SAD} model \citep{Ferreira-2006,Marcel-2019,Marcel-2020} the \textit{hard} time lags would reflect the viscous time scale in the disc, given that the energy carried by mass accretion rate fluctuations will dissipate at the transition radius between the standard and the jet-dominated parts of the accretion disc \citep{Ferreira-2022}.

In the model of Comptonisation in a relativistic jet/outflow \citep{Reig-2003,Giannios-2004,Reig-2015,Reig-2021,Reig-2018,Kylafis-2018,Kylafis-2020} the \textit{hard} lags of the BBN reflect the difference in the travel time of the photons emitted from the disc that reach the observer directly and the disc photons that are inverse-Compton scattered in the jet/outflow before reaching the observer. 

The Lense-Thirring precession model of the QPOs \citep{Ingram-2009,Ingram-2016,Nathan-2022} assumes that, as it precesses, a hot torus located inside the inner radius of the accretion disc illuminates the surface of the disc asymmetrically and produces the variability at the QPO frequency. This precession leads to a \textit{soft} delay between the modulated flux of the hard component from the precessing torus and the emission from the disc (especially the iron line). 

In the time-dependent Comptonisation model, \texttt{VKOMPTH} \citep[][see also \citealt{Karpouzas-2021,Garcia-2022,Zhang-2022b,Zhang-2023,Rawat-2023,Ma-2023b,Rout-2023,Zhang-2023b}]{Karpouzas-2020,Garcia-2021,Bellavita-2022}, the lags of the QPO are either \textit{hard} or \textit{soft} depending on the ``net'' delay\footnote{In steady state both processes are at work; the net delay results from the solution of the time-dependent Kompaneets equation \citep{Kompaneets-1957}.} between the photons from the disc and those that are inverse-Compton scattered in the corona, or the direct corona photons and those photons that return and are re-processed and re-emitted by the disc. 

\citet[][see also \citealt{Ma-2023}]{Ma-2021} assume that the lags of the QPO are produced in a small-scale precessing jet. In this case, the energy of the emitted photons decreases with height in the jet, and the \textit{soft} lags at the QPO frequency represent the difference of the time of arrival of photons produced at different heights of the jet. 

While conceiving mechanisms that produce time delays is relatively straightforward, mechanisms that produce phase delays are less obvious. A possible phase delay is mentioned in a study of the type-C QPO in GRS 1915+105 \citep{Lin-2000}, where the authors discuss the phase delay between the fundamental and the second harmonic of the QPO. Constant phase lags for each variability process, without discussing the mechanism that produces them, is also considered by \cite{Nowak-1999b} to explain the phase-lag shelves and the coherence function of GX 339--4. 

\subsection{How should we search for variability?}
\label{search}

In this paper we show that some variability components in LMXBs are not detected significantly in the PS, and that one needs to resort to the CS to detect them. Our findings prompt the question of the most effective method for detecting variability in these sources. 

The significance of a Lorentzian component is proportional to the rms amplitude of the Lorentzian squared times the source count rate \citep{vanderKlis-1989b}. The rms amplitude of most variability components increases with energy \citep[e.g.][]{Berger-1996,Mendez-2001,Gierlinski-2005,Zdziarski-2005,Zhang-2020},
whereas the observed count rate decreases at high energies, both because of the shape of the spectrum of these sources and the drop of sensitivity of X-ray detectors. At low energies the source count rate usually decreases also because the effective area of X-ray detectors drops and because of the effect of the interstellar absorption. Based on all this, and although it is not possible to give a procedure that fits all possible cases, a generic approach would be something along these lines: 

\begin{enumerate}
    \item Fit the PS in the broadest possible energy band with a model consisting of a number of Lorentzian functions to identify the strongest variability components. One could also fit the PS in several narrow bands over the full energy range of the detector, because some components may be more significant in the full band while others may be more significant in some of the narrow bands. 
    \item Produce CS of two bands covering, respectively, the low- and the high-energy parts of the energy spectrum. While there is no general rule to select the bands, one could consider dividing the energy spectrum such that each band has more or less the same count rate and the combined bands cover the full band of the detector. 
    \item Fit the full band PS and the two-bands CS with the same Lorentzian model that fits the PS, with the frequency and FWHM of each Lorentzian free to vary but linked in the PS and CS. The Lorentzian functions in the Real and Imaginary part of the CS need to be multiplied by, respectively, $\cos{\left[g_i\left(\nu;p_{j,i}\right)\right]}$ and $\sin{\left[g_i\left(\nu;p_{j,i}\right)\right]}$ assuming a model of the phase lags to get the parameters $p_{j,i}$. If necessary add new Lorentzians to the model.
    \item Repeat the previous step for (some of) the narrow-band PS and the CS of (some of) the narrow bands with respect to the full band
    \footnote{In doing all this, one has to take into account that, on the one hand the full-band PS and the PS of the narrow bands, and on the other hand the two-bands CS and the CS of the narrow bands with respect to the full band, are correlated \citep{Ingram-2019b} and hence do not provide independent information to assess the significance of some of the signals.}. 

\end{enumerate}

\section{Conclusions}
\label{conclusions}

We propose a new method to measure the energy-dependent phase and time lags in X-ray binaries, and use it to fit the power and cross spectra (respectively, PS and CS) of a number of sources. As we show in \S\ref{examples}, this procedure is capable of unveiling signals that are not significantly detected in the PS alone, opening up new possibilities to understand the properties of the accretion flow around neutron-star and black-hole X-ray binaries. Specifically:

\begin{itemize}
    \item We show that, mathematically, the PS and CS of X-ray binaries can be fitted with the same combination of Lorentzian functions, with each Lorentzian in the CS multiplied by the cosine (sine) of a function of Fourier frequency, $\Delta\phi(\nu)_{xy,i}=g_i(\nu;p_{j,i})$ with parameters $p_{j,i}$, with the centroid frequency and FWHM of each of the Lorentzians being the same in the PS and CS. 
    
    \item We successfully fit the PS and the CS of the black-hole binaries GX 339--4, GRS 1915+105 and MAXI J1820+070 assuming that, for each Lorentzian $i$ and parameters $k_i$, either $g_i(\nu;k_i) = 2\pi k_i$ (constant phase lags) or $g_i(\nu;k_i) = 2\pi k_i \nu$ (constant time lags).

    \item We find that there is a significant shoulder at a slightly higher frequency than that of (i) the fundamental of the type-C QPO in, respectively, GX 339--4 and GRS1915+105, and of (ii) the second harmonic of the type-C QPO in GRS 1915+105. The shoulder is sometimes significant only in the CS, and would go undetected if one only analysed the PS of these sources. 

    \item Contrary to previous reports, the frequency of the type-C QPO in an observation of GRS 1915+105 is consistent with being independent of energy. The apparent change of the QPO frequency with energy that was previously reported can be explained by a QPO feature that consists of two components, the QPO and a QPO shoulder, each of them with an rms amplitude spectrum that depends differently upon energy. This model is statistically better and requires less parameters than the model in which the QPO frequency depends upon energy.
    
    \item While the PS of an observation of the black-hole binary MAXI J1820+070 can be fitted with four Lorentzians, the simultaneous fit to the PS and CS requires seven Lorentzians, all of them significant.

    \item We find a narrow QPO in an observation of MAXI J1820+070 that is very significant in the Imaginary part of the CS, but is not significantly detected in the PS. This ``imaginary'' QPO causes a sharp drop in the coherence function at the QPO frequency.

    \item All the above shows that, by measuring the lags in the traditional way, one is bound to miss variability components for which the Real part of the CS is small compared to the Real part of other components in the same frequency interval. 
    
    \item We argue that because of this, and because so far we have only used the PS to detect variability components, we have missed signals with large positive or negative lags a significant component of the cross vector along the Imaginary axis) in these sources. 

    \item We conclude that, as it was previously shown for the PS, in the CS of X-ray binaries the so-called broadband noise component is in fact the combination of individual Lorentzian functions with more or less well-defined time scales. The transfer function of models used to fit the broadband variability in X-ray binaries must account for this. 
    
\end{itemize}

\section*{Acknowledgements}
Unfortunately, TMB passed away before this paper was accepted. We will miss the insightful discussions with him about these and other topics. MM acknowledges the research programme Athena with project number 184.034.002, which is (partly) financed by the Dutch Research Council (NWO). F.G. is a researcher of CONICET and acknowledges support from PIP 0113 (CONICET), PICT-2017-2865 (ANPCyT) and PIBAA 1275 (CONICET). TMB acknowledges financial contribution from PRIN-INAF 2019 N.15. MM, FG and TMB thank the Team Meeting at the International Space Science Institute (Bern) for fruitful discussions. D.A. acknowledges support from the Royal Society. K.A. acknowledges support from Tamkeen under the NYU Abu Dhabi Research Institute grant CASS. We thank the referee, Michiel van der Klis, for comments that helped us improve the manuscript. This research has made use of NASA’s Astrophysics Data System.

\section*{Data Availability}

This research has made use of data obtained from the High Energy Astrophysics Science Archive Research Center (HEASARC),
provided by NASA’s Goddard Space Flight Center.


\bibliographystyle{mnras}
\bibliography{mendez}{}


\appendix

\section{Simulations of phase lags of the QPO: The traditional vs. our method}
\label{simulations}

In this subsection we carry out simulations to compare the results that we obtain with the new and the traditional methods as a function of the strength of the QPO relative to the strength of other variability components. We note that for these simulations we assume that the PS and CS consist of a linear combination of Lorentzian functions, as described in \S\ref{mathematical}. Here we only discuss the case of the constant phase-lag model, but we find similar results using the constant time-lags model.

We take the best-fitting model to the PS and CS of the observation of GRS 1915+105 in \S\ref{1915_1} as the basis for our simulation. To make the comparison more straightforward, we simulate the data considering a simplified version of the model shown in Figure~\ref{fig2}; in particular, we ignore the weak shoulders of the QPO fundamental at $\sim 2$ Hz and of the second harmonic at $\sim 4$ Hz. We further assume that the QPO has a phase lag of $-0.10$ rad, and the low- and high-frequency broad Lorentzians that contribute to the variability at the QPO frequency (these are the Lorentzians in the middle panel of Figure~\ref{fig2} peaking at, respectively, $\sim 1$ Hz and $\sim 3.5$ Hz) have lags of $-0.20$ rad and $+0.30$ rad, respectively.  Finally, for the simulations we take the normalisation of the Lorentzian components in the PS and CS that we obtain from the best-fitting model. Once we have established the base model, we carry out Monte Carlo simulations of the power and cross spectra for a range of values of $A$, the normalisation of the QPO in the PS of band 1 (channels $0-13$), such that the ratio of the QPO power to the combined power of the other two components in that band, integrated over one FWHM of the QPO, is in the range $\sim 0.25 - 35$. We then do the same to simulate the PS of band 2, applying the same factor, in the range $\sim 0.25 - 35$, to the normalisation of the QPO, $B$, in band 2 (channels $14-249$) and compute the normalisation of the cross vector, $C = \sqrt{A B}$ (see \S\ref{mathematical}). For the simulations we assume that the powers in the PS and the amplitude squared in the CS are normally distributed. This is a reasonable approximation given that the simulated data are based on the average of at least 25 separate realisations of those Fourier quantities, which are also rebinned logaithmically in frequency (see \S\ref{examples}).

We find that when the power of the QPO is $\sim 25$\% of the power of the broad underlying components, the QPO is $4.2\sigma$ and $4.4\sigma$ significant in, respectively, band 1 and band 2; under these circumstances our method gives a phase lag of the QPO of $-0.047 \pm 0.068$ rad, already consistent with the input value of the simulation of $-0.10$ rad. On the contrary, the lags obtained with the traditional method are $ 0.165 \pm 0.011$ rad, more than $20\sigma$ different from the input value of the simulation, but close to the lags of the underlying high-frequency broad Lorentzian. Even when the QPO is $\sim 8$ times stronger than the broad underlying components, the lags with the traditional method are $-0.068 \pm 0.005$, more than $6\sigma$ different from the input value of the simulation, whereas the lags with our method are $-0.104 \pm 0.006$. Only when the QPO is $\sim 25$ times stronger than the broad underlying components, the traditional lags of the QPO are $-0.091 \pm 0.003$ rad, consistent (albeit just at the $3\sigma$ level) with the input value of the simulation. This last result shows emphatically that, if the PS and CS can be decomposed into a number of Lorentzian functions, and a QPO appears in a frequency range in which there are other variability components, the traditional lags of the QPO are biased towards the lags of those components, even when the QPO is significantly stronger than the other variability component.

It is worth noting that, while the values of the lags in the traditional method are significantly different from the input value of the simulation, the errors of the traditional lags are much smaller than those obtained with our method. This is easy to understand from the fact that in our method the errors of the lags of each Lorentzian reflect the uncertainties of the parameters of all the other Lorentzians in the model. Furthermore, each individual Lorentzian component is weaker, and hence less significant, than the sum of all components, which is what one measures when one averages the CS over the FWHM of a QPO. Because of this, if the PS and CS consist truly of the sum of individual Lorentzian components, the traditional method underestimates the relative errors of the Real and Imaginary parts of the CS and hence of the lags \citep[see also][]{Peirano-2022,Alabarta-2022}.

All the above shows that in cases in which a QPO appears on top of a broad component in the PS, while the lags obtained using the traditional method appear to be very precise, they are very inaccurate. Both the inaccuracy and the apparent high precision of the measurements are an artefact of the contamination of the underlying components. On the contrary, while the lags obtained using our method have larger errors than those of the traditional lags, they are significantly more accurate and should therefore be used instead of the traditional lags.

\section{The broadband variability in MAXI J1820+070 fitted with the constant time-lags model}
\label{Appendix-A}

%
%
\begin{figure*}
\centering
\includegraphics[width=0.5\textwidth]{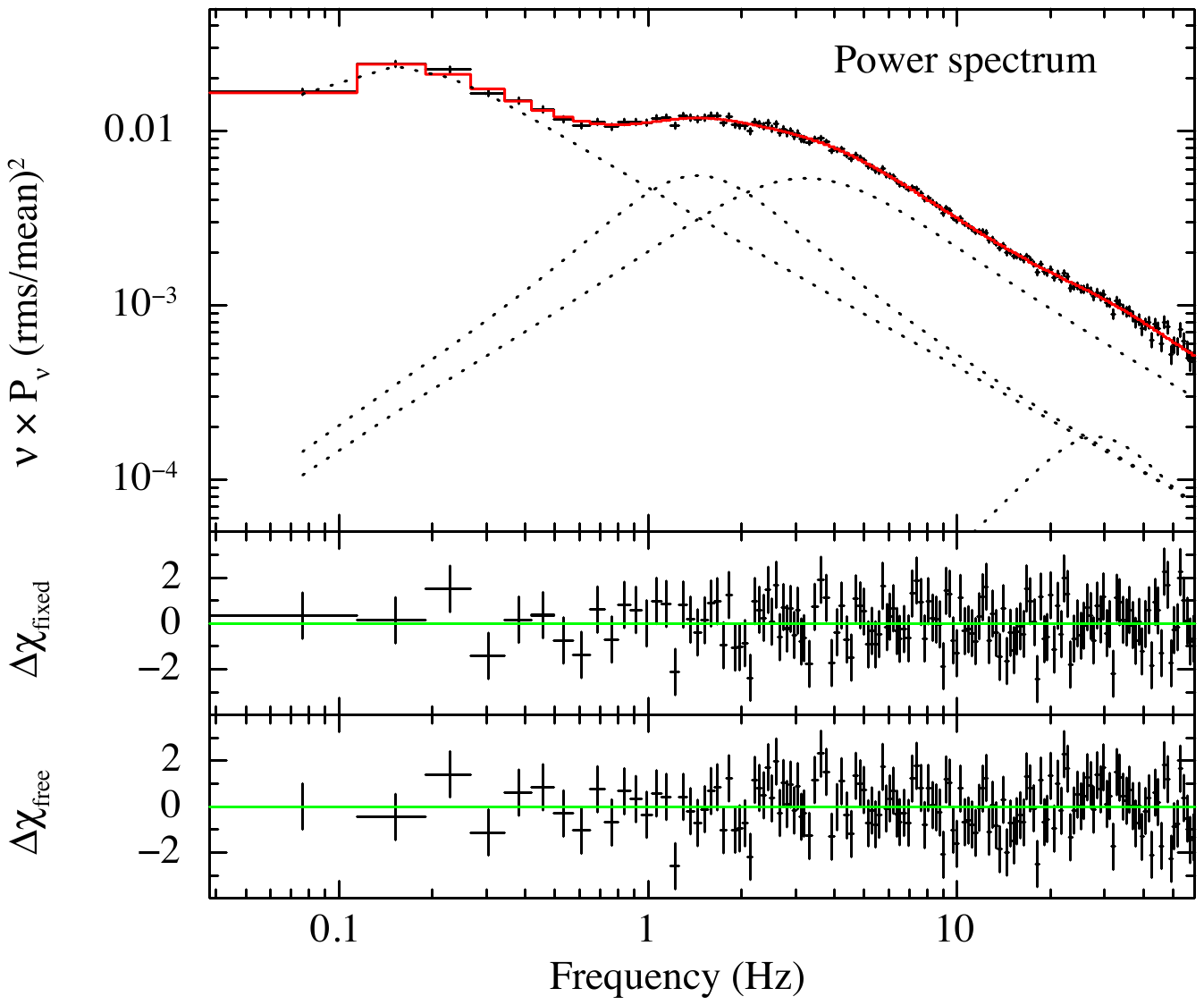}
\hspace{-1cm}
\includegraphics[width=0.5\textwidth]{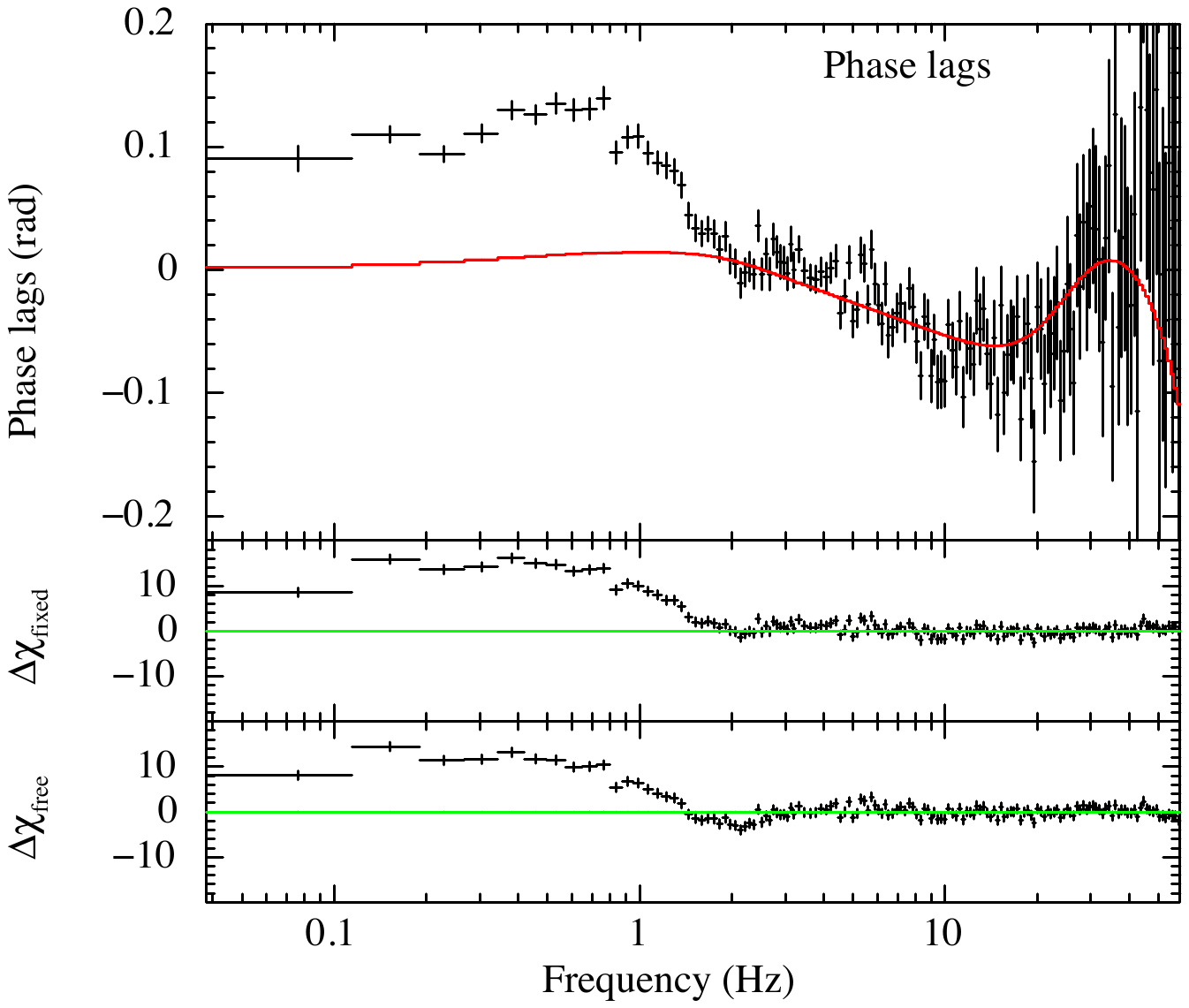}\\
\includegraphics[width=0.5\textwidth]{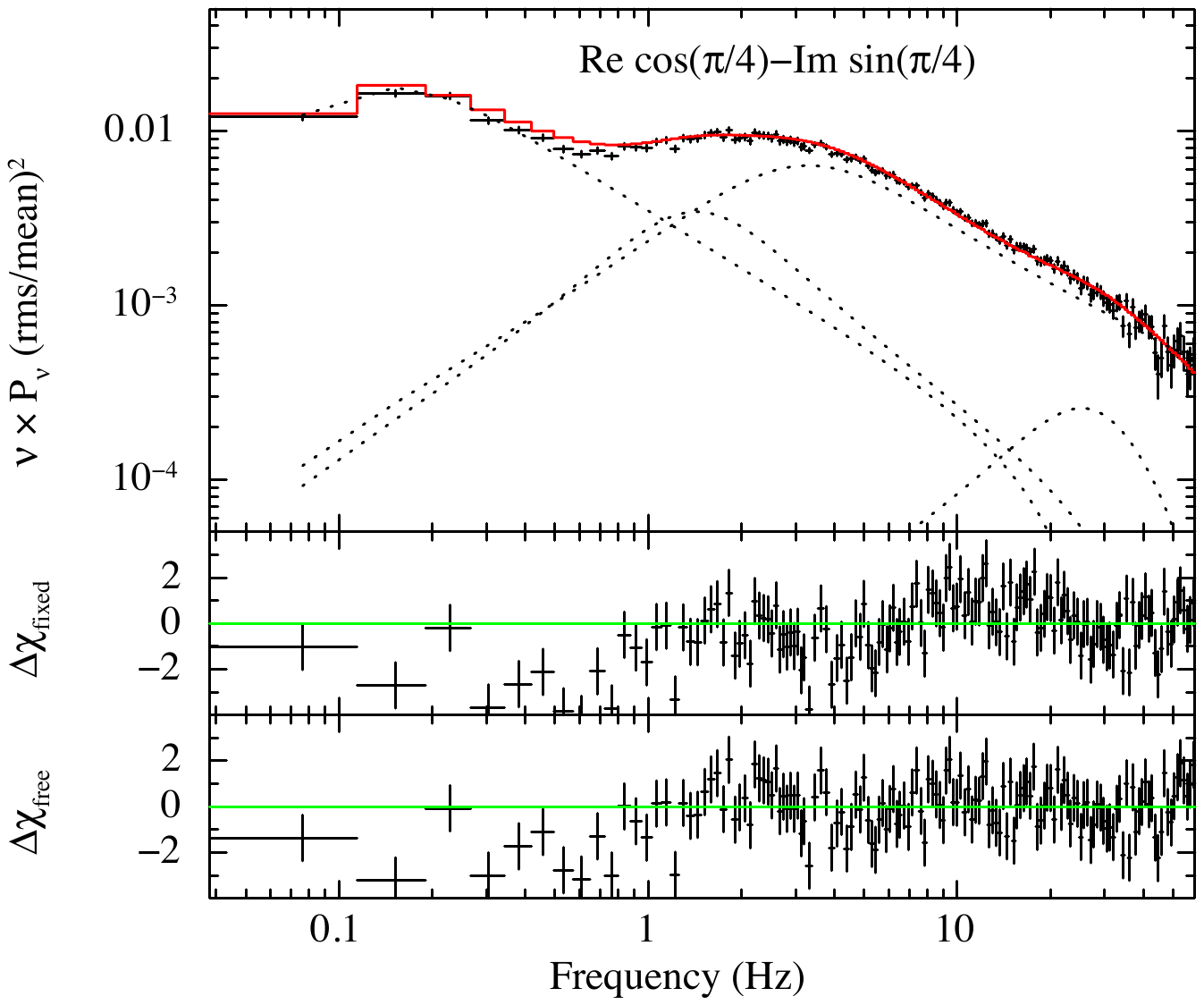}
\hspace{-1cm}
\includegraphics[width=0.5\textwidth]{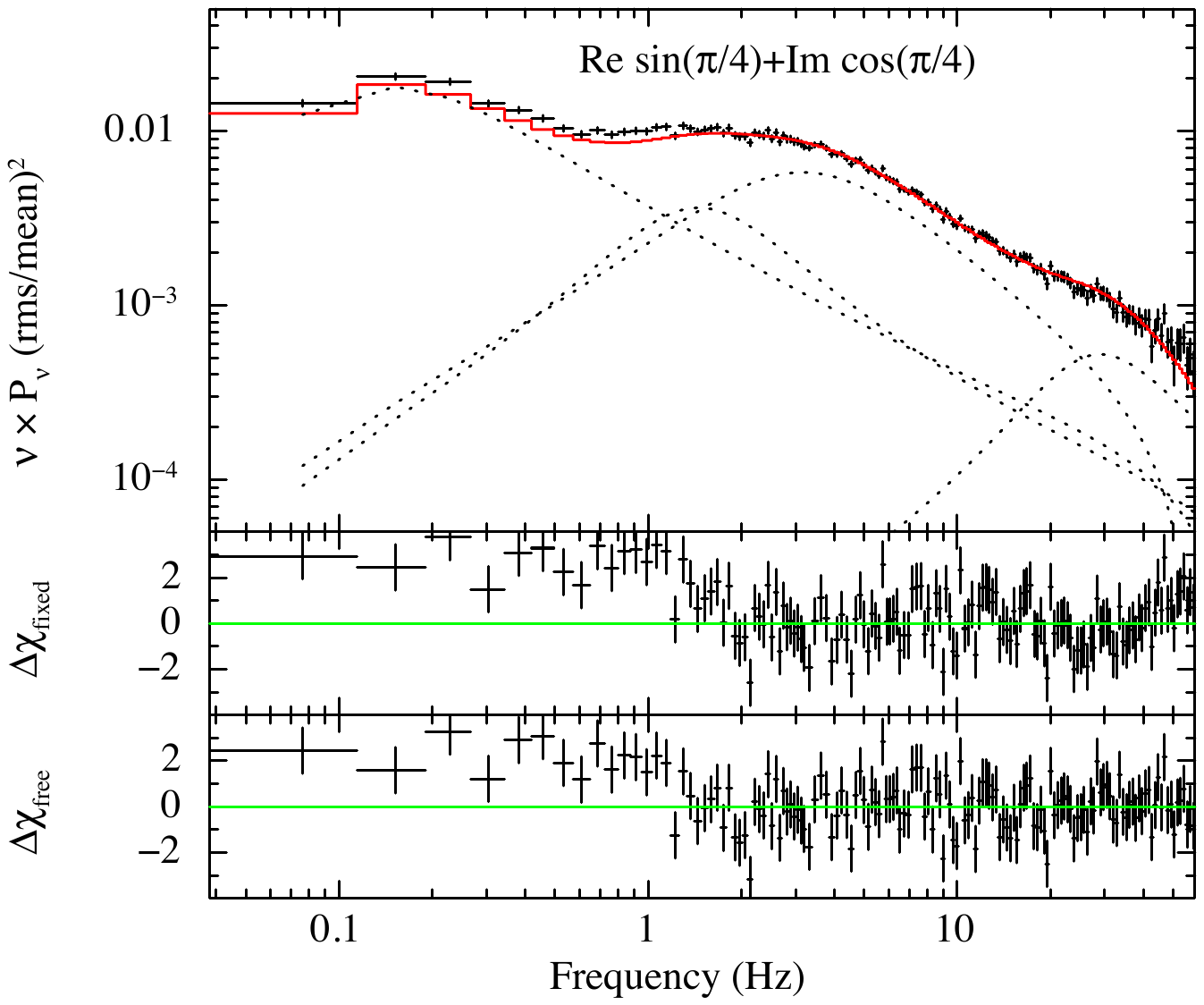}\caption{Same as Figure~\ref{fig4} but now assuming the constant time-lags model (see Table~\ref{tab:definitions} and \S\ref{phaselagmodel}). Notice the difference in the scale of the residuals plot of the lags in this Figure and in Figure~\ref{fig4}.}
\label{figA1}
\end{figure*}

\textit{Constant time-lags model:} As we did in \S\ref{1820}, in Figure~\ref{figA1} we show the fit to the PS and the CS with four Lorentzians, assuming the constant time-lags model (the top left and the two bottom panels). The panel at the top right shows the phase-lag frequency spectrum with the derived model. As in Figure~\ref{fig4}, the top sub-panels show the data and the model and the middle sub-panels show the residuals in the case in which we first fit the PS and then fix the frequency and FWHM of each Lorentzian to the values that we obtain from the PS. The joint fit to the PS and CS with this model is very bad, with $\chi^2=724.9$ for 438 dof, while the derived model of the lags gives $\chi^2=2704.5$ for 150 dof. The bottom sub-panels show the residuals when we let the frequency and the FWHM of each of the four Lorentzian components free but linked across the three spectra. In this case the fit gives $\chi^2 =587.5$ for 430 dof, while the the derived model of the lags gives $\chi^2=1686.6$ for 150 dof.

%
%
\begin{figure*}
\centering
\includegraphics[width=0.5\textwidth]{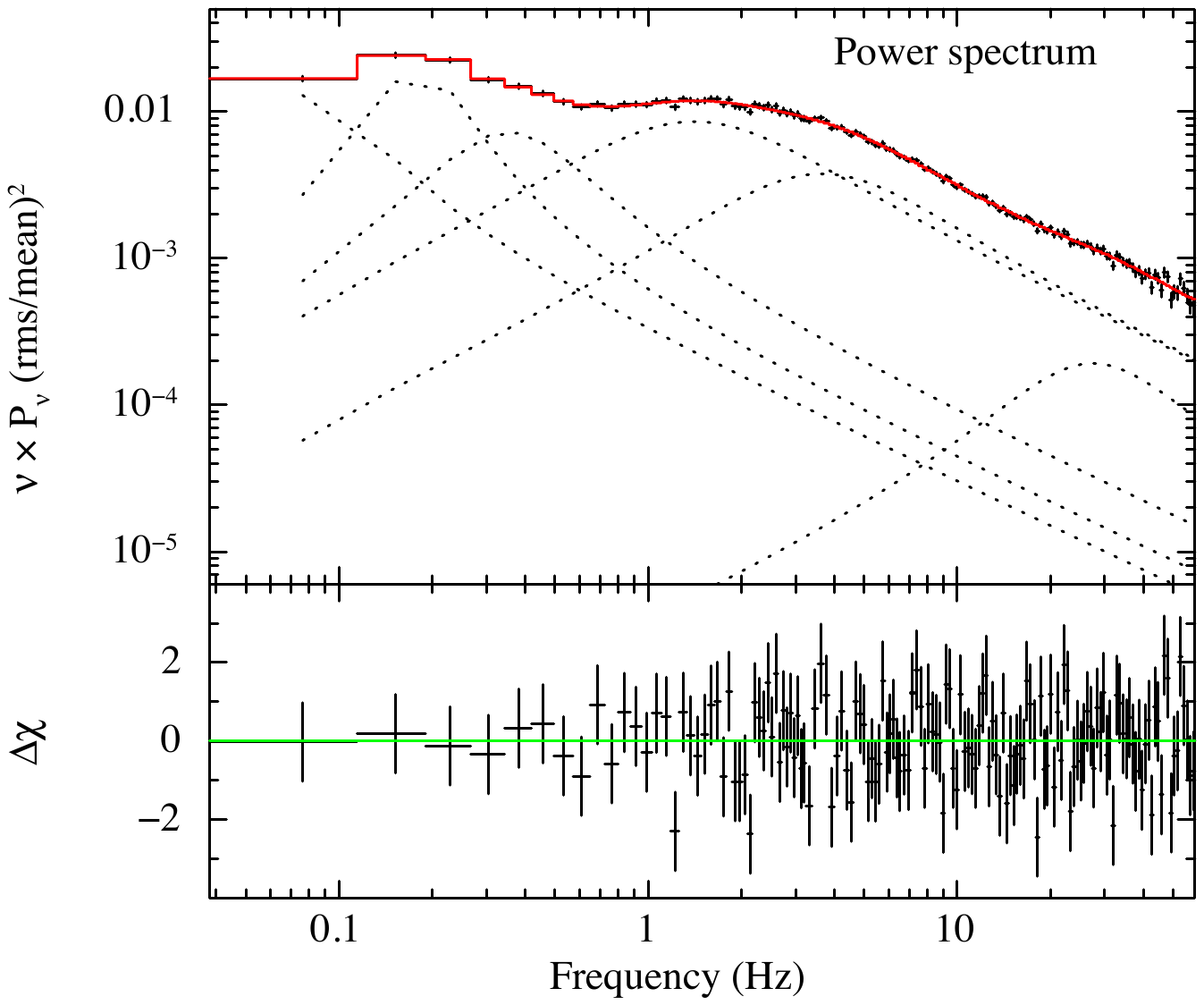}\hspace{-1cm}
\includegraphics[width=0.5\textwidth]{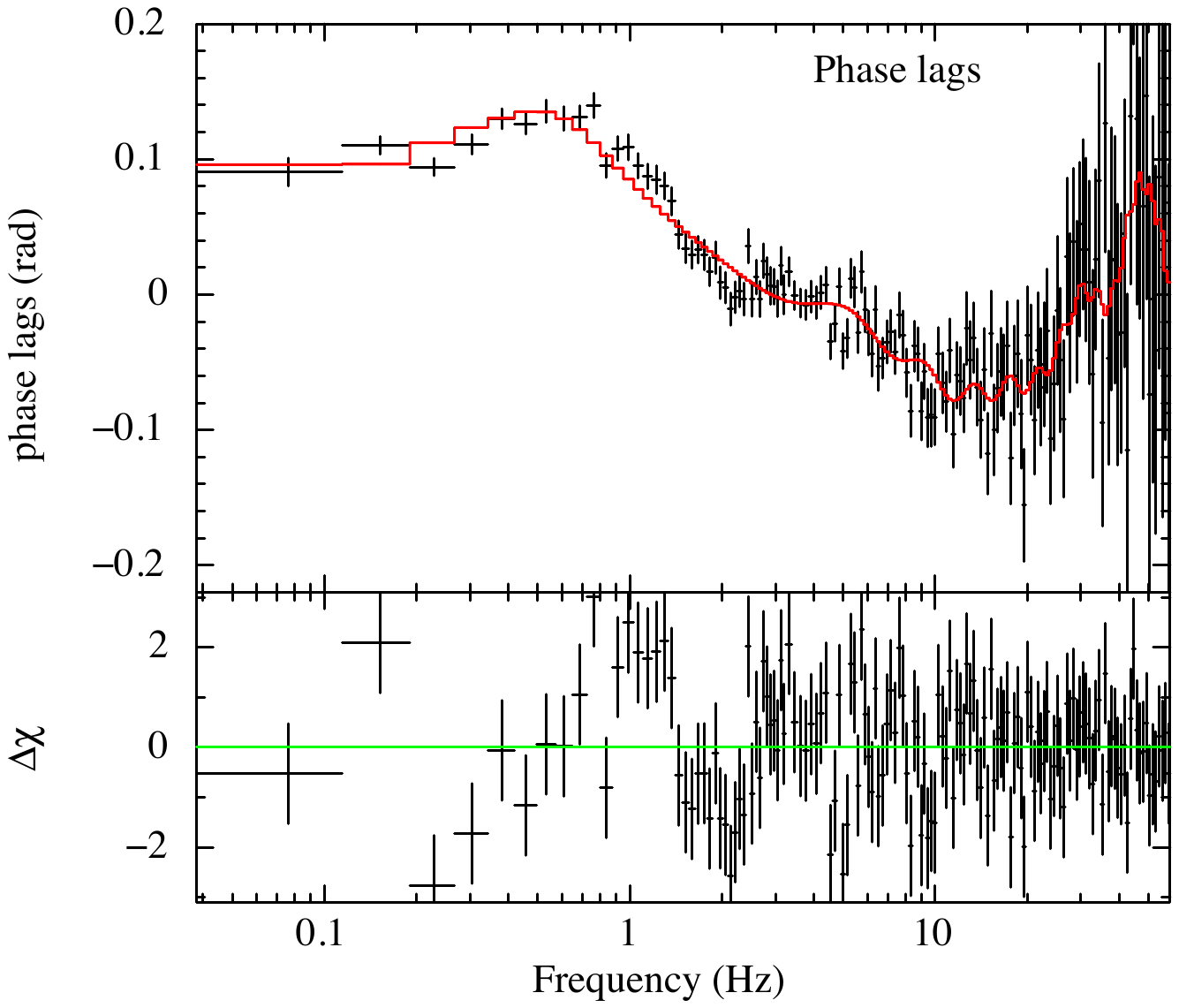}\\
\includegraphics[width=0.5\textwidth]{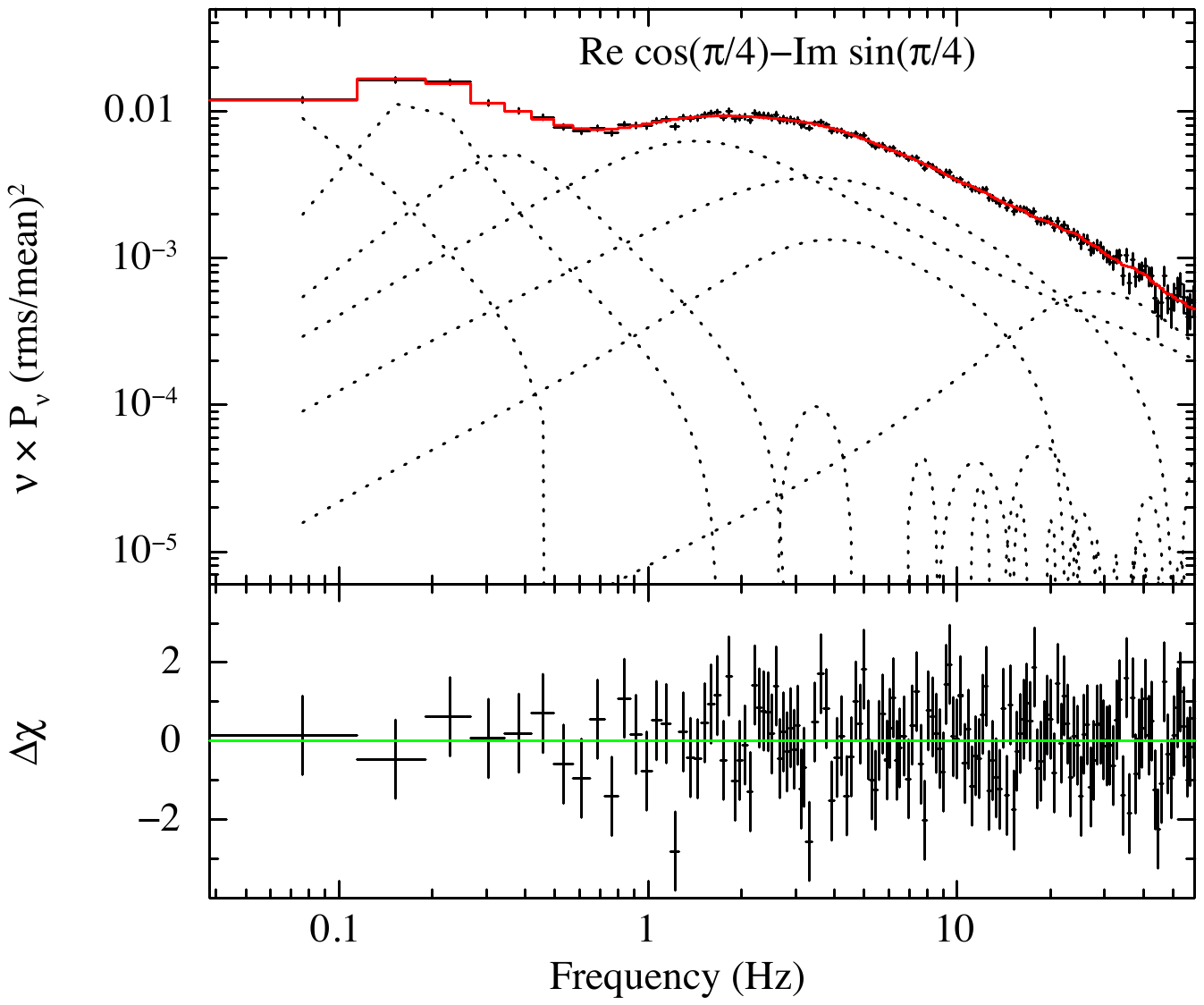}\hspace{-1cm}
\includegraphics[width=0.5\textwidth]{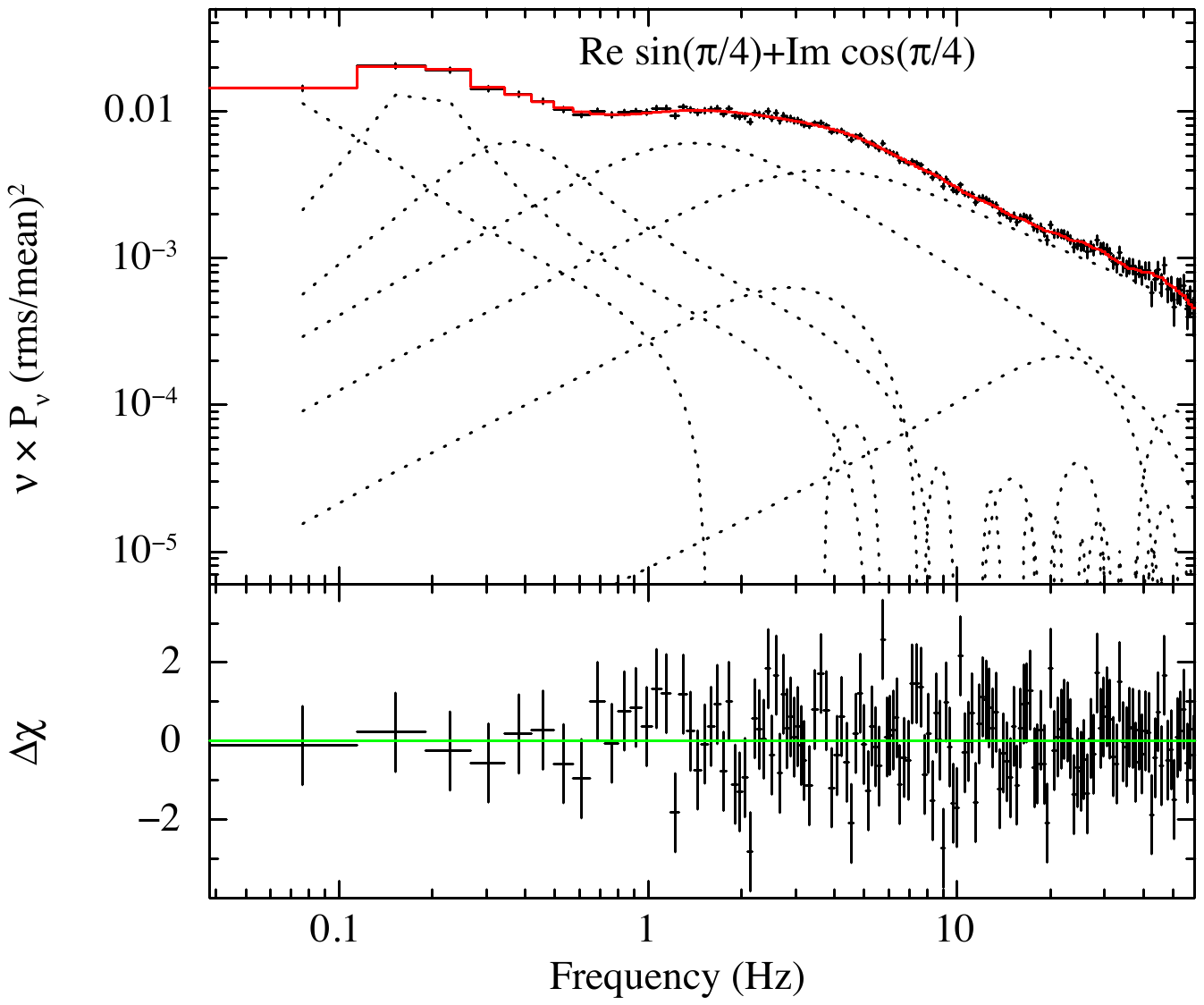}
\caption{Same as Figure~\ref{fig5} but now assuming the constant time-lags model.}
\label{figA2}
\end{figure*}

%
%
\begin{figure}
\centering
\includegraphics[width=0.5\textwidth]{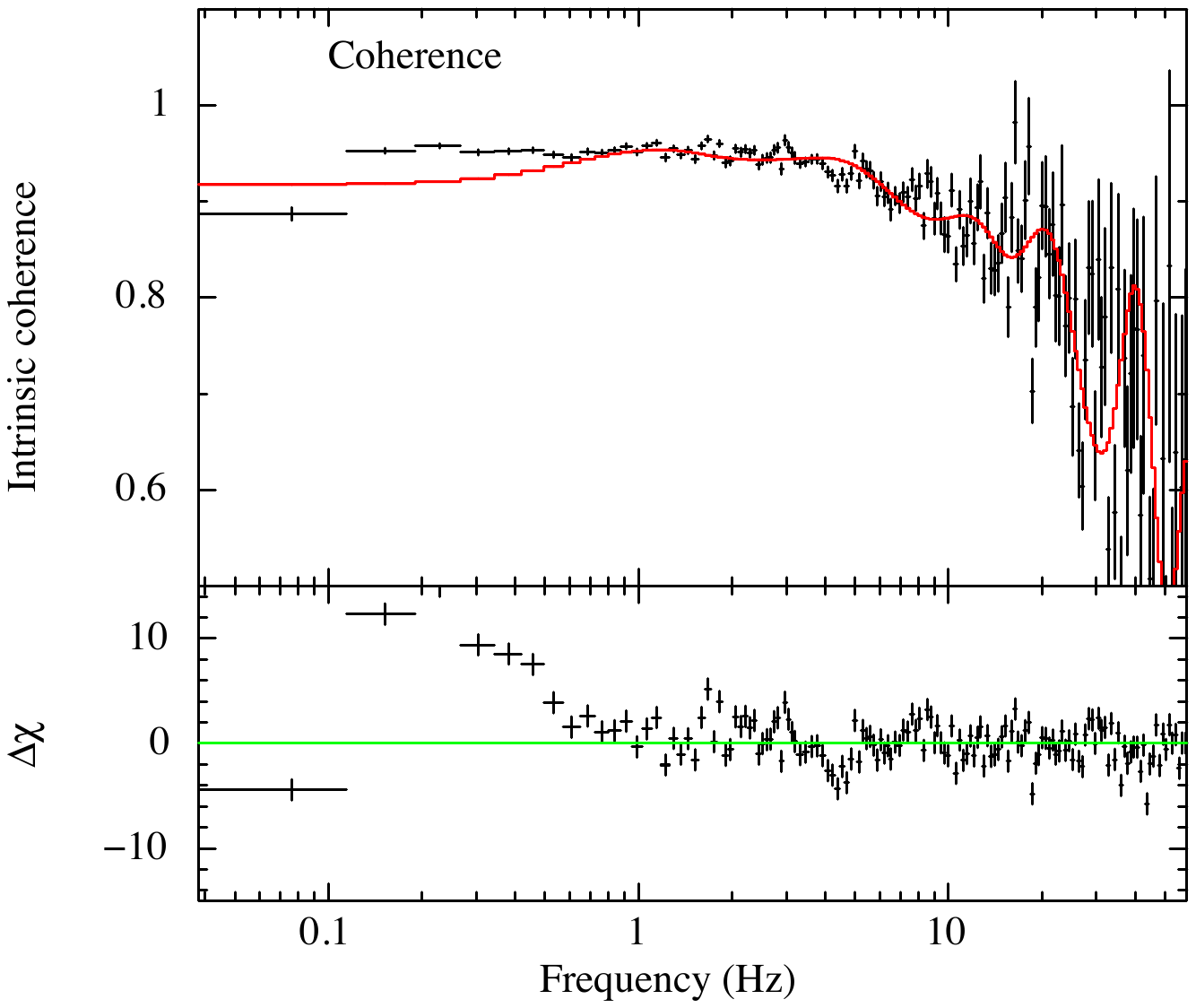}\\
\caption{Intrinsic coherence function of the same observation of MAXI J1820+070. The plot is the same as that in Figure~\ref{fig6}, except that here we assume the constant time-lags model (right panel).
\label{figA3}}
\end{figure}

As in the case of the constant phase-lags models (\S\ref{1820}, Fig.~\ref{fig4}), when the frequency and FWHM of the Lorentzian components are fixed to the values obtained from the PS, the fit to the CS and the derived model for the phase lags show large residuals (significantly larger than those in the case of the constant phase-lags model).

Figure~\ref{figA2} shows the data fitted with a model consisting of seven Lorentzians assuming the constant time-lags model. The panels show the same information as the panels in Figure~\ref{fig5}. The fit gives $\chi^2 = 417$ for 415 dof, marginally better than in the case of the constant phase-lags model. The derived model of the lags gives $\chi^2= 205$ for 150 dof, worse than for the constant phase-lags model.

In this case, some of the components have centroid frequencies and time lags such that the phase lags wrap as explained in \S\ref{phaselagmodel}. For instance, the oscillations in the bottom panels of Figure~\ref{figA2} are due to Lorentzians 1, 2 and 3 with centroid frequencies of $\sim 0.05$ Hz, $\sim 0.165$ Hz and $\sim 0.29$ Hz that have, respectively, time lags of $\sim 242$ ms, $\sim 73$ ms and $\sim 45$ ms. For instance, the phase lags of the first of those Lorentzians rotate at a rate of $\sim 1.5$ rad per Hz such that, as explained in \S\ref{phaselagmodel}, the phase angle of the CV wraps $\sim 14$ times over the $0.07629-60$ Hz frequency range leading to the oscillations seen in the Figure.

We report the best-fitting parameters in Table~\ref{tab:timelags}. Notice that the frequency and FWHM of some of the Lorentzians are different in this case compared to the fit with the constant phase-lags model. This can be seen from comparing the values in Tables~\ref{tab:phaselags} and \ref{tab:timelags}, and looking at the plots in Figures~\ref{fig5} and \ref{figA2}. For this reason, and because the Lorentzians in the Tables are sorted by their centroid frequency, it is not possible to compare the properties of all the Lorentzians obtained from the fits with the two types of models. As in the case of the constant phase-lags model, in this case there is again a Lorentzian (number 5) that is not significant in the PS but is significant in the CS.

In Figure~\ref{figA3} we plot the observed intrinsic coherence with the derived model for the constant time-lags model. The largest deviations appear again at $0.2$ Hz (see \S\ref{1820} for an explanation).

%
%
\begin{table*}
\centering
\caption{Same as Table~\ref{tab:phaselags}, assuming the constant time-lags model}
\label{tab:timelags}
\begin{tabular}{cccccccc}
\hline
\textbf{Component} & \textbf{$\nu_0$ (Hz)} & \textbf{FWHM (Hz)} & \textbf{rms$_{\rm PS}$ (\%)} & \textbf{significance$^{*}$} & \textbf{time lags (ms)} & \textbf{rms$_{\rm CS}$ (\%)} & \textbf{significance$^{*}$}\\ 
\hline
Lorentzian 1 & $  0.05  \pm   0.01 $	& $  0.063  \pm   0.008 $	& $  15.8  \pm   0.8 $	& $   10.0 $	&$ 242  \pm    4 $	& $  16.8  \pm   0.8 $	& $   11.3 $ \\
Lorentzian 2 & $  0.165  \pm   0.005 $	& $  0.12  \pm   0.02 $	& $  14.3  \pm   1.1 $	& $    7.1 $	&$  73  \pm    3 $	& $  14.9  \pm   1.0 $	& $    7.8 $ \\
Lorentzian 3 & $  0.29  \pm   0.02 $	& $  0.39  \pm   0.03 $	& $  10.7  \pm   0.9 $	& $    5.8 $	&$  45  \pm    2 $	& $  11.4  \pm   0.9 $	& $    6.3 $ \\
Lorentzian 4 & $  0.74  \pm   0.06 $	& $  2.4  \pm   0.2 $	& $  14.2  \pm   0.7 $	& $    9.9 $	&$  -1.9  \pm    0.8 $	& $  14.4  \pm   0.9 $	& $    8.7 $ \\
Lorentzian 5 & $  1.3  \pm   0.3 $	& $  6.8  \pm   0.3 $	& $  < 2.2^\dagger     $	& $   < 1  $	&$   2.5  \pm    0.5 $	& $  11.9  \pm   0.9 $	& $    6.5 $ \\
Lorentzian 6 & $  2.1  \pm   0.3 $	& $  5.9  \pm   0.2 $	& $   9.2  \pm   0.9 $	& $    4.7 $	&$ -17  \pm    1 $	& $   5.7  \pm   0.5 $	& $    4.8 $ \\
Lorentzian 7 & $ 21.0  \pm   2.1 $	& $ 35.3  \pm   3.9 $	& $   1.9  \pm   0.1 $	& $    7.0 $	&$  -2.9  \pm    0.9 $	& $   3.3  \pm   0.4 $	& $    4.4 $ \\
\hline
$\chi^2/\mathrm{dof}$ & &&&&&& $417/415$\\
\hline
\multicolumn{8}{l}{\small$^{*}$ In units of $\sigma$.}\\
\multicolumn{8}{l}{\small$^{\dagger}$ 95\% upper limit.}\\
\multicolumn{8}{l}{\small The rms amplitude of the power and cross spectra of each Lorentzian are integrated from zero to infinity.}\\
\end{tabular}
\end{table*}

\section{A QPO in the Imaginary part of the cross spectrum of MAXI J1820+070: the constant time-lags model}
\label{Appendix-B}

In Figure~\ref{figB2} we plot the observed intrinsic coherence with the model derived from the simultaneous fit of the two PS and the CS using the constant time-lags model. As in \S\ref{1820_b}, the significant drop in the observed intrinsic coherence at $\sim 2$ Hz, where there is a QPO that is only present in the Imaginary part of the CS, is well described by the model derived from the fit to the two PS and the CS.

%
%
\begin{figure*}
\centering
\includegraphics[width=0.5\textwidth]{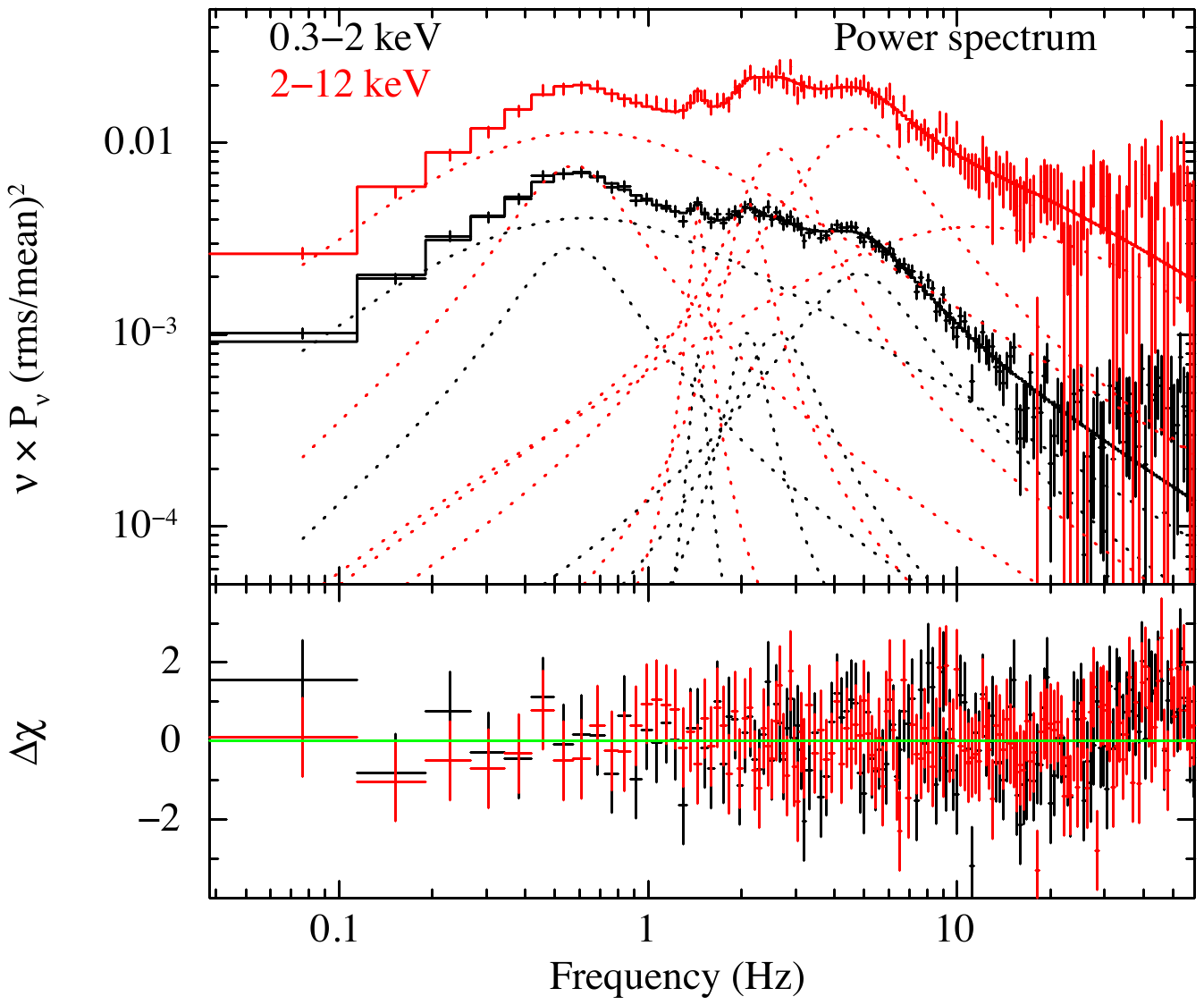}\hspace{-1cm}
\includegraphics[width=0.5\textwidth]{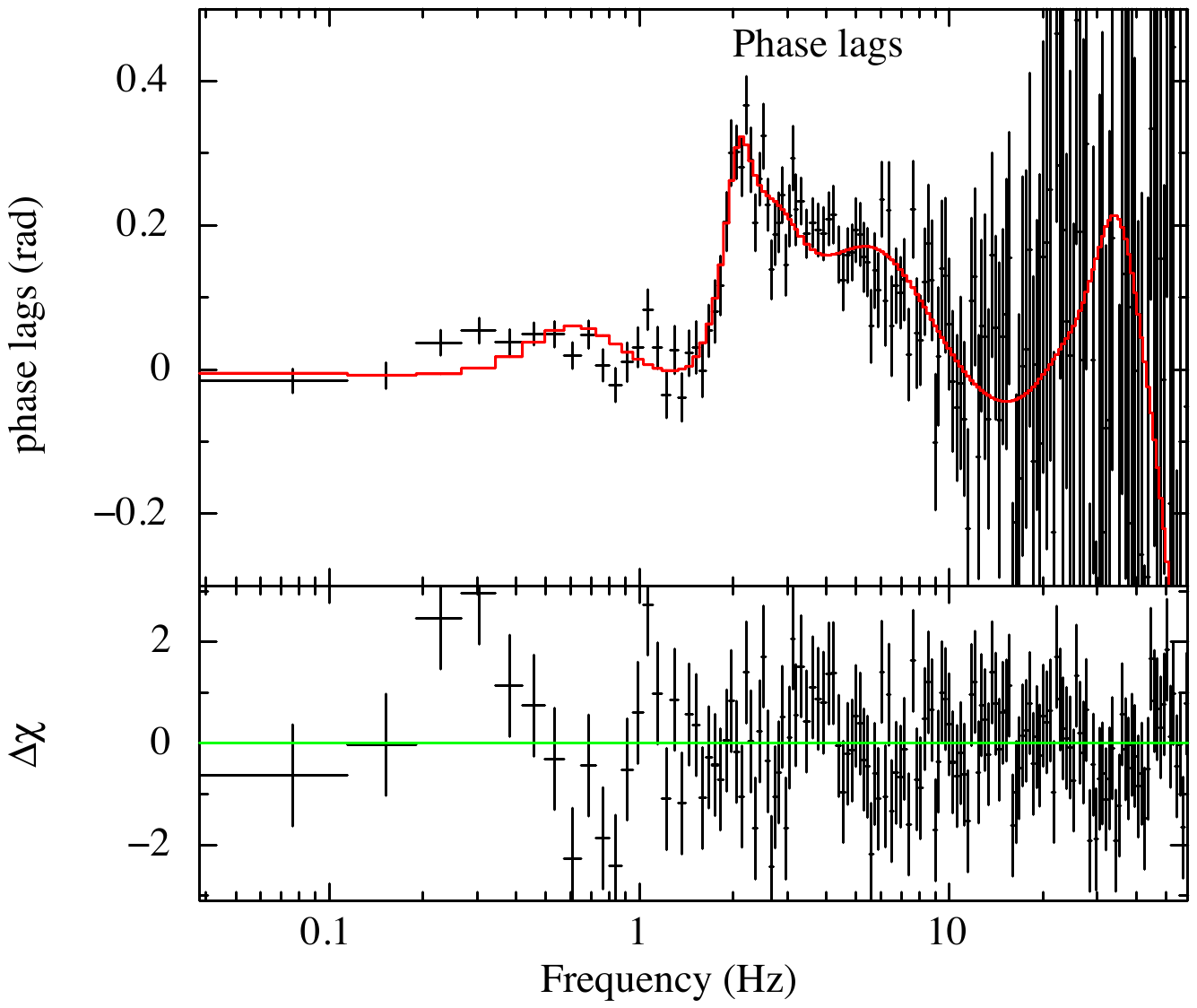}\\
\includegraphics[width=0.5\textwidth]{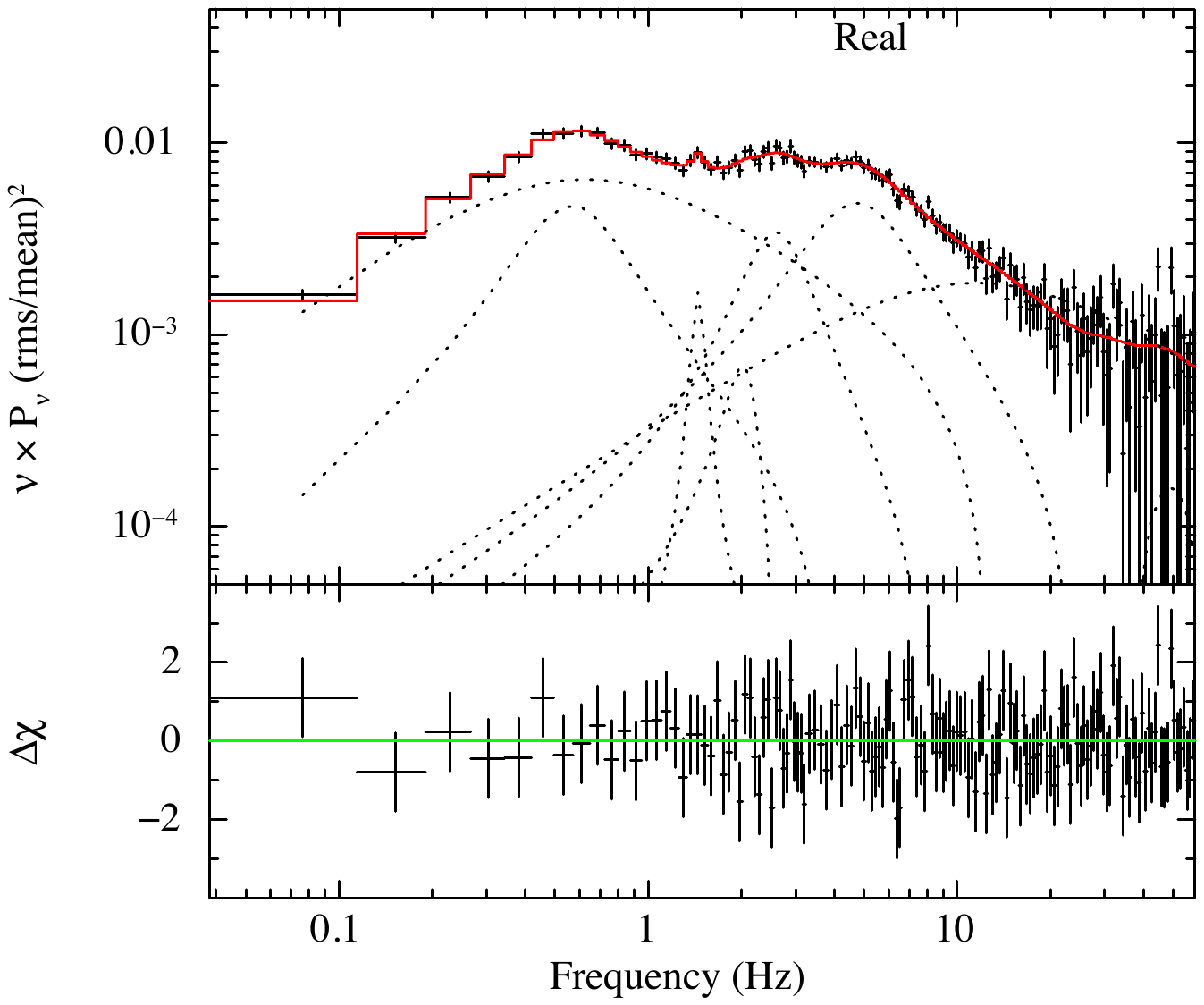}\hspace{-1cm}
\includegraphics[width=0.5\textwidth]{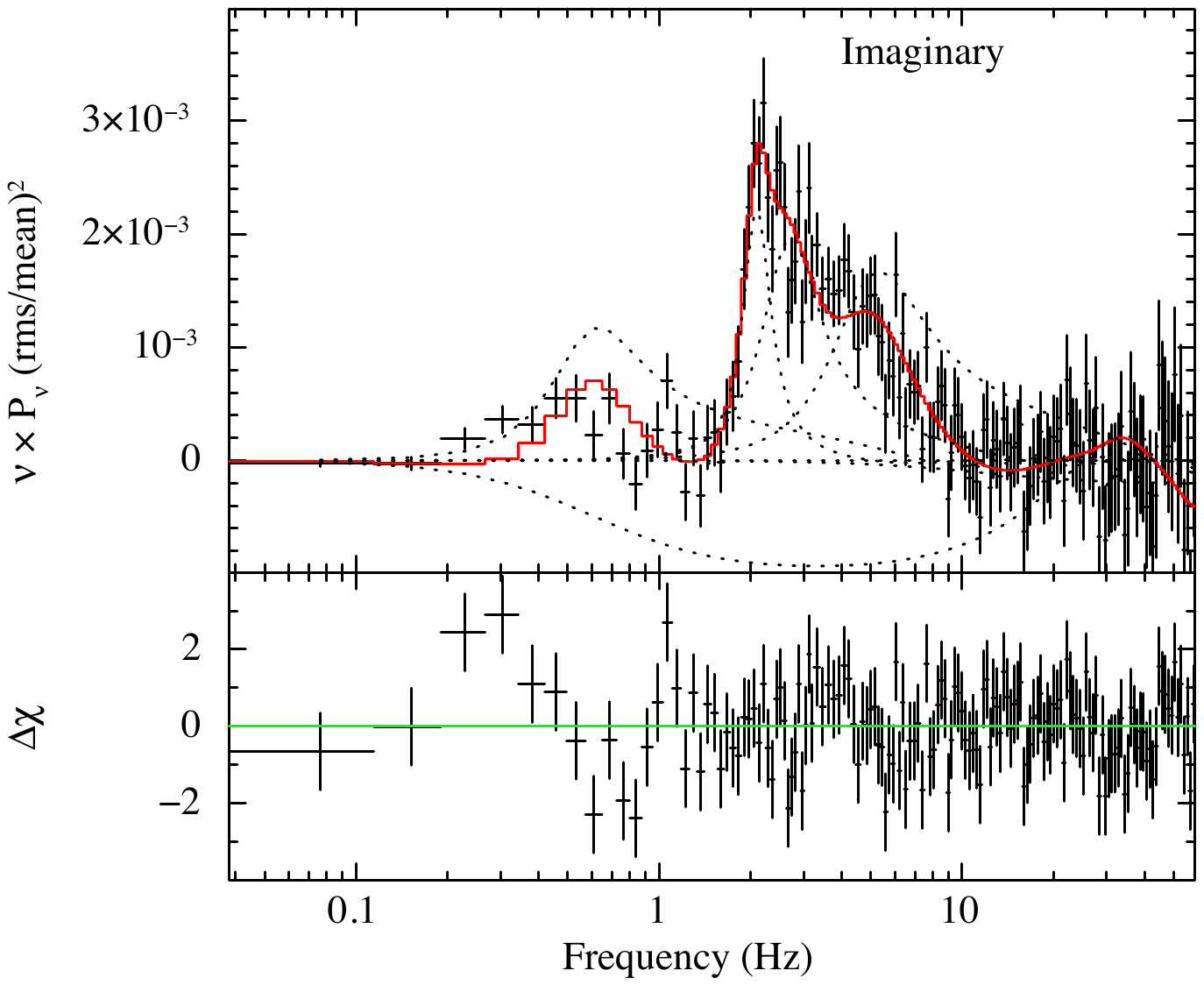}
\caption{Same as Figure~\ref{fig7} but now assuming the constant time-lags model.}
\label{figB1}
\end{figure*}

%
%
\begin{figure}
\centering
\includegraphics[width=0.5\textwidth]{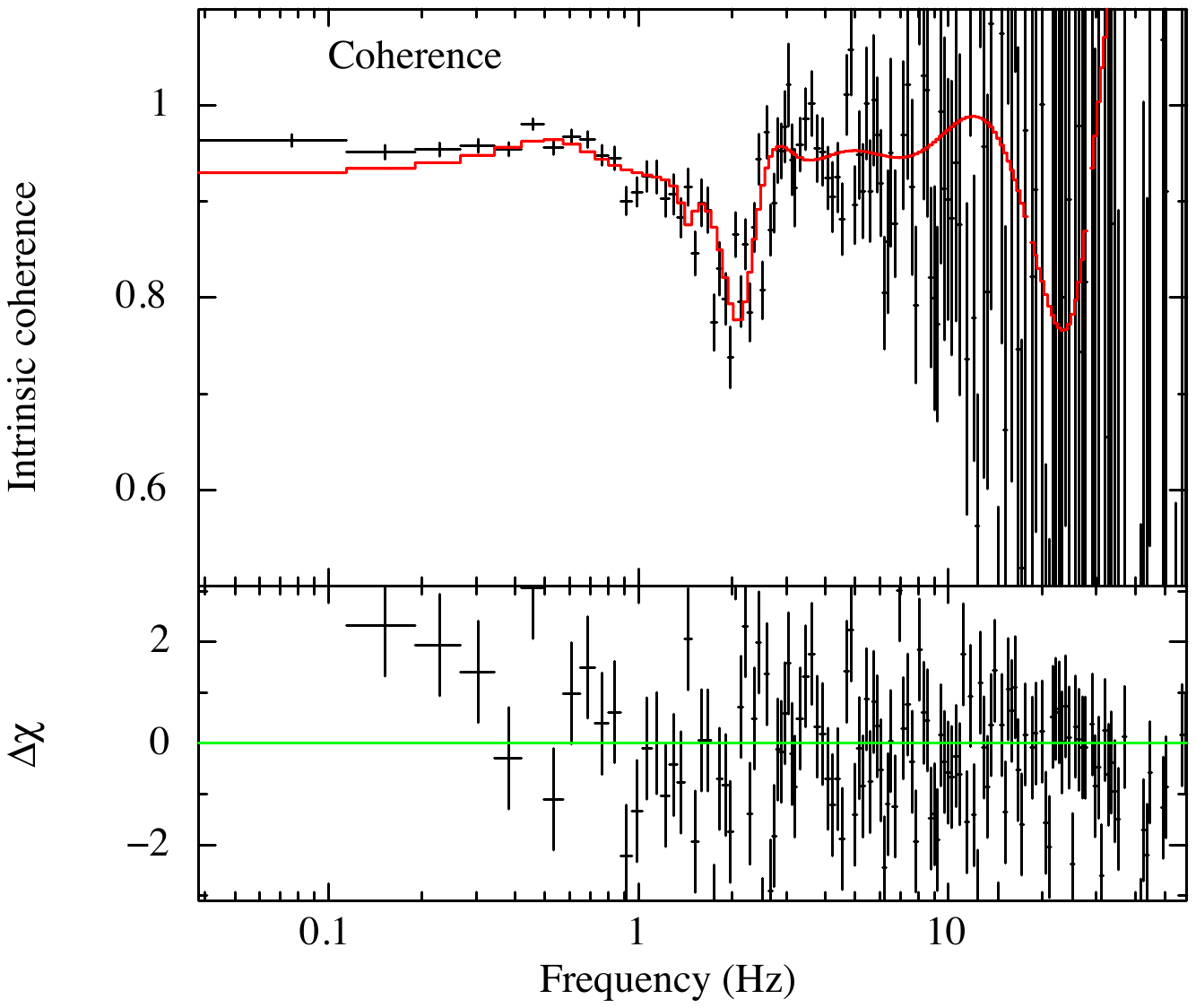}\\
\caption{Intrinsic coherence function of the same observation of MAXI J1820+070 shown in Figure~\ref{figB1} with the derived model obtained from the fit to the PS and CS assuming the constant time-lags model.
}
\label{figB2}
\end{figure}

\bsp	
\label{lastpage}

\end{document}